\documentclass[11pt,letterpaper]{article}

\usepackage[utf8]{inputenc}
\usepackage{amsmath,amsfonts,amssymb,mathtools,amsthm}
\usepackage{xspace}

\usepackage{hyperref}
\usepackage[margin=1in]{geometry}
\usepackage[shortlabels]{enumitem}

\hypersetup{colorlinks=true,allcolors=blue}
\usepackage[nameinlink]{cleveref}

\usepackage{multirow}
\usepackage{thm-restate}
\usepackage{comment}

\usepackage{algorithm}
\usepackage[noend]{algpseudocode}
\usepackage{xcolor}

\usepackage{framed}
\usepackage{url}
\usepackage{appendix}

\usepackage{booktabs}
\usepackage{graphicx}
\usepackage{subcaption}
\usepackage[normalem]{ulem} 
\usepackage{color-edits}

\usepackage[normalem]{ulem}

\crefname{algocf}{Algorithm}{Algorithms}
\Crefname{algocf}{Algorithm}{Algorithms}

\newcommand{\R}{\mathbb{R}}
\newcommand{\Z}{\mathbb Z}

\newcommand{\E}{\mathbb{E}}

\newcommand{\poly}{\operatorname{poly}}
\newcommand{\polylog}{\operatorname{polylog}}
\newcommand{\norm}[1]{\lVert #1\rVert}

\newcommand{\ip}[2]{\langle #1,#2\rangle}

\newcommand{\diam}{\operatorname{diam}}

\newcommand{\eps}{\varepsilon}

\newcommand{\OPT}{\operatorname{opt}}
\newcommand{\opt}{\operatorname{opt}}

\newcommand{\dist}{\operatorname{dist}}

\newcommand{\sgn}{\operatorname{sgn}}

\newcommand{\Nzero}{\mathbb N_0}
\newcommand{\bfv}{\mathbf f}
\newcommand{\bbv}{\mathbf b}
\newcommand{\bF}{\mathbf F}
\newcommand{\btau}{\boldsymbol\tau}

\DeclareMathOperator{\cost}{cost}

\def\compactify{\itemsep=0pt \topsep=0pt \partopsep=0pt \parsep=0pt}

\theoremstyle{plain}
\newtheorem{theorem}{Theorem}[section]
\newtheorem{lemma}[theorem]{Lemma}
\newtheorem{claim}[theorem]{Claim}
\newtheorem*{claim*}{Claim}

\newtheorem{fact}[theorem]{Fact}

\theoremstyle{definition}
\newtheorem{definition}[theorem]{Definition}

\theoremstyle{remark}
\newtheorem{remark}[theorem]{Remark}

\newcommand{\tO}{\tilde{O}}

\DeclareMathOperator{\supp}{supp}

\newcommand{\cX}{\mathcal X}

\newcommand{\cV}{\mathcal V}

\newcommand{\xfar}{x^{\mathrm{far}}}

\newcommand{\ProblemName}[1]{\textsc{#1}}
\newcommand{\kMedian}{\ProblemName{$k$-Median}\xspace}
\newcommand{\kMeans}{\ProblemName{$k$-Means}\xspace}

\newcommand{\kzC}{\ProblemName{$(k,z)$-Clustering}\xspace}

\crefname{theorem}{Theorem}{Theorems}
\crefname{lemma}{Lemma}{Lemmas}
\crefname{claim}{Claim}{Claims}
\crefname{corollary}{Corollary}{Corollaries}
\crefname{definition}{Definition}{Definitions}
\crefname{section}{Section}{Sections}
\crefname{subsection}{Section}{Sections}
\crefname{subsubsection}{Section}{Sections}

\let\varepsilon\varepsilon 

\title{Adversarially Robust Streaming Algorithms for $(k, z)$-Clustering}
\author{\hspace{0.5in}
Chengzhi Guo\thanks{Peking University. Email: \texttt{2501112030@stu.pku.edu.cn}.} \and
Shaofeng H.C. Jiang\thanks{Peking University. Email: \texttt{shaofeng.jiang@pku.edu.cn}.} \and
Robert Krauthgamer\thanks{Weizmann Institute of Science. 
    The Harry Weinrebe Professorial Chair of Computer Science.
    Work partially supported by the Israel Science Foundation grant \#1336/23.
    Email: \texttt{robert.krauthgamer@weizmann.ac.il}
}
\hspace{0.5in}
\and
David P. Woodruff\thanks{Carnegie Mellon University and Google Research. Email: \texttt{dwoodruf@cs.cmu.edu}. Author is supported in part by Office of Naval Research award number N000142112647 and a Simons Investigator Award. } 
\and
Shenghao Xie\thanks{Carnegie Mellon University. Email: \texttt{shenghax@cs.cmu.edu}. Author is supported in part by Office of Naval Research award number N000142112647 and a Simons Investigator Award. } 
\and
Samson Zhou\thanks{Texas A\&M University. Email: \texttt{samsonzhou@gmail.com}. 
Author is supported in part by NSF CCF-2335411.}
}

\begin{document}

\begingroup
\pagestyle{empty}
\begin{titlepage}
    \maketitle
    \thispagestyle{empty}
    \begin{abstract}
We design adversarially robust algorithms for dynamic geometric streams that $(1 + \epsilon)$-approximate \kMedian and \kMeans in Euclidean space $\mathbb{R}^d$, using $\tO(kd \poly(\epsilon^{-1}))$ space. 
This result significantly improves a recent $O(d\log n)$-approximation for \kMedian [Gribelyuk, Lin, Woodruff, Yu, and Zhou, FOCS 2026], and is furthermore the first algorithm whose space is linear in $kd$, even in the oblivious (non-adaptive) setting. 
We develop several new techniques which may be of independent interest, including a new data-oblivious Euclidean primitive called hierarchical Lipschitz  assignment, a robust dimensionality reduction for clustering, and a robust streaming algorithm for inner-product queries.
\end{abstract}

     \tableofcontents
\end{titlepage}
\endgroup

\section{Introduction}
\label{sec:intro} 
Clustering problems such as \kMedian and \kMeans are fundamental to computer science and data analysis, and therefore algorithms for these problems have been studied extensively. 
We consider \kzC in Euclidean spaces $\mathbb{R}^d$ (for fixed integer $z \geq 1$), where given a dataset $P \subseteq \mathbb{R}^d$ and a parameter $k$, one seeks a center set $C \subseteq \mathbb{R}^d, |C|=k$ that minimizes the objective
\begin{align*}
\cost(P, C) := \sum_{x \in P} (\dist(x, C))^z ,
\end{align*}
where $\dist(x,y) := \|x-y\|$ denotes Euclidean distance and $\dist(x, C) := \min_{c \in C} \dist(x, c)$.
This formulation captures the famous problems of \kMeans $(z = 2)$ and \kMedian $(z = 1)$.

In the model of dynamic geometric streams, introduced by Indyk~\cite{stoc/Indyk04}, the input $P$ is presented as a stream of point \emph{insertions and deletions} from the integer grid $[\Delta]^d$, and the algorithm has to produce an output after each update. 
As usual, we denote $n=|P|$ and assume for simplicity that $\Delta=\poly(n)$, and our primary goal is to design space-efficient algorithms. 
In the clustering literature, the output of a streaming algorithm is usually an \emph{$\varepsilon$-coreset}~\cite{stoc/Har-PeledM04}, which is actually stronger than reporting a solution. 
A coreset is a (reweighted) subset $P'\subseteq P$, such that the cost of every center set $C$ evaluated on $P'$ is a $(1 + \varepsilon)$-approximation to that on the dataset $P$. 
Such a summary often provides a stronger guarantee than reporting a single approximate solution, since a coreset can be used to evaluate arbitrary center sets and recover an approximately optimal one. 

In this work, we focus on streaming algorithms for \kzC that are \emph{adversarially robust}, meaning that each stream update can depend on the algorithm's outputs so far, i.e., an adversary can choose the input to be \emph{adaptive} to the output stream. 
This is in contrast to the classical model of an \emph{oblivious adversary} that chooses the entire input stream before the algorithm is run. 
The adaptive-adversary setting is important in applications where decisions based on the algorithmic outputs influence how future data is generated or collected. 
For example, placing temporary facilities at the reported cluster centers can influence future customer demand, making the input stream adaptive. 
This setting was introduced to the streaming model by \cite{pods/Ben-EliezerY20,jacm/Ben-EliezerJWY22}, and is known to have fundamental importance in other computational models, such as dynamic algorithms~\cite{Wajc20,BeimelKMNSS22,BravermanDPSW25} and data structures~\cite{CherapanamjeriN20,CohenLNSSS22,CherapanamjeriSWZZZ23,AhmadianC24}. 

Designing adversarially robust streaming algorithms is a central challenge that has attracted substantial attention~\cite{MitrovicBNTC17,AvdiukhinMYZ19,pods/Ben-EliezerY20,BravermanHMSSZ21,AjtaiBJSSWZ22,jacm/Ben-EliezerJWY22,Ben-EliezerEO22,BeimelKMNSS22,ChakrabartiGS22,AssadiCGS23,PengR23,WoodruffZZ23b,FengJW24,WoodruffZ24,Ben-EliezerOS26,GalKSY26,LinSWXZ26}; see~\cite{WoodruffZ26} for a survey.
A major obstacle is that linear sketching, a powerful technique for dynamic streams, faces fundamental limitations against adaptive adversaries, even for basic estimation problems~\cite{stoc/HardtW13,GribelyukLWYZ24,GribelyukLWYZ25}.
Consequently, the problem of frequency moment estimation, which is foundational to the area of streaming algorithms, has been a long-standing open question until a recent breakthrough mitigated it via a novel non-linear sketching approach~\cite{stoc/GLWYZ26,focs/GLWYZ26}. 
Our work on clustering, which is a geometric optimization problem, can be viewed as a natural next step of going beyond frequency statistics. 

For clustering problems, there is currently a huge gap between data streams generated by an oblivious adversary and an adaptive adversary. 
For oblivious adversaries, the rich line of results on coresets~\cite{stoc/Har-PeledM04,dcg/Har-PeledK07,siamcomp/Chen09,stoc/FeldmanL11,siamcomp/FeldmanSS20,stoc/Cohen-AddadSS21,stoc/Cohen-AddadLSS22,focs/0001CPSS24,stoc/Huang0024} readily implies algorithms for insertion-only streams via the merge-and-reduce framework~\cite{stoc/Har-PeledM04}. 
These streaming algorithms output an $\varepsilon$-coresets using space that is comparable to the coreset size, which is nearly optimal. 
For dynamic streams (which include deletion of points), the merge-and-reduce framework is inapplicable, yet both \kMedian and \kMeans have algorithms, via more ad-hoc methods, that output $\varepsilon$-coresets using $O(k \poly(\varepsilon^{-1}d \log \Delta))$ space~\cite{icml/BravermanFLSY17,SongYZ18}. 
For the adaptive-adversary setting there is only one known algorithm, that achieves a much larger approximation ratio $O(d \log \Delta)$ using $\poly(kd \log \Delta)$ space, works only for \kMedian, and only finds a solution rather than a coreset~\cite{focs/GLWYZ26}. 

Bridging this gap between oblivious and adaptive adversaries introduces significant technical difficulties.
Previous clustering algorithms for oblivious-adversary streaming return not only a center set but also a coreset~\cite{icml/BravermanFLSY17,SongYZ18}.
This is not possible with an adaptive adversary, as it can delete the coreset points and force the algorithm to eventually report all the data points, which clearly requires linear (in $n$) space, and therefore we must seek an alternative to the coreset approach. 
Furthermore, existing methods for clustering in dynamic streams are based on linear sketches and sampling, which usually cannot work with adaptive adversary. 
Hence, we need fundamentally new techniques.

\subsection{Results}

We provide (in \Cref{thm:main1}) an adversarially robust streaming algorithm
that achieves $(1 + \varepsilon)$-approximation using $\tilde{O}(kd \poly(\varepsilon^{-1}))$ space
for both \kMedian and \kMeans. 
This guarantee follows immediately from our main result (\Cref{thm:main_intro}),
which gives an adversarially robust streaming algorithm for maintaining a \emph{cost oracle}. 
Informally, a cost oracle is a data structure that estimates the clustering cost of any center set $C$ given as a query;
see \Cref{def:cost-oracles} for the formal definition. 
A cost oracle is clearly stronger than an approximate solution (a center set),
as achieved in \Cref{thm:main1}, because one can enumerate all feasible solutions,
query the oracle for their cost, and report the one of minimum cost.
Throughout, $\tO(\cdot)$ hides logarithmic factors in all parameters,
i.e., $\poly\log(\varepsilon^{-1}kdm \Delta)$ where $m$ is the stream length
(hence $n\leq m$).

\begin{theorem}
\label{thm:main1}
There are adversarially robust algorithms for \kMedian and \kMeans
that process dynamic geometric streams using $\tO(kd \poly(\varepsilon^{-1}))$ space
and maintain a center set and a cost estimate, such that with constant probability, these achieve $(1 + \varepsilon)$-approximation at all times.
\end{theorem}

\begin{definition}[$\varepsilon$-Oracle]\label{def:cost-oracles}
Given $\varepsilon\in(0,1)$, an \emph{$\varepsilon$-oracle} for a dataset $P \subseteq \mathbb{R}^d$ is a function $\widehat{\cost}(\cdot)$ defined over all center sets $C$, such that
\[\forall C \subseteq \mathbb{R}^d, |C| \leq k,\qquad\widehat{\cost}(C) \in (1\pm \varepsilon)\cdot \cost(P, C).\]
\end{definition}

This notion has been considered before, e.g.\ in~\cite{tit/ZhuTHH25}, and is tightly related to the well-known notion of $\varepsilon$-coreset~\cite{stoc/Har-PeledM04}: every $\varepsilon$-coreset is also an $\varepsilon$-oracle, and the main difference is that an $\varepsilon$-coreset must correspond to a (reweighted) subset $S\subseteq P$ and satisfy $\widehat{\cost}(C) := \cost(S, C)$, whereas an $\varepsilon$-oracle can be any function $\widehat{\cost}(\cdot)$. 
As mentioned above, this is a necessary relaxation since returning a coreset, or a subset of data points in general, cannot be robust to an adaptive adversary, unless its space is linear.

\begin{theorem}[See \Cref{thm:main}]
\label{thm:main_intro}
There are adversarially robust algorithms for \kMedian and \kMeans
that process dynamic geometric streams using $\tO(kd \poly(\varepsilon^{-1}))$ space
and maintain a function $\widehat{\cost}$ such that, with constant probability, $\widehat{\cost}$ is an $\varepsilon$-oracle for the input stream at all times.
\end{theorem}

\begin{table}[t]
      \centering
      \caption{Algorithms for \kzC in dynamic geometric streams}
      \label{tab:result}
      \begin{tabular}{lllll}
           \toprule
           adversary & problem & guarantee & space & reference \\
           \midrule
           \multirow{2}{*}{oblivious} & \kMedian & $\varepsilon$-coreset & $\tO(k d^4 \varepsilon^{-2})$ &  \cite{icml/BravermanFLSY17} \\
            & \kMeans & $\varepsilon$-coreset & $\tO(k \poly(d \varepsilon^{-1}))$ &  \cite{SongYZ18} \\
           \midrule
           \multirow{4}{*}{adaptive} & \kMedian & $O(d \log \Delta)$-oracle & $\tO(k \poly(d))$ &  \cite{focs/GLWYZ26} \\
            &\kMedian & $\varepsilon$-oracle & $\tO(kd \poly(\varepsilon^{-1}))$ &  \Cref{thm:main} \\
            & \kMeans & $\varepsilon$-oracle & $\tO(kd \poly(\varepsilon^{-1}))$ &  \Cref{thm:main} \\
            & \kzC    & $\varepsilon$-oracle & $\tO(k (d \varepsilon^{-1})^{O(z)})$ &  \Cref{thm:main} \\
           \bottomrule
      \end{tabular}
\end{table}

\paragraph{Comparison with Prior Results.}
Overall, our algorithms bridge the gap between oblivious and adaptive adversaries for streaming algorithms for clustering. 
Moreover, our bounds improve in a useful parameter regime even on the existing bounds for the oblivious setting. 
Specifically, our result significantly improves over the known $O(d \log \Delta)$-approximation using $\poly(kd \log \Delta)$ space for \kMedian in the adaptive setting~\cite{focs/GLWYZ26}, and is also the first to achieve a nontrivial approximation for \kMeans. 
Our space bound has linear dependence on $kd$, which is optimal because the description of an $\varepsilon$-oracle for any finite approximation $\varepsilon$, already requires $\Omega(kd)$ space~\cite{tit/ZhuTHH25}.
Our result is the first to achieve space linear in $kd$ for maintaining an $\varepsilon$-oracle or coreset in dynamic streams, even in the oblivious setting. 
Prior algorithms~\cite{icml/BravermanFLSY17,SongYZ18}, however, have a better polynomial dependence on $\varepsilon^{-1}$.
Finally, our result works also for all $z \geq 1$ (as in \Cref{thm:main}), although for $z > 2$ the dependence on $d$ becomes $d^{O(z)}$, as opposed to linear for $z \leq 2$, while the dependence on $k$ remains linear. 
We summarize these results and comparisons in \Cref{tab:result}.

\paragraph{Polynomial-Time Implementation.}
For \kMedian, we devise also a polynomial-time adversarially robust streaming algorithm that achieves constant-factor approximation using space nearly linear in both $k$ and $d$. 
A constant-factor approximation is a natural target for time-efficient algorithms, since in high dimension, achieving $(1+\varepsilon)$-approximation for sufficiently small fixed $\varepsilon>0$ is NP-hard, even in the offline setting~\cite{socg/AwasthiCKS15,cohen2022johnson}.

\begin{theorem}[See \Cref{poly:thm:centers}]
\label{thm:poly_intro}
There is an adversarially robust algorithm for \kMedian that processes dynamic geometric streams 
using $\tO(kd)$ space and $\poly(kd\log(m\Delta))$ update time, 
and with constant probability, it maintain an $O(1)$-approximate center set at all times.
\end{theorem}

\subsection{Technical Overview}
\label{sec:tech_overview}
At a high level, our algorithm consists of three main ingredients, which constitute our main technical contributions: 
(1) A new data-oblivious deterministic geometric primitive in Euclidean space, called \emph{hierarchical Lipschitz assignment}. 
Serving as an important backbone for both our algorithm and analysis, 
this primitive generalizes standard tools, such as randomly-shifted quadtrees, 
and connects closely to separating decompositions~\cite{focs/Bartal96}. 
(2) An adversarially robust streaming algorithm for evaluating inner-product queries, which is an important streaming primitive. 
(3) An adversarially robust dimension reduction technique tailored for clustering, which circumvents a well-known lower bound against general robust dimension reduction~\cite{stoc/HardtW13}. 
Metric decomposition, inner-product estimation, and dimensionality reduction are fundamental tools for classical (non-robust) algorithms with a wide range of applications, c.f., \Cref{sec:related}, 
and we thus expect our robust counterparts to be of independent interest as modular building blocks across geometric streaming. 
Below, we outline how these three pieces fit together before examining each technique in detail.

Let $\cX := [\Delta]^d$ be the domain of the input points. 
The whole approach is based on a fairly straightforward idea:
let $\mu \in \mathbb{R}_{\geq 0}^{\cX}$ be the frequency vector that encodes the dataset $P$, and define $\cost(\mu, C) := \sum_{x \in \cX} \mu_x \dist^z(x, C) = \cost(P, C)$, which we can write it as an inner product by
$\cost(\mu, C) = \langle\mu, \dist^z(\cdot, C)\rangle$, 
where the two vectors lie in $\mathbb{R}^{\cX}$. 
This inner-product formulation opens the door to exploring sufficient conditions for estimating it.

\paragraph{New Technique: Robust Inner-Product Estimation.}
We give an adversarially robust streaming algorithm for estimating this inner product. 
Specifically, consider a fixed family of query vectors $\mathcal{G}$ where $\|g\|_\infty \leq 1$ for all $g\in \mathcal{G}$. 
The input is an adaptive sequence of $m$ updates, which defines a sequence of frequency vectors $a_0 = \vec{0}, a_1, \ldots, a_m$, and at each time step $t \in [m]$, the algorithm returns a function $\widehat{I}_t$ such that $|\widehat{I}_t(g) - \langle a_t, g \rangle | \leq \varepsilon \|a_t\|_1$ for all $g \in \mathcal{G}$.
Notice that the queries $g$ must have a bounded $\ell_\infty$ norm, and the estimates achieve aditive error $\|a_t\|_1$.

In the oblivious setting, the same bound follows from known techniques.
Indeed, assume for simplicity that all $g\in\mathcal{G}$ are 0-1 vectors.
Then one can draw uniform samples from the mass of $a_t$, say using $\ell_1$-samplers~\cite{siamcomp/JayaramW21,focs/SwartworthWZ25}, and report the fraction of samples that lie in the support of $g$, scaled appropriately by $\|a_t\|_1$.
It is easy to see that the additive error in this method is proportional to $\|a\|_1$. 
The bound on $\|g\|_\infty$ is natural, as otherwise a single coordinate of $g$ can dominate the inner product, and discovering it in streaming is hard.

To achieve robustness, we leverage the fact that we are only aiming for an additive error. 
A motivating case is when all the vectors $a_t$ have the same norm, say $\|a_t\|_1 = \kappa > 0$. 
We then break the entire sequence of updates into only $O(\log m)$ intervals, apply for each interval $(i, j]$ the non-robust result that estimates $\langle a_i - a_j, g \rangle$, 
and report the sum of these estimates.
Crucially, the additive error in each interval is bounded by  
$\varepsilon \|a_i - a_j\|_1 \leq \varepsilon (\|a_i\|_1 + \|a_j\|_1) \leq 2\varepsilon \kappa$, 
hence the total error is at most $O(\log m)\cdot \varepsilon \kappa$, which can be mitigated by picking a smaller $\varepsilon$. 
The robustness comes from the fact that each interval of updates cannot be altered by later updates of the adversary, and hence its non-robust estimate is accurate regardless of later updates. 
To maintain these intervals dynamically using low space, 
we employ a dyadic partition of the time horizon $[m]$, 
which guarantees the $O(\log m)$ bound on the number of intervals used in each output. 
It remains to handle the general case, for which we break the stream differently, now according to the scales $\kappa$ (not into intervals), as discussed in \Cref{sec:inner_prod}.

\paragraph{Overall Framework: Clustering via Inner-Product Form.}
To estimate the clustering cost $\cost(\mu, C) = \langle \mu, \dist(\cdot, C) \rangle$, we cannot readily apply the above inner-product estimation, because both $\|\dist(\cdot, C)\|_\infty$ and $\|\mu\|_1$ 
are not bounded by $O(1)$ and $O(\OPT)$, respectively.
Our strategy is to reduce the problem to another inner-product formulation,
where we these bounds do hold,
by leveraging the linearity of the space $\mathbb{R}^{\cX}$. 
Specifically, we define a data-oblivious linear operator $A$,
together with a sparse \emph{reference} vector $\nu \in \mathbb{R}^{\mathcal{X}}$
that is maintained by a subroutine robust to adaptive adversary,
such that
\begin{equation}
  \label{eqn:nu}
  |\supp(\nu)| \leq \tO(k),
  \qquad
  \|A(\mu - \nu)\|_1 \leq \tO(\OPT),
\end{equation}
and for each query $C$
represented by a data-oblivious vector $g_C := \dist(\cdot, C)$, 
we have
\begin{equation}
      \label{eqn:inner_prod}
      \cost(\mu, C) = \langle A (\mu - \nu), g_C \rangle + \langle A \nu, g_C\rangle.
\end{equation}
We estimate the first term $\langle A(\mu - \nu), g_C\rangle$ within additive error $\varepsilon\cdot\OPT$ using our robust inner-product estimation,
whose error is proportional to $\|A(\mu - \nu)\|_1$, which in turn is upper bounded by \eqref{eqn:nu}.
For $z = 1$ (i.e., \kMedian) it turns out that $\|g_C\|_\infty \leq 1$, whereas for general $z$, we will clip $g_C$ to $\hat{g}_C$ with $\|\hat{g}_C\|_\infty \leq 1$  such that the error from clipping can be eventually bounded against $\varepsilon\cdot \OPT$. 
We evaluate the second term $\langle A\nu, g_C \rangle$ without any error, by using a sparse recovery algorithm (cf.~\cite{ciss/Cormode006}) to extract $\nu$,\footnote{Even randomized sparse recovery algorithms are adversarially robust 
because their output is the exactly same (except when they fail), hence it cannot leak the algorithm's randomness to the adversary.
}
and computing the coordinates of $g_C$ on the fly using low space.

In what follows, we focus on $z = 1$, which is more intuitive,
and the additional steps for general $z$ are discussed in \Cref{sec:tech_general_z}.

\paragraph{New Technique: Hierarchical Lipschitz Assignment (HLA).}
The definition of $A, g_C$ and $\nu$ crucially rely on our new geometric tool, which we call a hierarchical Lipschitz assignment (HLA). 
Below, we describe it informally and then give the formal definition  (specialized to $z = 1$, as mentioned above).
In \Cref{sec:tech_HLA} we explain how to compute it using low space 
and also discuss its relation to other geometric notions.

An HLA has $L$ dyadic scales $s_1,\ldots,s_L$, meaning that each $s_{\ell} = \frac{1}{2} s_{\ell-1}$. 
Each scale $s_\ell$ defines a set of representatives $W_\ell\subseteq\mathbb R^{d}$  and \emph{fractionally assigns} each (potential data point) $x \in \mathbb{R}^d$ to nearby (data-oblivious) representatives $a\in W_\ell$; this fractional assignment is denoted $F_\ell(a, x)\ge 0$. 
As the word hierarchical indicates, there is connection between the scales: 
Each assignment $F_\ell$ can be transformed into a coarser $F_{\ell-1}$, via a fractional assignment of each representative $a\in W_l$ to nearby representatives $b\in W_{\ell-1}$; this fractional assignment is denoted $U_\ell(b, a)\ge 0$. 
For simplicity, we write throughout $\ell\in[L]$, even when $\ell=1$ is clearly irrelevant (because $\ell-1$ is used).

\begin{definition}[Hierarchical Lipschitz assignment, for $z = 1$]
\label{def:HLA_1}
Consider $L$ dyadic scales $s_1,\ldots,s_L$ with $s_{\ell} = \frac{1}{2} s_{\ell-1}$ for all $\ell\in[L]$ and a set of representatives $W_\ell\subseteq\mathbb R^{d}$ for each $\ell\in[L]$. 
A \emph{hierarchical Lipschitz assignment (HLA)} has, for each $\ell\in[L]$, 
a pair of maps $F_\ell : W_\ell \times \R^{d}  \to [0, 1]$ and $U_\ell : W_{\ell - 1} \times W_{\ell} \to [0, 1]$, 
such that $\sum_a F_\ell(a, x) = \sum_b U_\ell(b, a) = 1$, and the following holds:
\begin{itemize}
\compactify
\item 
Refinement Property:
\vspace{-0.8\baselineskip}
\[\forall a,x,\qquad F_{\ell-1}(a,x)=\sum_b U_\ell(a,b) F_\ell(b,x).\]
\vspace{-0.6\baselineskip}
\item 
Locality Property:
\vspace{-0.8\baselineskip}
\begin{align*}
  \forall a,x,  \qquad
  &F_\ell(a,x)>0 \ \rightarrow\ \dist(a, x) \le O(\sqrt{d})\cdot s_\ell ,
  \\
  \forall a,b, 
  \qquad
  &U_\ell(b,a)>0 \ \rightarrow\ \dist(a, b) \le O(\sqrt{d})\cdot s_\ell.
\end{align*}
\vspace{-0.6\baselineskip}
\item 
Lipschitz Property:
\vspace{-0.8\baselineskip}
\begin{equation}
\label{eqn:lip_1}
\forall x, c, \qquad
\sum_{a\in W_\ell}
|F_\ell(a,x)-F_\ell(a,c)|
\le
O\left(
\frac{\|x-c\|_1}{s_\ell}
\right).
\end{equation}
\vspace{-0.6\baselineskip}
\end{itemize}
\end{definition}
Informally, an HLA is a transportation plan, that routes the mass of any data point in $x\in\R^d$ to a hierarchy of representatives in a manner that is consistent over the scales, local (transports only to nearby representatives), and varies smoothly over $x\in\R^d$. 
Observe that we can write $F_\ell$ and $U_\ell$ as matrices (formally, 
we must restrict the domain $\R^d$ to its finite subset $\cX$), 
and then the Refinement Property can be written as $F_{\ell-1}=U_\ell F_\ell$. 

The Refinement and Locality Properties above will be sufficient for \eqref{eqn:inner_prod}, but the Lipschitz Property is the main non-trivial guarantee, and is crucially needed for \eqref{eqn:nu}. 
Observe that its right-hand side uses the $\ell_1$ norm, and in our Euclidean application we can bound it by the $\ell_2$ norm via $\|x - c\|_1 \leq \sqrt{d} \dist(x, c)$ (recall that $\dist(\cdot,\cdot)$ represents $\ell_2$ distance).
To interpret this guarantee in a clustering, 
think of $x$ as a data point assigned to a center point $c$. 
Then this property relates the difference between their fractional assignments $\sum_a |F_\ell(a, x) - F_\ell(a, c)|$ 
to the contribution of $x$ to the clustering cost.
We remark that fractional assignment, rather than integral assignment, is necessary for the Lipschitz Property to hold (see \Cref{remark:HLA}).

\paragraph{Defining $A$ and $g_C$: Telescoping Sum.}
Suppose we have a space-efficient HLA, i.e., every $F_\ell$ and $U_\ell$ can be evaluated using low space, 
and let us use it to define $A$ and $g_C$ (again for $z = 1$ for ease of discussion). 
The idea is to distribute the mass $\mu_x$ of each point $x \in \mathcal{X}$ to the representatives $a \in W_\ell$ for every $\ell$, fractionally according to $F_\ell(a, x)$. 
The vectors $g_C$ and $A\mu$ have coordinates indexed by pairs $(\ell, a)$ for $\ell\in [L]$ and $a\in W_\ell$; 
we define the linear operator $A$ as a matrix given by 
\begin{equation}
      \label{eqn:A}
      \forall x \in \mathcal{X}, \qquad A_{(\ell, a), x} := s_\ell F_\ell(a, x),
\end{equation}
which implies 
\begin{align*}
      (A\mu)_{\ell, a} = s_\ell \sum_{x \in \mathcal{X}} F_\ell(a, x) \mu_x,
\end{align*}
and we define
\begin{align*}
      (g_C)_{\ell, a} :=
      \frac{1}{s_\ell} \Big[\dist(a, C) - \sum_{b \in W_{\ell - 1}} U_\ell(b, a) \dist(b, C)\Big].
\end{align*}
For simplcity, we ignore in this overview edge cases, such as $\ell = 0$.

We can verify $\langle A\mu, g_C\rangle = \cost(\mu, C)$,
which is the main requirement in \eqref{eqn:inner_prod},
by direct calculation that relies (only) on the Refinement Property of HLA.
Observe that $g_C$ has the form of a telescoping sum,
which makes its $\ell_\infty$-norm bounded.
For intuition, suppose that $F_\ell$ is an integral assignment,
that is, each $x$ is assigned to a unique representative $x_\ell$ at each scale $\ell$ (with $x_L = x$),
then the Locality Property implies
$\dist(x_\ell, x_{\ell - 1}) \leq s_\ell$.
We can thus break $\dist(x, C)$ into a telescoping sum
$\dist(x, C) = \sum_{\ell} [\dist(x_\ell, C) - \dist(x_{\ell - 1}, C)]$,
where each summand is bounded by
$| \dist(x_\ell, C) - \dist(x_{\ell - 1}, C) | \leq \dist(x_\ell, x_{\ell - 1}) \leq s_\ell$.
Hence, we can define $g_C'$ that corresponds to these terms
and has bounded $\ell_\infty$-norm, namely,
its coordinates are indexed by $(x,\ell)\in \R^{\cX} \times [L]$, where 
$(g_C')_{x, \ell} := \frac{1}{s_\ell} [\dist(x_\ell, C) - \dist(x_{\ell - 1}, C)]$.
The definition of $A'$ is naturally $(A'\mu)_{x, \ell} = s_\ell \cdot \mu_x$,
and ensures that $\langle A'\mu, g_C' \rangle = \cost(\mu, C)$.
Our $A$ and $g_C$ is a generalization of this idea,
specifically to work with the fractional assignment of the HLA. 

The idea of a telescoping sum, which underlies our definition of $g_C$,
was introduced in~\cite{icml/BravermanFLSY17} for $z=1$ (\kMedian)
in the oblivious setting.
Apart from the new HLA notion 
(that wefurtehr generalize to $z > 1$, see \Cref{sec:tech_general_z}), 
our subsequent steps are also quite different from~\cite{icml/BravermanFLSY17}: They use sampling to build a coreset 
and the telescopic sum is mostly needed to analyze the error,
whereas we introduce another step of a sparse $\nu$ with the nontrivial guarantee in \eqref{eqn:nu},
which requires additional property from $A$ (which eventually depends on properties of HLA).

\paragraph{Sparse Reference $\nu$: Bounded Transportation via Lipschitz Property.}
Next we discuss the sparse vector $\nu$. 
It suffice to show there exists one that satisfies \eqref{eqn:nu}, because the algorithm can then employ a procedure for $\ell_1$-norm estimation to approximate $\|A(\mu - x)\|_1$ for every sparse query $x$, and take the minimizer as our vector $\nu$. 
We must use here an adversarially robust version of $\ell_1$-norm estimation, which was recently obtained in~\cite{focs/GLWYZ26}. 
However, since the number of queries $x$ can be $\Delta^{O(kd)}$, 
to make sure that all these queries succeed simultaneously, 
the known algorithm~\cite{focs/GLWYZ26} requires an additional factor of $k$ in the space complexity, exceeding our claim for space bound linear in $k$.
Fortunately, it can be shown that the framework supports many queries $x$ without the additional factor $k$, provided that $\mu$ is not updated by the stream. 
In particular, while $\mu$ is adaptive to previous algorithmic outputs, the set of queries $x$ is crucially non-adaptive.

To define a sparse $\nu$, we use an optimal \kMedian solution for $\mu$,
denoted $C^* := \{c_1^*, \ldots, c_k^*\} \subseteq \mathbb{R}^d$.
Recall that the frequency vector $\mu \in \mathbb{R}_{\geq 0}^{\cX}$
encodes the input dataset $P\subset \cX$. 
Let $P_i^* \subseteq P$ denote the $i$-th cluster in this optimal solution,
and let $\mu_i$ be its corresponding frequency vector.
For each cluster $P_i^*$,
we define a $1$-sparse vector $w_i \in \mathbb{R}_{\geq 0}^{\mathcal{X}}$, 
which has a single non-zero coordinate that corresponds to $c_i^*$
and has value $\sum_{x \in P_i} (\mu_i)_x$, i.e., the total mass of $\mu_i$.
We define $\nu := \sum_{i} w_i$,
and it suffices to show that 
\begin{equation}
  \label{eqn:Amui}
  \forall i\in[k],
  \qquad
  \|A(\mu_i - w_i)\|_1 \leq \tO(\cost(\mu_i, c_i^*)),
\end{equation}
because summing this over $i$ yields \eqref{eqn:nu}.
By construction, the total mass in $w_i$ is the same as in $\mu_i$,
and thus the vector $\lambda := \mu_i - w_i$
has total mass $\sum_{x} \lambda_x = 0$,
or equivalently, 
\begin{equation}
      \label{eqn:equal_mass_1}
      \langle \lambda, \mathbf{1} \rangle = 0.
\end{equation}
To bound $\|A(\mu_i - w_i)\|_1 = \|A\lambda\|_1$,
we view it as a transportation cost and write it, using \eqref{eqn:A}, as
\begin{equation}
\label{eqn:Alambda1}
      \|A\lambda\|_1
      = \sum_{\ell} \sum_{a \in W_\ell} |s_\ell \sum_{x \in \cX} F_\ell(a, x) \lambda_x|
      = \sum_{\ell} \sum_{a \in W_\ell} s_\ell |\langle F_\ell(a, \cdot), \lambda\rangle|.
\end{equation}
The problem reduces to bounding a term of the form $|\langle a, b \rangle|$
(in our case $|\langle F_\ell(a, \cdot), \lambda \rangle|$),
and we describe a general strategy:
introduce a vector $a'$ that is orthogonal to $b$, i.e., $\langle a', b \rangle = 0$,
and then 
\[|\langle a, b \rangle| = |\langle a - a', b \rangle|\leq \langle |a - a'|, |b| \rangle,\]
where the absolute value on vectors $a - a'$ and $b$ is applied \emph{coordinate-wise}.
Picking a good $a'$ can be roughly viewed as a near-neighbor search problem:
think for example about $b = \mathbf{1}$,
then we seek $a'$ that is orthogonal to $b$
and its $\ell_1$ distance to $a$ is bounded.

Our strategy to bounding $|\langle F_\ell(a, \cdot), \lambda \rangle|$
is to pick $a'$ using the information we have about the clustering problem.
Specifically, we pick the vector $a' := F_\ell(a, c_i^*) \mathbf{1}$,
for which $\langle a', \lambda \rangle = 0$ due to \eqref{eqn:equal_mass_1}.
We then get, using the Lipschitz Property \eqref{eqn:lip_1} of HLA
in the second inequality below, that
\begin{align*}
    \sum_{a \in W_\ell} |\langle F_\ell(a, \cdot), \lambda \rangle|
    &\leq \sum_{a \in W_\ell} \langle |F_\ell(a, \cdot) - F_\ell(a, c_i^*)|, |\lambda| \rangle \\
    &= \sum_{a \in W_\ell} \sum_{x \in \cX} |F_\ell(a, x) - F_\ell(a, c_i^*)| \cdot |\lambda_x|  \\
&\leq \sum_{x \in \cX} O\left(\frac{\sqrt{d}\|x - c_i^*\|_2}{s_\ell}\right) \cdot (\mu_i + w_i)_x  \\
    &\leq O(\sqrt{d})  \cost(\mu_i, c_i^*) / s_\ell.
\end{align*}
Plugging this bound into \eqref{eqn:Alambda1},
we obtain \eqref{eqn:Amui}, as desired.

\paragraph{New Technique: Robust Dimension Reduction.}
The above discussion leads overall to an $\tO_{\varepsilon, z}(k \poly(d))$-space algorithm,
whereas our goal is $\tO(kd)$.
A standard approach to improve the dependence on $d$ to linear,
is to apply dimension reduction:
use $\tO(d)$ space, which is comparable to reading a single data point,
to reduce the dimension $d$ to $\tO(1)$ for downstream algorithms,
and then run a known $\poly(d)$-space algorithm (applied in the low dimension),
using overall $\tO(d)$ space.
While there exists $(1 + \varepsilon)$-approximate dimension reduction for clustering~\cite{siamcomp/MakarychevMR19},
it works only for oblivious adversary.
In fact, dimension reduction can be a hard task in adaptive streaming in general;
for example, no dimension reducing linear sketch can preserve and report the norm
of the current adaptive input point~\cite{stoc/HardtW13},
whereas the standard method for dimension reduction for clustering~\cite{siamcomp/MakarychevMR19,NarayananSIZ21,compgeom/JiangKS24,stoc/HuangJKY25,innovations/JiangKSSY26} is based on Gaussian linear maps~\cite{JL84}.

We bypass this issue by computing multiple linear maps, instead of just one,
and averaging their outputs.
This method preserves the distance $\dist(x, C)$ simultaneously
for all $x \in \mathcal{X}$ and all $C \subseteq \mathcal{X}$ with $|C| \leq k$,
with high probability.
It resolves the issue of adaptive adversary, because its success event
means it works simultaneously for all possible inputs in the domain $\mathcal{X}$.
Consider first a single Gaussian linear map with target dimension roughly $O(\log k)$;
it preserves the distance of every fixed pair $(x, C)$ separately
with constant probability.
The natural next step is to take a union bound over all pairs $(x, C)$,
whose number is roughly $\Delta^{O(kd)}$.
Using only one map, known bounds for Gaussian maps
require target dimension $O(kd \log \Delta)$,
which exceeds the original dimension $d$ and is therefore useless.
Instead, we draw independently multiple maps,
and for every pair $(x, C)$ we \emph{average} their all estimates for $\dist(x, C)$.
This averaging allows for measure concentration arguments,
and then it suffices to run $O(kd \log \Delta)$ independent downstream algorithms,
which increases the overall space complexity by this affordable factor.

However, this method raises another issue:
if each downstream algorithm requires space that is linear in $k$,
then running this algorithm independently under each of the $\tO(kd)$ maps
would lead to $k^2$ total space.
To finally obtain the claimed space bound, which is linear in $k$,
we define a single instance of $(A, g_C, \nu)$,
and use it with all the $\tO(kd)$ maps. 
Luckily, our HLA is data-oblivious and deterministic,
hence it is possible to use this single HLA in the target space of all the maps,
and the algorithm considers only a single $(A, g_C, \nu)$
that works when the Guassian maps are combined (by averaging their estimates).
We note that this is also the reason we need to use averaging instead of taking median to aggregate the independent maps,
as median is not linear and is not compatible with the linear $A, g_C, \nu$.
We remark that randomized structures, such as randomly-shifted quadtrees,
may work for a single projection and still require space linear in $k$,
however they might require $k$ independent repetitions
to work for all projections simultaneously,
and thus the space would increase by factor $k$.

\subsubsection{Major New Steps for General \texorpdfstring{$z$}{z}}
\label{sec:tech_general_z}

Our algorithm for general $z$ follows the same framework as for $z = 1$,
i.e., to establish \eqref{eqn:nu} and \eqref{eqn:inner_prod}.
It still based on a hierarchical Lipschitz assignment (HLA),
however we need to revise its definition to handle general $z$, as follows.

\begin{restatable}[Order-$z$ hierarchical Lipschitz assignment]{definition}{defHLA}
\label{def:HLA}
Let $z\geq 1$.
Consider $L$ dyadic scales $s_1, \ldots s_L$
(meaning that each $s_{\ell} = \frac{1}{2} s_{\ell-1}$)
and a set of representatives $W_\ell\subseteq\mathbb R^{d}$ for each $\ell\in[L]$.
An \emph{order-$z$ hierarchical Lipschitz assignment (HLA)} has, for each $\ell\in[L]$, a pair of maps 
$F_\ell : W_\ell \times \R^{d}  \to [0, 1]$ and $U_\ell : W_{\ell - 1} \times W_{\ell} \to [0, 1]$,
such that $\sum_a F_\ell(a, x) = \sum_b U_\ell(b, a) = 1$,
and the following holds:
\begin{itemize} \compactify
\item Refinement Property:
$F_{\ell-1}=U_\ell F_\ell$.

\item Locality Property:
\begin{align*}
  \forall a,x,  \qquad
  &F_\ell(a,x)>0 \ \rightarrow\ \dist(a, x) \le O(z\sqrt{d})\cdot s_\ell ,
  \\
  \forall a,b,
  \qquad
  &U_\ell(b,a)>0 \ \rightarrow\ \dist(a, b) \le O(z\sqrt{d})\cdot s_\ell.
\end{align*}

\item Order-$z$ Lipschitz Approximation:
For every $c\in\R^{d}$,
there exist degree-$(z-1)$ polynomials
$\{\pi_{\ell,a,c}:\mathbb R^{d}\to\mathbb R\}_{a \in W_\ell}$,
such that 
\begin{equation}
\label{eqn:lip}
\forall x, \qquad
\sum_{a\in W_\ell}
\bigl|F_\ell(a,x)- F_\ell(a,c) - \pi_{\ell,a,c}(x)\bigr|
\le
O(2^z)
\left(
\frac{\|x-c\|_1}{s_\ell}
\right)^z.
\end{equation}
\end{itemize}
\end{restatable}
The first two properties are similar to the case $z = 1$ (\Cref{def:HLA_1}).
The main difference is the third property,
as the right-hand side is now raised to power $z$,
which is stronger because the interesting case is when $x$ and $c$ are close.
Another difference is that the Lipschitz condition has a low-degree ``calibration term'' $\pi$.
A calibration term is necessary even for $z = 2$,
as otherwise $x \mapsto F_\ell(a, x)$ must be constant (independent of $x$),
thereby violating locality constraint (see \Cref{remark:HLA}).
Intuitivlely, one can think of $F_\ell(a, x) - F_\ell(a, c)$
as a degree-$z$ polynomial in $x - c$,
and in the interesting case where $x$ and $c$ are close,
the terms with degree smaller than $z$
can exceed the desired $\|x - c\|_1^z$ bound,
and then the calibration term $\pi$ can cancel them out.

\paragraph{Generalized Definition of $A$ and $g_C$.}
We generalize $A$ and $g_C$ by raising to power $z$ in the natural places.
We define
\begin{equation}
      \label{eqn:A_z}
      \forall x \in \mathcal{X}, \qquad A_{(\ell, a), x} := s_\ell^z F_\ell(a, x),
\end{equation}
and 
\begin{align*}
      (g_C)_{\ell, a} :=
      \frac{1}{s_\ell^z} \Big[ \dist^z(a, C) - \sum_{b \in W_{\ell - 1}} U_\ell(b, a) \dist^z(b, C) \Big] .
\end{align*}

\paragraph{Existence of Sparse $\nu$.}
We omit the discussion on \eqref{eqn:inner_prod},
and focus on the major new step,
which is the existence of a sparse vector $\nu$ that satisfies \eqref{eqn:nu}.
We still follow the framework of the case $z = 1$,
where we take optimal clustering of the input $\mu$,
in which the centers are $C^* := \{c_1^*, \ldots, c_k^*\}$
and the clusters are $P_1^*, \ldots, P_k^*\subset P$,
and we define $\mu_i$ be the frequency of cluster $P_i^*$.
It again suffices to identify a sparse frequency vector $w_i \in \mathbb{R}_{\geq 0}^{\mathcal{X}}$,
such that
\begin{equation}
      \label{eqn:Amui_z}
      \|A(\mu_i - w_i)\|_1 \leq \tO(1) \cost(\mu_i, c_i^*).
\end{equation}

The next step, to define $w_i$, differs different from the case $z = 1$,
since letting $w_i$ be the point $c_i^*$ weighted appropriately does not work,
and we eventually let $w_i$ have roughly $d^{O(z)}$ points in its support.
Before defining $w_i$, we state some fundamental properties
that will eventually imply \eqref{eqn:Amui_z}:
\begin{equation}
      \label{eqn:w_i_sparse}
      \supp(w_i) \subseteq P_i^*, \quad |\supp(w_i)| \leq O(d^{z - 1}),
\end{equation}
\begin{equation}
      \label{eqn:w_i_le}
      \cost(w_i, c_i^*) \leq \cost(\mu_i, c_i^*)
\end{equation}
and for every degree-$(z - 1)$ polynomial $q : \mathbb{R}^d \to \mathbb{R}$,
\begin{equation}
  \label{eqn:null_z}
  \sum_{x \in P_i^*} (\mu_i)_x q(x) = \sum_{x \in P_i^*} (w_i)_x q(x) 
  \quad \Leftrightarrow \quad 
  \langle \mu_i - w_i, q \rangle = 0.
\end{equation}

Let us now show how these properties of $w_i$ imply \eqref{eqn:Amui_z}.
Set $\lambda := \mu_i - w_i$,
then by \eqref{eqn:A_z},
\begin{equation*}
      \|A(\mu_i - w_i)\|_1
      = \|A\lambda\|_1
      = \sum_{\ell} \sum_{a \in W_\ell} |s_\ell^z \sum_{x \in P_i^*} F_\ell(a, x) \lambda_x |
      = \sum_{\ell} s_\ell^z \sum_{a \in W_\ell} |\langle F_\ell(a, \cdot), \lambda \rangle |
\end{equation*}
To further bound $|\langle F_\ell(a, x), \lambda_x \rangle|$,
we follow the technique we used for $z = 1$,
namely, we bound $|\langle a, b \rangle|$ using a vector $a'$ that is orthogonal to $b$:
\begin{equation*}
  \langle a', b \rangle = 0
  \quad \Rightarrow \quad 
  |\langle a, b \rangle|  
  \leq \langle |a - a'|, |b| \rangle,
\end{equation*}
where the absolute value on $a - a'$ and $b$ are coordinate-wise.
Similarly to the case $z = 1$,
we pick $a'$ so that we can leverage the Lipschitz property \eqref{eqn:lip} of HLA.
Specifically,
let $\pi_{\ell, a, c_i^*}$ be the degree-$(z - 1)$ calibration polynomial on $x$
guaranteed in \eqref{eqn:lip},
and pick
\begin{equation*}
      a' := F_\ell(a, c_i^*) \cdot \mathbf{1} + \pi_{\ell, a, c_i^*}(\cdot).
\end{equation*}
Then $\langle a', \lambda\rangle = 0$ due to \eqref{eqn:null_z},
and we get
\begin{align*}
      \sum_{a \in W_\ell} |\langle F_\ell(a, \cdot), \lambda \rangle |
      &\leq \sum_{a \in W_\ell} \langle |F_\ell(a, \cdot) - (F_\ell(a, c_i^*) \cdot \mathbf{1} + \pi_{\ell, a, c_i^*})|, |\lambda| \rangle | \\
      &= \sum_x |\lambda_x| \sum_{a \in W_\ell}  |F_\ell(a, x) - (F_\ell(a, c_i^*) + \pi_{\ell, a, c_i^*})(x)|  \\
      &\leq \sum_x ((\mu_i)_x + (w_i)_x) \cdot O(2^z d^{z / 2} / s_\ell^z) \|x - c_i^*\|_2^z \\
      &\leq O(2^z d^{z / 2}/ s_\ell^z) \cost(\mu_i, c_i^*) ,
\end{align*}
where the second inequality is by the Lipschitz Property \eqref{eqn:lip},
and the last inequality uses \eqref{eqn:w_i_le}.

To prove the existence of $w_i$ that satisfies the above properties,
we employ linear programming (LP) that, once again, is used only in the analysis.
We write the following LP with variables $(w_i)_x$ for $x \in P_i^*$:
\begin{align*}
      \text{min } & \sum_{x \in P_i^*} (w_i)_x \|x - c_i^*\|_2^z & \\
      \text{s.t. } & \sum_{x \in P_i^*} (w_i)_x (x - c_i^*)^{[\alpha]}
      = \sum_{x \in P_i^*} (\mu_i)_x (x - c_i^*)^{[\alpha]} & \forall \alpha \in \{0, \ldots, z - 1\}^d, \|\alpha\|_1 \leq z - 1, 
\end{align*}
where $x^{[\alpha]} := \prod_{i \in [d]} x_i^{\alpha_i}$
denotes the monomial in $x \in \mathbb{R}^d$ with degrees according to the vector $\alpha$.
The constraints imply that every feasible $w_i$ satisfies \eqref{eqn:null_z},
because $\mu_i$ and $w_i$ agree on all monomials of $x$ up to degree $z - 1$
(and thus also on all monomials of $x - c_i^*$). 
Now if we take $w_i$ to be any optimal solution,
then \eqref{eqn:w_i_le} holds immediately 
because $\mu_i$ is a feasible solution and $w_i$ is an optimal one.
Lastly, since there are exactly $\binom{d + z - 1}{d} \leq O(d^{z - 1}) $ constraints (corresponding to distinct $\alpha$'s),
there must exist an optimal solution where only $O(d^{z - 1})$ coordinates
are nonzero.
We pick it to be our $w_i$, and this implies the sparsity \eqref{eqn:w_i_sparse}.

\subsubsection{Construction of HLA}
\label{sec:tech_HLA}

We first give an alternative interpretation of an HLA.
The fractional assignment $F_\ell$ (for a given $\ell$) may be viewed as
a probability distribution of a randomized geometric hashing
$h_\ell : \mathbb{R}^d \to \mathbb{R}^d$,
where we define $F_\ell(a, x) := \Pr[h_\ell(x) = a]$.
The Locality Property is equivalent to $\dist(h_\ell(x), x) \leq s_\ell$.
In the case $z = 1$, the Lipschitz Property is (see \eqref{eqn:lip_1}) 
\begin{equation}
      \label{eqn:h_z_1}
      \sum_a \big| \Pr[h_\ell(x) = a] - \Pr[h_\ell(c) = a] \big| 
      \leq O(1) \frac{\|x - c\|_1}{s_\ell}.
\end{equation}
This is tightly related to a well-known notion called separating decomposition~\cite{focs/Bartal96},
whose guarantee has the form (when clusters have diameter bounded by $s_\ell$):
\begin{equation}
      \label{eqn:sep_decomp}
      \Pr[h_\ell(x) \neq h_\ell(c)] \leq O(1) \frac{\|x - c\|_1}{s_\ell}.
\end{equation}
This separating-decomposition guarantee \eqref{eqn:sep_decomp}
\emph{implies} our Lipschitz guarantee as in \eqref{eqn:h_z_1}, because
\begin{align*}
      \sum_a \big| \Pr[h_\ell(x) = a] - \Pr[h_\ell(c) = a] \big|
      \leq \E\Big[\sum_a |\mathbb{I}(h_\ell(x) = a) - \mathbb{I}(h_\ell(c) = a)|\Big]
      = 2 \Pr[h_\ell(x) \neq h_\ell(c)].
\end{align*}
Therefore, by plugging in a simple construction of a randomly-shifted quadtree,
which achieves \eqref{eqn:sep_decomp} (see e.g.~\cite{HP11})
and also has the Refinement Property,
we readily obtain an HLA when $z = 1$.

We can push this interpretation further to define HLA also for $z > 1$.
We focus for simplicity on $z = 2$, $d = 1$ (one-dimensional line)
and unit scale $s_\ell = 1$.
In dimension $d=1$, the randomly-shifted quadtree approach
is simply to pick random shift $S$ and define $h(x) := \lfloor x - S \rfloor$, 
which maps $x$ to the left endpoint of the (randomly shifted) unit interval that contains $x$.
We first show why picking $S$ uniformly from $[0, 1]$,
which is the standard random shift, does \emph{not} work for us.
In that case,
\begin{align}
    F(0, x) = \Pr[h(x) = 0] = \Pr[S \in (x -1, x]]
    =\begin{cases}
        x & \text{if } 0 \leq x \leq 1 \\
        2 - x & \text{if } 1 < x < 2 \\
        0 & \text{otherwise},
    \end{cases}
    \label{eqn:triangular}
\end{align}
a probability density that has triangular shape.
Consider $c = 1$ and $x = 1 + \eta$ for a small $\eta > 0$;
then $|F(0, x) - F(0, c)| = \eta$, and we thus need a calibration term.
The calibration term should have degree $z  - 1 = 1$,
i.e., linear of the form $\pi(x) =\kappa x + \beta$.
When $x = c = 1$, it must be zero to satisfy the Lipschitz Property \eqref{eqn:lip},
and therefore it must be $\pi(x) = \kappa (x - 1)$.
When $x = 1 + \eta$ or $x = 1 - \eta$,
the value of $|F(0, x) - F(0, c) - \pi(x)|$
is $\eta|1 + \kappa|$ or $\eta|1 - \kappa|$, 
and at least one of these will be $\Omega(\eta) \gg \eta^2$
as required in the RHS of \eqref{eqn:lip}.

The distribution that works for $S$ is to let $S := V_1 + V_2$
where both $V_1,V_2$ are picked uniformly from $[0, 1]$,
which is the sum of two independent shifts.
Intuitively, this is smoother than the triangular probability density \eqref{eqn:triangular};
indeed, our counter-example is built around the corner $c = 1$. 
Technically, this distribution introduces to $\Pr[h(x) = a]$ the necessary quadratic terms.
Specifically, let $\rho(u)$ be the density function of $S$,
then we can verify that $|\rho(u) - \rho(v)| \leq |u - v|$ which is Lipschitz.
Now $|F(a, x) - F(a, c)|$ roughly corresponds to an intergration of $\rho(u) - \rho(v)$,
which has both quadratic and linear terms in $|u - v|$,
and the calibration term $\pi$ is used to cancel out the (larger) linear term
so that only the (smaller) quadratic term remains,
thereby guaranteeing the order-$2$ Lipschitz Property.

For general $z$, the distribution of $S$ is the sum of $z$ iid uniform shifts.
The analysis is similar, where one first establish some high-order Lipschitz consdition for the density $\rho(u)$. The integration gives order-$z$ terms,
and the calibration term cancels the (larger) low-order terms.
This idea also generalizes to dimension $d > 1$
where one uses a fixed quadtree and a random shift in $\mathbb{R}^d$.
Finally, for space efficiency,
we notice that $\rho(u)$ has a closed form and can be computed in low space.

\subsubsection{Polynomial-Time \texorpdfstring{$O(1)$}{O(1)}-Approximation for \texorpdfstring{$z = 1$}{z=1}}

For \kMedian, we obtain constant-factor approximation with polynomial update time,
while still using near-linear $kd$ space and being adversarially robust.
This is not directly based on the cost oracle framework, 
but it still builds on the ingredients from above, including robust dimension reduction, inner product estimation and HLA.
Instead of enumerating over a family of candidate center sets, which  takes exponential time, we start from an arbitrary candidate set $C$ of at most $k$ centers and repeatedly improve it. 
We apply random hashing and $\ell_1$ sampling to identify clusters poorly covered by centers in $C$ 
and recovers new candidate centers in those clusters in polynomial time.
Given an upper bound $S$ on the clustering cost of $C$, 
our approach produces a new candidate set $C'$ of at most $k$ centers with clustering cost $O(\opt)+S/10$,
and therefore repeating this logarithmically many times yields $O(1)$-approximation.
A more detailed overview can be found in \Cref{poly:sec:polynomial}.

A notable new ingredient is a time-efficient construction of HLA for $z = 1$.
In the HLA construction from above (the one that is only space-efficient),
the fractional assignment vector $F_\ell(\cdot,x)$ can have exponentially many nonzero entries,
which is fine because we do not store them explicitly and evaluating $F_\ell(a, x)$ for each $(a, x)$ takes little space.
However, to achieve polynomial update time, we need to explicitly compute the entire fractional assignment of $x$.
For \kMedian ($z = 1$), we develop an efficient construction of HLA by sharing the fractional assignment across each dimension,
where each $F_\ell(\cdot,x)$ has at most $d+1$ nonzero entries
and can be computed in $O(d^2)$ time using $\widetilde O(d)$ space.
Crucially, this more efficient HLA has the Refinement, Locality, and Lipschitz Properties required for $z=1$.
See \Cref{sec:HLA_z_1} for details.

\subsection{Related Work}
\label{sec:related}

\paragraph{Classical (Oblivious) Dynamic Geometric Streams.}
We review algorithms in the classical oblivious-adversary setting under the same dynamic geometric streaming model of \cite{stoc/Indyk04} that we study in this paper. 
Since its introduction, this model has been extensively studied across a wide range of geometric optimization problems.
Prior works in this model largely assume an oblivious adversary and build on randomly shifted hierarchical grids (quadtrees) to solve a wide range of geometric optimization problems. 
Early algorithms either incurred exponential dependence on the dimension $d$ to maintain $\varepsilon$-coresets of size $\tilde{O}(k\varepsilon^{-O(d)})$ for \kMedian and \kMeans as well as $(1+\varepsilon)$-approximations for Euclidean minimum spanning tree, max-cut,
and facility location~\cite{FrahlingS05, FrahlingIS08,soda/CzumajLMS13}, or relied on tree metric embeddings in high dimensions to obtain $O(d\log^2\Delta)$-approximations for facility location, $O(d \log \Delta)$-approximation for minimum spanning tree and min-cost (bi-chromatic) matching~\cite{stoc/Indyk04}. 
Subsequent works on high-dimensional dynamic streams developed linear sketches for Earth Mover's Distance (EMD) and Wasserstein-$z$ optimal transport, yielding constant-factor approximations to \kMedian via sparse recovery~\cite{BackursIRW16} and two-pass algorithms for $(k,z)$-clustering~\cite{Cohen-AddadWZ23}. 
To achieve $(1+\varepsilon)$-approximations in high dimensions with $\mathrm{poly}(d)$ space in a single pass, Braverman et al.~\cite{icml/BravermanFLSY17} introduced a telescoping-sum cost decomposition over hierarchical grids to obtain an $\varepsilon$-coreset for \kMedian in $\tilde{O}(kd^4\varepsilon^{-2})$ space. 
Because squared distances suffer unbounded distortion across quadtree boundaries, Song, Yang, and Zhong~\cite{SongYZ18} simulated Chen's partition framework~\cite{siamcomp/Chen09} and sensitivity sampling~\cite{stoc/FeldmanL11} over randomly shifted grids via dynamic point-cell recovery structures, obtaining $\tilde{O}(k\,\mathrm{poly}(d\varepsilon^{-1}))$-space coresets for \kMeans and $M$-estimators, along with applications to geometric Max-Cut and average pairwise distance. 
Crucially, all of these oblivious dynamic geometric streaming algorithms rely on linear sketches, random sampling over grids, or explicit coreset recovery, which fail against adaptive adversaries.

\paragraph{Dimension Reduction for Clustering and Geometric Streams.}
Dimensionality reduction via Johnson-Lindenstrauss (JL) linear projections~\cite{JL84} is a standard tool for reducing high-dimensional overhead in Euclidean clustering~\cite{CohenEMMP15, siamcomp/MakarychevMR19, CharikaW25}, as well as related geometric problems studied in dynamic streams such as facility location and single-linkage clustering~\cite{NarayananSIZ21, stoc/HuangJKY25}, Wasserstein barycenters~\cite{IzzoSZ21}, $k$-center~\cite{compgeom/JiangKS24}, and discrete-center clustering~\cite{innovations/JiangKSSY26}. 
Specifically, Makarychev, Makarychev, and Razenshteyn~\cite{siamcomp/MakarychevMR19} showed that projecting to $O(\varepsilon^{-2}\log(k/\varepsilon))$ dimensions preserves the $(k, z)$-clustering cost for any fixed partition. 
However, standard oblivious JL maps only provide a ``for-each'' guarantee and fail against adaptive adversaries that can align subsequent point insertions with the kernel of the projection matrix~\cite{stoc/HardtW13}. 
Our robust dimension reduction in \Cref{sec:robust:dim} bypasses this barrier by averaging multiple independent Gaussian projections to establish a uniform ``for-all'' guarantee over the candidate domain. \section{Preliminaries}
\label{sec:prelim}

We write $[n]\coloneqq\{1,\ldots,n\}$.
For $u\in\R$, write $(u)_+\coloneqq\max\{u,0\}$.
As in the model description, the input is from $[\Delta]^d$,
and we write $\mathcal{X} := [\Delta]^d$.
A vector in $\R^S$, for a finite set $S$, assigns one real weight
to each element of $S$; its coordinates are indexed by $S$.
For $x\in S$, the vector $e_x$ has value one at coordinate $x$
and zero at every other coordinate.
While the input frequency vector is nonnegative at any time,
we may work with general real vector $\lambda \in \mathbb{R}^{\mathcal{X}}$ on which we can still evaluate the cost function
$\cost(\lambda, C) := \sum_{x \in \mathcal{X}} \lambda_x \dist^z(x, C)$.
For nonnegative $\mu \in \mathbb{R}^{\cX}_{\geq 0}$,
let $\OPT(\mu) := \min_{C \subseteq \mathbb{R}^d : |C| \leq k} \cost(\mu, C)$,
and when the context is clear we may simply write $\OPT$.

The adversary observes all output of the algorithm which is reported cost oracles and center sets for \Cref{thm:poly_intro},
and may choose subsequent updates and queries from that transcript; the
private states of the sketches are not exposed.
We allow exact real arithmetic during updates and queries and
$O(1)$ real registers; bit-space bounds count stored bit strings
and discrete workspace.

 \section{Robust Dimensionality Reduction}
\label{sec:robust:dim}

\begin{restatable}{lemma}{lemrobustprojection}
\label{lemma:robust:projection}
Given $0 < \varepsilon <1$, $z \geq 1$
and a ground set $X \subseteq \mathbb{R}^d$,
let $\widetilde d:= \poly(z\varepsilon^{-1} \log k)$ and 
$n_g:= (z\varepsilon^{-1}k\log |X|)^{O(z)}$.
Then there exists a collection of $n_g$ matrices $G_1, \ldots, G_{n_g}\in\mathbb{R}^{\widetilde{d}\times d}$
that can be computed from a shared random seed using $O(zk\log|X| \log(\varepsilon^{-1}zkd\log |X|))$ space, such that with high constant probability,
\begin{equation}
\label{eq:projection-guarantee}
\forall x \in X, C \subseteq X, |C| \leq k, \qquad
\frac{1}{n_g}\sum_{j=1}^{n_g}\dist(G_jx,G_jC)^z=(1\pm\varepsilon)\dist(x,C)^z.
\end{equation}
\end{restatable}

\begin{proof}
We show \eqref{eq:projection-guarantee} via independent Gaussian maps $G_1,\ldots,G_{n_g}$, whose entries are independent $N(0,1/\widetilde d)$ random variables.
We first show that, the gurantee of \eqref{eq:projection-guarantee}
holds for a single $(x, C)$ with probability at least $1 - |X|^{-O(k)}$.
We then use Nisan's pseudorandom generator~\cite{combinatorica/Nisan92}
to show a small-space version of this claim.
Finally, we apply a union bound over all $(x, C)$ tuples on the domain $X$ to 
conclude the required for-all guarantee in \eqref{eq:projection-guarantee}.

For a fixed point $x$ and center set $C$,
we give upper bound and lower bound separately
for  \eqref{eq:projection-guarantee}.
We setup several global parameters as follows.
\[
 \eta:=O(\eps/z),\qquad b:=O(k\log|X|).
\]

\paragraph{Lower Bound.}
Fix $x\in X$ and a nonempty center set $C\subseteq X$ of size at most $k$. Let $\mathcal E_j$ be the event
\[
 \dist(G_jx,G_jc)\geq(1-\eta)\dist(x,c)
 \qquad\text{for every }c\in C.
\]
Gaussian norm concentration gives, for each $c\in C$,
\[
 \Pr\!\left[\dist(G_jx,G_jc)<(1-\eta)\dist(x,c)\right]
 \leq \exp\!\left(-\Omega(\eta^2\widetilde d)\right).
\]
A union bound over the centers and the choice of $\widetilde d$ imply
\[
 \Pr[\lnot\mathcal E_j]
 \leq k\exp\!\left(-\Omega(\eta^2\widetilde d)\right)\leq O(\eps).
\]
Condition on $\mathcal E_j$ happens,
\[
 \dist(G_jx,G_jC)
 =\min_{c\in C}\dist(G_jx,G_jc)
 \geq(1-\eta)\min_{c\in C}\dist(x,c)
 =(1-\eta)\dist(x,C).
\]
By the Chernoff bound, except with probability $\exp(-\Omega(\eps n_g))\leq e^{-b}$, at least $(1-O(\eps))n_g$ of the events $\mathcal E_j$ hold,
and if this event so happens, we have
\begin{equation}\label{eq:lower}
 \begin{aligned}
 \frac1{n_g}\sum_{j=1}^{n_g}\dist(G_jx,G_jC)^z
 &\geq\frac1{n_g}\sum_{j:\mathcal E_j}\dist(G_jx,G_jC)^z\\
 &\geq(1-O(\eps))(1-\eta)^z\dist(x,C)^z\\
 &\geq(1-O(\eps))\dist(x,C)^z,
 \end{aligned}
\end{equation}
where the last inequality uses $(1-\eta)^z\geq1-z\eta$.
Therefore, the lower bound of \eqref{eq:projection-guarantee} holds with probability at least $1 - \exp(-\Omega(\eps n_g))\geq 1 - e^{-b}$.

\paragraph{Upper Bound.}
Fix $x\in X$ and $C\subseteq X$ of size at most $k$, and let $c^*\in C$ be a nearest center of $x$ in the original space. Since
\[
 \dist(G_jx,G_jC)\leq\dist(G_jx,G_jc^*)
 \qquad\text{and}\qquad
 \dist(x,c^*)=\dist(x,C),
\]
it suffices to bound the average of $\dist(G_jx,G_jc^*)^z$ for this single center $c^*$. Define
\[
 Y_j:=\frac{\dist(G_jx,G_jc^*)}{\dist(x,C)}.
\]
Gaussian maps give $Y_j^2\sim\chi^2_{\widetilde d}/\widetilde d$.
We will prove that, except with probability $O(e^{-b})$,
\begin{equation}\label{eq:upper}
 \frac1{n_g}\sum_{j=1}^{n_g}Y_j^z\leq1+O(\eps).
\end{equation}
This immediately gives the required upper bound on the projected clustering cost.

First we bound the expectation. For $1\leq z\leq2$, Jensen's inequality and $\E Y_j^2=1$ give $\E Y_j^z\leq1$. For $z>2$, put $q:=\lceil z/2\rceil$. The chi-square moment formula gives
\[
 \E Y_j^{2q}=\prod_{a=0}^{q-1}\left(1+\frac{2a}{\widetilde d}\right).
\]
Therefore,
\begin{equation}\label{eq:mean}
 \E Y_j^z
 \leq(\E Y_j^{2q})^{z/(2q)}
 \leq\exp\!\left(\frac{z(q-1)}{2\widetilde d}\right)
 \leq\exp\!\left(\frac{z^2}{4\widetilde d}\right)
 \leq1+O(\eps).
\end{equation}
The last inequality follows from the value of $\widehat{d}$.

We next apply a sub-Weibull concentration inequality~\cite{zhang2022sharper} to bound the tail probability of $Y_j^z$,
which we state as follows. 
or $\alpha>0$, write
\[
 \norm{V}_{\psi_\alpha}
 :=\inf\{a>0:\E\exp((|V|/a)^\alpha)\leq2\}.
\]
For independent mean-zero variables with $\max_j\norm{V_j}_{\psi_\alpha}\leq K$ and $0<\alpha\leq1$, the sub-Weibull inequality (see Theorem 3.1 of~\cite{kuchibhotla2022moving}) states that
\begin{equation}\label{eq:subweibull}
 \Pr\!\left[
  \left|\frac1m\sum_{j=1}^mV_j\right|
  >\kappa_\alpha K
       \left(\sqrt{\frac{s}{m}}+\frac{s^{1/\alpha}}m\right)
 \right]\leq O(e^{-s})\qquad(s\geq1),
\end{equation}
where $\kappa_\alpha\leq O(1/\alpha)^{1/\alpha}$.

To apply the inequality \eqref{eq:subweibull},
we observe that the chi-square moment-generating function gives
\[
 \E e^{Y_j^2/4}
 =\left(1-\frac1{2\widetilde d}\right)^{-\widetilde d/2}
 \leq \sqrt{2},
\]
so $\norm{Y_j}_{\psi_2} \le 2$. Thus, with
\[
 \alpha:=\min\{1,2/z\},\qquad V_j:=Y_j^z-\E Y_j^z,
\]
we have $\norm{V_j}_{\psi_\alpha}\leq2^{O(z)}$. For $z\geq2$, this follows from $\norm{Y_j^z}_{\psi_{2/z}}=\norm{Y_j}_{\psi_2}^z$ and~\eqref{eq:mean}; for $1\leq z\leq2$, use $Y_j^z\leq1+Y_j^2$ and $\E Y_j^z\leq1$. In particular, $\kappa_\alpha\norm{V_j}_{\psi_\alpha}\leq z^{O(z)}$.
Hence, we apply~\eqref{eq:subweibull} with $m=n_g$ and $s=b$. By our choices of $b$ and $n_g$, its deviation threshold is $O(\eps)$. Combining this with~\eqref{eq:mean} proves~\eqref{eq:upper}. Therefore, except with probability $O(e^{-b})$,
\[
 \frac1{n_g}\sum_{j=1}^{n_g}\dist(G_jx,G_jC)^z
 \leq\frac1{n_g}\sum_{j=1}^{n_g}\dist(G_jx,G_jc^*)^z
 \leq(1+O(\eps))\dist(x,C)^z.
\]

\paragraph{Generation from a Short Seed.}
In conclusion, the above argument shows that the guarantee in \eqref{eq:projection-guarantee}
holds for a single but arbitrary $(x, C)$ with probability $ 1- O(e^{-b})$.
Now, we show that this guarantee can be achieved without using the fully random $G_j$'s but only a small seed, via Nisan's PRG~\cite{combinatorica/Nisan92}.
Recall that in Nisan's PRG~\cite{combinatorica/Nisan92},
it simulates a randomized test for a fixed input provided random access,
that reads $R$ public random words once and uses $S$ words of workspace from a seed of $O((S+b)\log R)$ words, with an additive change of at most $e^{-b}$ in its acceptance probability.

Now, we can regard the randomness defining the $G_j$'s as public randomness, read as a tape of matrix entries in order. The length of this tape is
\[
 R=O(n_g\widetilde d\,d).
\]
For each fixed point $x$ and center set $C$, the lower and upper events above can be checked in one pass using $S=O(k)$ words of workspace:
store the current row sums and squared distances for $x$ and each center $c \in C$, and accumulated sums. Nisan's generator, with error $e^{-b}$ per check, therefore gives a seed of
\[
 O\bigl((S+b)\log R\bigr)
 =O\!\left(zk\log|X|
                \log\frac{zkd\log|X|}{\eps}\right)
\]
words.

To finish the proof, we need to do a union bound over all $(x, C)$ tuples,
where there are at most $|X|^{k+1}$ of them.
After the union bound we have that the failure probability
is at most $O(e^{-b} |X|^{k + 1}) \leq O(1)$ by our choice of $b$, as desired.
\end{proof} \section{Construction of Hierarchical Lipschitz Assignments}
We restate the definition of Lipschitz hierarhical assignment and state an efficient construction of HLA in \Cref{lemma:HLA}.
We prove \Cref{lemma:HLA} by constructing the assignment in one
dimension, then taking products across coordinates.

\defHLA*

\begin{remark}
    \label{remark:HLA}
We remark that one must consider the fractional assignment, instead of integral, in order for this Lipschitz property to hold.
Indeed, if the assignment is integral (equivalent to defining the assignment by mapping $h_\ell : \mathbb{R}^d \to W_\ell$),
then $F_\ell(\cdot,\cdot) \in \{0, 1\}$;
now, item 3 implies $\lim_{x \to c} |F_\ell(a, x) - F_\ell(a, c)| = 0$,
so it must be the case that $F_\ell(a, x) = F_\ell(a, c) = 1$ or $F_\ell(a, x) \ F_\ell(a, c) = 0$, for all $x \to c$.
Because $x$ and $c$ can be arbitrary in $\mathbb{R}^d$,
then $x \mapsto F_\ell(a, x)$ must be constant over $\mathbb{R}^d$, and hence it must violate the locality (item 2).

Similarly, this error term $\pi$ is necessary even for $z = 2$,
as otherwise $x \mapsto F_\ell(a, x)$ must be constant (without dependence on $x$), thereby violating locality.
To see this, suppose otherwise that \eqref{eqn:lip} always holds with $\pi = 0$.
Now consider any $x, c$ and subdivide the segment from $x$ to $c$ into $n$ pieces,
such that the $j$-th endpoint is $x_j := c + \frac{j}{n} (x - c)$.
Then $\|F_\ell(\cdot, x) - F_\ell(\cdot, c)\|_1 \leq \sum_{i = 0}^{n-1} \|F_\ell(\cdot, x_i) - F_\ell(\cdot, x_{i + 1})\|_1 \leq n \cdot O_\ell(\|(x - c) / n\|_1^2)
=O_\ell(\|x - c\|_1^2) / n$,
which tends to $0$ when $n \to \infty$.
This implies $F_\ell(a, x) = F_\ell(a, c)$ for every $x, c$.
\end{remark}

\begin{lemma}
  \label{lemma:HLA}
    For every integer $z, L, d \geq 1$ there exists an HLA with
    $W_\ell=s_\ell\mathbb Z^d$ for every $\ell\in[L]$.
    The HLA can be described by $\tilde{O}(d)$ space,
    and evaluating the value of $F, U$ for any input takes
    the same order of space.
    The running time for evaluation is $O(\poly(z)dL)$.
\end{lemma}

We prove \Cref{lemma:HLA} in two steps. 
In \Cref{sec:one-dimensional}, we construct the assignment in one dimension, verify locality and refinement, and derive formulas for evaluating the entries in small space. 
Then, \Cref{lem:one-dimensional} supplies the estimates needed for the Lipschitz approximation property. In \Cref{sec:general-dimension}, we take products across coordinates and combine the one-dimensional estimates into the required approximation bound.

\subsection{Proof of \texorpdfstring{\Cref{lemma:HLA}}{Lemma}: One Dimension}
\label{sec:one-dimensional}

\paragraph{Construction.}
For $1\le\ell\le L$, let $W_\ell\coloneqq s_\ell\Z$.
Let $V_1,\ldots,V_z$ be independent uniform
random variables on $[0,1)$, and set
\[
  S_L\coloneqq\sum_{j=1}^z V_j.
\]
For $2\le\ell\le L$, in decreasing order, draw
$K_\ell\sim\operatorname{Binomial}(z,1/2)$ independently of all
previously drawn variables, and set
\begin{equation}
  S_{\ell-1}\coloneqq\frac{S_\ell+K_\ell}{2}.
  \label{eq:shift-coupling}
\end{equation}
At every level, $S_\ell$ has the distribution of a sum of $z$ independent
uniform variables on $[0,1)$. Indeed, let $V$ be uniform on $[0,1)$
and let $B$ be an independent fair bit. Then $(V+B)/2$ is uniform on
$[0,1)$. Applying this fact independently to the $z$ summands proves
the claim by induction.

For this sampled hierarchy, define $h_\ell:\R\to W_\ell$
for $1\le\ell\le L$ and $p_\ell:W_\ell\to W_{\ell-1}$
for $2\le\ell\le L$ by
\begin{equation}
\begin{aligned}
  h_\ell(x)&\coloneqq s_\ell\left\lfloor\frac{x}{s_\ell}-S_\ell\right\rfloor
    &&(x\in\R),\\
  p_\ell(s_\ell v)&\coloneqq s_{\ell-1}\left\lfloor\frac{v-K_\ell}{2}\right\rfloor
    &&(v\in\Z).
\end{aligned}
\label{eq:hash-parent}
\end{equation}
The same sampled maps are used for all input points. Define their
probability matrices, with the domains in \Cref{def:HLA}, by
\begin{equation}
  F_\ell(a,x)\coloneqq\Pr[h_\ell(x)=a],
  \qquad
  U_\ell(b,a)\coloneqq\Pr[p_\ell(a)=b].
  \label{eq:probability-matrices}
\end{equation}
These are deterministic matrices of probabilities. Each column has
finite support and sums to one, since $0\le S_\ell<z$ and
$K_\ell\in\{0,\ldots,z\}$.

\paragraph{Locality and Refinement.}
Fix $1\le\ell\le L$ and $x\in\R$.
Since flooring changes the value by less than one, $0\le S_\ell<z$, $0\le K_\ell\le z$, and $s_{\ell-1}=2s_\ell$, both $h_\ell$ and $p_\ell$ move each input by at most $O(z)s_\ell$
in every realization. Therefore, their probability matrices $F_\ell$ and $U_\ell$ satisfy locality.

To prove refinement, use $s_{\ell-1}=2s_\ell$ and
\eqref{eq:shift-coupling} to obtain
\[
\begin{aligned}
  \left\lfloor\frac{x}{s_{\ell-1}}-S_{\ell-1}\right\rfloor
  &=\left\lfloor\frac{x/s_\ell-S_\ell-K_\ell}{2}\right\rfloor\\
  &=\left\lfloor\frac{\lfloor x/s_\ell-S_\ell\rfloor-K_\ell}{2}\right\rfloor.
\end{aligned}
\]
The second equality holds because $K_\ell$ is an integer. Consequently,
$h_{\ell-1}(x)=p_\ell(h_\ell(x))$ for every realization of the hierarchy.
Moreover, $p_\ell$ depends only on $K_\ell$, which is independent of $S_\ell$
and hence of $h_\ell$. Therefore, for every $b\in W_{\ell-1}$,
\[
\begin{aligned}
  F_{\ell-1}(b,x)
  &=\sum_{a\in W_\ell}\Pr[h_\ell(x)=a,\ p_\ell(a)=b]\\
  &=\sum_{a\in W_\ell}F_\ell(a,x)U_\ell(b,a),
\end{aligned}
\]
as required.

\paragraph{Evaluating the Entries.}
To obtain the algorithmic guarantee in \Cref{lemma:HLA},
we express its probability entries by explicit formulas that can be evaluated in small space.
For each integer $0\le n\le z$ and $t\in\R$, define
\[
  \beta_n(t)\coloneqq
  \Pr\!\left[t-1<\sum_{j=1}^nV_j\le t\right].
\]
For $u\in\R$, let $u_+\coloneqq\max\{u,0\}$.
For $1\le\ell\le L$, $v\in\Z$, and $x\in\R$, the distribution
of $S_\ell$ gives
\[
  F_\ell(s_\ell v,x)=\beta_z(x/s_\ell-v)
\]
To efficiently evaluate $F_\ell(s_\ell v,x)$, we prove that $\beta_z(t)$ admits the following explicit formula:
\begin{equation}
    \beta_z(t)=\frac1{z!}\sum_{j=0}^{z+1}
       (-1)^j\binom{z+1}{j}(t-j)_+^z,
    \label{eq:entry-formula}
\end{equation}

This formula follows by repeated integration of the interval indicator. The empty sum gives
$\beta_0(t)=\mathbf1_{[0,1)}(t)$. Conditioning on the last uniform
variable gives $\beta_n(t)=\int_0^1\beta_{n-1}(t-u)\,du$ for $1\le n\le z$.
For $n=1$, this is
$\beta_1(t)=t_+-2(t-1)_++(t-2)_+$. For $2\le n\le z$, integrate each term
using
\[
  \int_0^1(t-u-j)_+^{n-1}\,du
  =\frac{(t-j)_+^n-(t-j-1)_+^n}{n}
\]
and combine adjacent coefficients by Pascal's identity. This proves
\eqref{eq:entry-formula} by induction.

For $2\le\ell\le L$ and $u,v\in\Z$, the parent entry is
\begin{equation}
\begin{aligned}
  U_\ell(s_{\ell-1}u,s_\ell v)
  &=2^{-z}\left[\binom z{v-2u}+\binom z{v-2u-1}\right]\\
  &=2^{-z}\binom{z+1}{v-2u},
\end{aligned}
\label{eq:parent-entry}
\end{equation}
with binomial coefficients zero outside their usual range, since
$\lfloor(v-K_\ell)/2\rfloor=u$ exactly when
$K_\ell\in\{v-2u,v-2u-1\}$. 
The formulas \eqref{eq:entry-formula} and \eqref{eq:parent-entry} can be evaluated using $O(z^2)$ arithmetic operations and $O(1)$ auxiliary words, giving the stated space and time bound in \Cref{lemma:HLA}.

\paragraph{One-dimensional Lipschitz Approximation.}
To prove the remaining Lipschitz approximation property, we establish Taylor estimates for the assignment at scale $1$. Using the random variables
$V_1,\ldots,V_z$ from the construction, define, for each integer
$0\le n\le z$ and $t\in\R$, the probability vector
\[
  \bfv_n(t)
  \coloneqq
  \left(\Pr\!\left[\left\lfloor t-\sum_{j=1}^nV_j\right\rfloor=v\right]
  \right)_{v\in\Z}.
\]
For $n=0$, this means
$(\bfv_0(t))_v=\mathbf 1_{\{\lfloor t\rfloor=v\}}$.
In every case, $\norm{\bfv_n(t)}=1$.

The following lemma supplies the one-dimensional approximation bound, and proving it for every truncation order $1 \le r \le z$ provides the estimates needed for tensorization.

\begin{lemma}
\label{lem:one-dimensional}
For every $c\in\R$, there are coefficient vectors
$\bbv_0(c),\ldots,\bbv_{z-1}(c)\in\ell_1(\Z)$,
with $\bbv_0(c)=\bfv_z(c)$, such that
\begin{equation}
  \norm{\bbv_j(c)}\le\frac{2^j}{j!}
  \qquad(0\le j<z),
  \label{eq:coefficient-bound}
\end{equation}
and, simultaneously for every integer $1\le r\le z$ and $\delta\in\R$,
\begin{equation}
  \left\|\bfv_z(c+\delta)
       -\sum_{j=0}^{r-1}\bbv_j(c)\delta^j\right\|_1
  \le\frac{2^r}{r!}|\delta|^r.
  \label{eq:one-dimensional-remainder}
\end{equation}
\end{lemma}

\begin{proof}
We first bound the derivatives of $\bfv_z$, then apply Taylor's formula.
Conditioning on the last uniform variable gives, for $1\le n\le z$
and $t\in\R$,
\begin{equation}
  \bfv_n(t)
  =\int_0^1\bfv_{n-1}(t-v)\,dv
  =\int_{t-1}^{t}\bfv_{n-1}(u)\,du.
  \label{eq:averaging}
\end{equation}
All integrals and derivatives are coordinatewise. On a bounded interval
of inputs, only finitely many coordinates can be nonzero.

First, every $\bfv_n$ with $1\le n\le z$ is $2$-Lipschitz.
For $a,b\in\R$ with $a<b$,
subtracting the integrals in \eqref{eq:averaging} gives
\[
  \bfv_n(b)-\bfv_n(a)
  =\int_a^b
    [\bfv_{n-1}(u)-\bfv_{n-1}(u-1)]\,du.
\]
Since each probability vector has norm $1$,
\begin{equation}
  \norm{\bfv_n(b)-\bfv_n(a)}
  \le\int_a^b(1+1)\,du
  =2|b-a|.
  \label{eq:basic-lipschitz}
\end{equation}
In particular, $\bfv_n$ is continuous for $1\le n\le z$.

For $2\le n\le z$, the integrand in \eqref{eq:averaging} is continuous.
The fundamental theorem of calculus therefore gives
\begin{equation}
  \bfv_n'(t)=\bfv_{n-1}(t)-\bfv_{n-1}(t-1).
  \label{eq:endpoint-derivative}
\end{equation}
Thus differentiation replaces an average by the difference of its two
endpoint values. Repeated application gives, for $0\le j\le z-1$,
\begin{equation}
  \bfv_z^{(j)}(t)
  =\sum_{k=0}^{j}(-1)^k\binom jk\,\bfv_{z-j}(t-k).
  \label{eq:finite-difference}
\end{equation}
Indeed, the case $j=0$ is immediate. For $j<z-1$, differentiate each
term using \eqref{eq:endpoint-derivative} and collect equal translates
using $\binom jk+\binom j{k-1}=\binom{j+1}{k}$. This step is valid because
$z-j\ge2$. All the asserted derivatives are continuous, since their
expressions involve $\bfv_n$ with $n\ge1$.

Taking norms in \eqref{eq:finite-difference}, we obtain
\begin{equation}
\begin{aligned}
  \norm{\bfv_z^{(j)}(t)}
  &\le\sum_{k=0}^{j}\binom jk\,\norm{\bfv_{z-j}(t-k)}\\
  &=\sum_{k=0}^{j}\binom jk=2^j.
\end{aligned}
\label{eq:derivative-bound}
\end{equation}
The factor $2^j$ is the sum of the absolute coefficients of the repeated
endpoint differences. Since $z-j\ge1$, applying
\eqref{eq:basic-lipschitz} to the same expression also gives,
for $t,u\in\R$,
\begin{equation}
\begin{aligned}
  \norm{\bfv_z^{(j)}(t)-\bfv_z^{(j)}(u)}
  &\le\sum_{k=0}^{j}\binom jk
      \norm{\bfv_{z-j}(t-k)-\bfv_{z-j}(u-k)}\\
  &\le\sum_{k=0}^{j}\binom jk\,2|t-u|
   =2^{j+1}|t-u|.
\end{aligned}
\label{eq:derivative-lipschitz}
\end{equation}

For $c\in\R$ and $0\le j<z$, define
$\bbv_j(c)\coloneqq\bfv_z^{(j)}(c)/j!$.
Then \eqref{eq:derivative-bound} proves \eqref{eq:coefficient-bound}.
For $r=1$, the remainder bound
\eqref{eq:one-dimensional-remainder} is simply
\eqref{eq:basic-lipschitz}.

Fix $c,\delta\in\R$ and an integer $2\le r\le z$.
First apply Taylor's formula only through
degree $r-2$. Its integral remainder gives
\begin{equation}
\begin{aligned}
  \bfv_z(c+\delta)
  ={}&\sum_{j=0}^{r-2}\frac{\delta^j}{j!}\bfv_z^{(j)}(c)\\
  &+\frac{\delta^{r-1}}{(r-2)!}
    \int_0^1(1-t)^{r-2}\bfv_z^{(r-1)}(c+t\delta)\,dt.
\end{aligned}
\label{eq:ordinary-integral-remainder}
\end{equation}
This formula follows from repeated applications of the fundamental
theorem of calculus. For example, when $r=2$, it is
\[
  \bfv_z(c+\delta)
  =\bfv_z(c)+\delta\int_0^1\bfv_z'(c+t\delta)\,dt.
\]
We next subtract the degree-$(r-1)$ Taylor term. It can be written with
the same integral weight, since
\[
\begin{aligned}
  &\frac{\delta^{r-1}}{(r-2)!}
    \int_0^1(1-t)^{r-2}\bfv_z^{(r-1)}(c)\,dt\\
  &\qquad=
    \frac{\delta^{r-1}}{(r-2)!}\cdot\frac1{r-1}
      \bfv_z^{(r-1)}(c)
    =\frac{\delta^{r-1}}{(r-1)!}\bfv_z^{(r-1)}(c).
\end{aligned}
\]
Subtracting this expression and the lower-degree terms from
\eqref{eq:ordinary-integral-remainder} gives
\begin{equation}
\begin{aligned}
  &\bfv_z(c+\delta)
    -\sum_{j=0}^{r-1}\frac{\delta^j}{j!}\bfv_z^{(j)}(c)\\
  &\quad=
    \frac{\delta^{r-1}}{(r-2)!}
    \int_0^1(1-t)^{r-2}
    [\bfv_z^{(r-1)}(c+t\delta)-\bfv_z^{(r-1)}(c)]\,dt.
\end{aligned}
\label{eq:integral-remainder}
\end{equation}
The bracket is a difference of derivative values, so
\eqref{eq:derivative-lipschitz} bounds its norm by $2^r t|\delta|$.
Consequently,
\[
\begin{aligned}
  \left\|\bfv_z(c+\delta)
      -\sum_{j=0}^{r-1}\bbv_j(c)\delta^j\right\|_1
  &\le\frac{2^r|\delta|^r}{(r-2)!}
       \int_0^1t(1-t)^{r-2}\,dt\\
  &=\frac{2^r}{r!}|\delta|^r,
\end{aligned}
\]
where $\int_0^1t(1-t)^{r-2}\,dt=1/[r(r-1)]$.
This proves \eqref{eq:one-dimensional-remainder} for every $1\le r\le z$.
\end{proof}

\paragraph{Completing One-dimensional Proof.}
We finish the proof of the Lipschitz approximation property by applying \Cref{lem:one-dimensional} at scale $s_l$ with truncation order $z$.
For $1\le\ell\le L$, $c\in\R$, and $v\in\Z$, define
$\pi_{\ell,s_\ell v,c}:\R\to\R$ by
\[
  \pi_{\ell,s_\ell v,c}(x)
  \coloneqq\sum_{j=1}^{z-1}(\bbv_j(c/s_\ell))_v
       \left(\frac{x-c}{s_\ell}\right)^j.
\]
The sum is empty when $z=1$. Since
$(F_\ell(s_\ell v,x))_{v\in\Z}=\bfv_z(x/s_\ell)$,
\Cref{lem:one-dimensional} with $r=z$ gives
\begin{equation}
  \sum_{v\in\Z}
  |F_\ell(s_\ell v,x)-F_\ell(s_\ell v,c)-\pi_{\ell,s_\ell v,c}(x)|
  \le O(2^z)\left(\frac{|x-c|}{s_\ell}\right)^z.
  \label{eq:one-dimensional-integer-order}
\end{equation}
The polynomials have degree at most $z-1$ and vanish at $x=c$.
Thus the construction satisfies all requirements of
\Cref{lemma:HLA} when $d=1$. We next extend
the same matrices to arbitrary dimension.

\subsection{Proof of \texorpdfstring{\Cref{lemma:HLA}}{Lemma}: General Dimension by Tensorization}
\label{sec:general-dimension}

We now take products of the one-dimensional assignments. Column normalization, locality, refinement, and efficient evaluation follow coordinate-wise; the remaining task is to combine the
one-dimensional Lipschitz approximation bounds while keeping total degree at most $z-1$.

\paragraph{Construction and the First Two Properties.}

Let $F_\ell^{(1)}$ and $U_\ell^{(1)}$ denote the one-dimensional
matrices constructed above. For $1\le\ell\le L$, let
$W_\ell\coloneqq s_\ell\Z^d$. Taking independent copies of the entire
one-dimensional hierarchy in each coordinate gives the probability matrices
\begin{equation}
  F_\ell(a,x)\coloneqq\prod_{j=1}^d F_\ell^{(1)}(a_j,x_j),
  \qquad
  U_\ell(b,a)\coloneqq\prod_{j=1}^d U_\ell^{(1)}(b_j,a_j).
  \label{eq:product-matrices}
\end{equation}
Here $F_\ell$ is defined for $1\le\ell\le L$, $a\in W_\ell$,
and $x\in\R^d$; $U_\ell$ is defined for $2\le\ell\le L$,
$b\in W_{\ell-1}$, and $a\in W_\ell$.
Their columns have finite support and sum to one. Coordinatewise
locality gives Euclidean distance at most $O(z)\sqrt d\,s_\ell$.
For $2\le\ell\le L$, $b\in W_{\ell-1}$, and $x\in\R^d$,
refinement follows by factoring the finite sum:
\[
\begin{aligned}
  \sum_{a\in s_\ell\Z^d}U_\ell(b,a)F_\ell(a,x)
  &=\prod_{j=1}^d\left(
       \sum_{a_j\in s_\ell\Z}U_\ell^{(1)}(b_j,a_j)F_\ell^{(1)}(a_j,x_j)
     \right)\\
  &=\prod_{j=1}^dF_{\ell-1}^{(1)}(b_j,x_j)
   =F_{\ell-1}(b,x).
\end{aligned}
\]
\paragraph{Implicit Storage and Entry Evaluation.}
Given access to the scales, formulas \eqref{eq:entry-formula}, \eqref{eq:parent-entry}, and \eqref{eq:product-matrices} specify the HLA using $O(1)$ additional words. 
Each entry is a product of $d$ one-dimensional factors, so evaluation takes $O(dz^2)$ time and $O(1)$ auxiliary words. Including the query coordinates gives $O(d)$ words of
space beyond the given scales, proving the resource bounds in \Cref{lemma:HLA}.

\paragraph{Tensorization of the  Lipschitz Approximation Bound.}
It remains to prove the Lipschitz approximation guarantee. 
Let $\bfv:\R\to\ell_1(\Z)$ take values in probability vectors.
For every $c\in\R$, let
$\bbv_0(c),\ldots,\bbv_{z-1}(c)\in\ell_1(\Z)$ be coefficient vectors
with $\bbv_0(c)=\bfv(c)$ satisfying
\eqref{eq:coefficient-bound} and \eqref{eq:one-dimensional-remainder}
with $\bfv$ in place of $\bfv_z$.
For $x\in\R^d$, define
\[
  \bF_d(x)\coloneqq\bigotimes_{j=1}^d\bfv(x_j),
  \qquad
  (\bF_d(x))_v=\prod_{j=1}^d(\bfv(x_j))_{v_j}
  \quad(v\in\Z^d).
\]
For vectors $a,b$, the identity used below is $\norm{a\otimes b}=\norm a\,\norm b$.

The following lemma provides the general-dimensional approximation bound, uses only the one-dimensional coefficient and remainder bounds.
The correction term $\pi_{\ell,a,c}(x)$ in \eqref{eqn:lip} must have total degree at most $z-1$. 
We bound the error of truncating the product by total degree, using the one-dimensional estimates.

\begin{lemma}[Tensorization]
\label{lem:tensorization}
For $c,u\in\R^d$ and an integer $1\le r\le z$, let $\btau_r(c,u)$ be obtained
by expanding
\[
  \bigotimes_{j=1}^d
  \left(\sum_{k=0}^{r-1}\bbv_k(c_j)u_j^k\right)
\]
and retaining only terms of total degree less than $r$ in $u$.
Then
\begin{equation}
  \norm{\bF_d(c+u)-\btau_r(c,u)}
  \le\frac{2^r}{r!}\norm u^r.
  \label{eq:tensor-remainder}
\end{equation}
Moreover, the homogeneous degree-$j$ part $\btau_r^{[j]}(c,u)$ satisfies
\begin{equation}
  \norm{\btau_r^{[j]}(c,u)}
  \le\frac{2^j}{j!}\norm u^j
  \qquad(0\le j<r).
  \label{eq:tensor-coefficients}
\end{equation}
\end{lemma}

\begin{proof}
We induct on $d$, proving \eqref{eq:tensor-remainder} simultaneously
for all $1\le r\le z$. For $d=1$, it is exactly the assumed
one-dimensional remainder bound.

For the induction step, let $c'\coloneqq(c_2,\ldots,c_d)\in\R^{d-1}$
and $u'\coloneqq(u_2,\ldots,u_d)\in\R^{d-1}$.
Grouping terms by their degree in $u_1$ gives
\begin{equation}
  \btau_r(c,u)
  =\sum_{j=0}^{r-1}\bbv_j(c_1)u_1^j
      \otimes\btau_{r-j}(c',u').
  \label{eq:truncated-product}
\end{equation}
After using degree $j$ in the first coordinate, the remaining coordinates
may contribute total degree less than $r-j$.

Let
\[
  \mathbf v\coloneqq\bF_{d-1}(c'+u'),
  \qquad
  \mathbf e\coloneqq\bfv(c_1+u_1)
        -\sum_{j=0}^{r-1}\bbv_j(c_1)u_1^j.
\]
Since $\mathbf v$ is a probability vector, $\norm{\mathbf v}=1$.
Subtracting \eqref{eq:truncated-product} from the exact product gives
\begin{equation}
\begin{aligned}
  \bF_d(c+u)-\btau_r(c,u)
  ={}&\mathbf e\otimes\mathbf v\\
  &+\sum_{j=0}^{r-1}\bbv_j(c_1)u_1^j
       \otimes[\mathbf v-\btau_{r-j}(c',u')].
\end{aligned}
\label{eq:product-error}
\end{equation}
Use the one-dimensional remainder bound for $\mathbf e$, the coefficient
bound for $\bbv_j(c_1)$, and the induction hypothesis for each remaining
factor. The tensor norm identity then yields
\[
\begin{aligned}
  \norm{\bF_d(c+u)-\btau_r(c,u)}
  &\le\frac{2^r|u_1|^r}{r!}
    +\sum_{j=0}^{r-1}\frac{2^j|u_1|^j}{j!}
       \frac{2^{r-j}\norm{u'}^{r-j}}{(r-j)!}\\
  &=\frac{2^r}{r!}\sum_{j=0}^{r}\binom rj
       |u_1|^j\norm{u'}^{r-j}\\
  &=\frac{2^r}{r!}(|u_1|+\norm{u'})^r
   =\frac{2^r}{r!}\norm u^r.
\end{aligned}
\]
This proves \eqref{eq:tensor-remainder}.

For \eqref{eq:tensor-coefficients}, expand the degree-$j$ terms and
apply the coefficient bound and the multinomial theorem:
\[
  \norm{\btau_r^{[j]}(c,u)}
  \le
  \sum_{\substack{\alpha_1,\ldots,\alpha_d\in\Nzero\\
                   \alpha_1+\cdots+\alpha_d=j}}
    \prod_{k=1}^d\frac{(2|u_k|)^{\alpha_k}}{\alpha_k!}
  =\frac{2^j}{j!}\norm u^j.
\]
\end{proof}

\paragraph{Completing General-dimensional Proof.}
We finish the proof of the Lipschitz approximation property in general dimension by applying \Cref{lem:tensorization} at scale $s_l$ with truncation order $z$.
Apply \Cref{lem:tensorization} with $\bfv\coloneqq\bfv_z$.
For $1\le\ell\le L$ and $x\in\R^d$, \eqref{eq:product-matrices} gives
\[
  (F_\ell(s_\ell v,x))_{v\in\Z^d}
  =\bigotimes_{j=1}^d\bfv_z(x_j/s_\ell)
  =\bF_d(x/s_\ell).
\]
For $1\le\ell\le L$, $c\in\R^d$, and $v\in\Z^d$, define
$\pi_{\ell,s_\ell v,c}:\R^d\to\R$ by
\[
  \pi_{\ell,s_\ell v,c}(x)
  \coloneqq\left(\btau_z\!\left(\frac{c}{s_\ell},\frac{x-c}{s_\ell}\right)\right)_v
    -F_\ell(s_\ell v,c).
\]
These polynomials have total degree at most $z-1$ and vanish at $c$.
The tensorization lemma gives
\[
  \sum_{v\in\Z^d}
  |F_\ell(s_\ell v,x)-F_\ell(s_\ell v,c)-\pi_{\ell,s_\ell v,c}(x)|
  \le O(2^z)\left(\frac{\lVert x-c\rVert_1}{s_\ell}\right)^z.
\]
This is precisely \eqref{eqn:lip}. Together with refinement, locality,
column normalization, and the resource bounds above, this proves
\Cref{lemma:HLA}.

\section{Robust Inner Product Estimation}
\label{sec:inner_prod}

We state our adversarially robust inner product estimation algorithm
in \Cref{lemma:linear_query}.
Our algorithm only needs to know $|\mathcal G|$ during initialization;
when evaluating a query, it accesses the coordinates of $g$ as needed.

\begin{lemma}
\label{lemma:linear_query}
Let $U$ be a domain and $\mathcal G\subseteq[-1,1]^U$ a fixed query family.
Given $\varepsilon,\delta\in(0,1)$ and $0<r_-\leq r_+$,
there exists an adversarially robust streaming algorithm that processes $m$ coordinate updates
to a frequency vector $x\in\mathbb R^U$,
where $r_-\leq\|x\|_1\leq r_+$ whenever $x\ne0$.
After each update, the algorithm reports a mapping
$\widehat I:\mathcal G\to\mathbb R$ such that, with probability at least $1-\delta$,
\begin{equation}
      \label{eqn:inner_prod_error}
      \forall g\in\mathcal G,\qquad
      |\widehat I(g)-\langle x,g\rangle|
      \leq\varepsilon\|x\|_1
\end{equation}
holds simultaneously after all updates.
Also, after each update, $\widehat I$ is linear: for any $g_1,g_2\in\R^U$ and any $a,b\in\R$, we have $\widehat I(ag_1+bg_2)=a\widehat I(g_1)+b\widehat I(g_2)$.
The algorithm uses space
\[
      \log|\mathcal G|\cdot
      \poly\!\left(\varepsilon^{-1}\log\frac1\delta
        \log\frac{r_+}{r_-}\log(m|U|)\right).
\]

\end{lemma}

The proof first establishes the error guarantee for a single fixed $g\in\mathcal G$,
and then takes a union bound over $\mathcal G$.
We start with the non-adaptive algorithm in \Cref{lemma:inner_prod_non}.
Here the update sequence is oblivious, no lower bound on the nonzero norms is needed,
and the probability guarantee is for each fixed query separately.
The algorithm uses signed $\ell_1$ samples and an estimate of $\|x\|_1$ to estimate $\langle x,g\rangle$.
The same samples and norm estimate define the mapping for every $g\in[-1,1]^U$.
The proof of \Cref{lemma:inner_prod_non} is deferred to \Cref{sec:proof_inner_prod_non}.
We remark that the linearity of $\widehat I$ in the statement is only needed in our polynomial-time result in \Cref{poly:sec:polynomial}.

\begin{restatable}{lemma}{lemmainnerprodnon}
      \label{lemma:inner_prod_non}
Let $U$ be a domain.
Given $\varepsilon,\delta\in(0,1)$ and $r_+>0$,
there exists a streaming algorithm that processes $m$ oblivious coordinate updates
to a frequency vector $x\in\mathbb R^U$,
such that $\|x\|_1\leq r_+$ after every update.
After each update, it reports a mapping
$\widehat I:[-1,1]^U\to\mathbb R$.
For every $g\in[-1,1]^U$,
with probability at least $1-\delta$,
\[
      |\widehat I(g)-\langle x,g\rangle|
      \leq\varepsilon\|x\|_1
\]
holds simultaneously after all updates.
Also, after each update, $\widehat I$ is linear: for any $g_1,g_2\in\R^U$ and any $a,b\in\R$, we have $\widehat I(ag_1+bg_2)=a\widehat I(g_1)+b\widehat I(g_2)$.
The algorithm uses space
\[
      O\!\left(\varepsilon^{-2}\log\frac{m}{\delta}\right)
      \cdot\polylog\!\left(\frac{m|U|r_+}{\varepsilon}\right).
\]
\end{restatable}

For a fixed query family $\mathcal G$, replacing $\delta$ by $\delta/|\mathcal G|$
gives simultaneous accuracy for all its queries and all times.
The resulting space dependence on $|\mathcal G|$ is logarithmic.

\paragraph{Proof Overview.}
Let $x^{(1)}, \ldots, x^{(m)}$ be the frequency vectors after each update.
Then the strategy is to break the input stream into substreams,
such that one can apply this non-adaptive algorithm in each substream independently,
and aggregate the result yields the final estimate for $\langle x, g\rangle |$.

More specifically, let us start with a motivating special case,
where we assume that $r\leq \|x^{(t)}\|_1 \leq 2r$ for some $r$ for all $t \in [m]$.
The key feature of this, is that the accumulated updates for every interval $[i, j] \subseteq [m]$,
has $\|x^{(j)} - x^{(i - 1)}\|_1 \leq 4r$ by triangle inequality.

To utilize this, first think about answering the inner product after we see the full update sequence, denoted as $S$.
Then the elements of $S$ are $(x^{(1)}, x^{(2)} - x^{(1)}, \ldots, x^{(m)} - x^{(m - 1)})$.
We break $S$ into $t := O(\log m)$ contiguous subsequences $S_1, \ldots, S_t$.
Then we run the non-adaptive algorithm \Cref{lemma:inner_prod_non} for each $S_i$ whose input is the subsequence of operations in $S_i$.
Then each part will have an $\epsilon \cdot 4r$ additive error,
and they sum to $O(\epsilon t  r)$ additive error in total, which we can afford by rescaling $\epsilon$ by a $t = O(\log m)$ factor.
The fact that this plan entirely operates on the update sequence
is crucial for robustness,
since once the operation sequence $S_i$ is revealed (for some $i$), 
the adversary cannot change it anymore.
This is different to operating on the elements in $U$, since the adversary can only append to the operation sequence whereas it can change the frequency of $U$ arbitrarily.

The remaining issue is that we need to report the inner products after every update, which may change the partition of $S$.
To this end, we use a greedy dyadic partition for $S$.
For example, we partition $[1, 15]$ with $[1, 8] \cup [9, 12] \cup [13, 14] \cup [15]$,
which may also be interpreted as the binary representation.
In general, over the course of $m$ updates, this dyadic partition always has $O(\log m)$ parts,
and for any $t \in [m]$,
once $t$ is not the start point of an interval, it will not be after all later updates.

Hence, for every $t$,
we can maintain a non-adaptive structure \Cref{lemma:inner_prod_non}
for $O(\log m)$ intervals with lengths $1, 2, \ldots, m$,
and remove once this $t$ stops being the start of some interval.
Again, when we need to reveal the result of an interval,
the interval is already fixed and the adversary can no longer make changes to it.
The general case can be handled by creating substreams with respect to $\|x^{(t)}\|_1$,
which may be estimated in a robust way by using the robust $\ell_1$-norm estimator~\cite{focs/GLWYZ26}.

\begin{proof}[Proof of \Cref{lemma:linear_query}]
We first prove the result for $r_-=1$.
A general lower bound $r_->0$ is handled by dividing all updates by $r_-$ and multiplying the answers by $r_-$;
the upper bound on the norm then becomes $r_+/r_-$.
Write $u_t=x^{(t)}-x^{(t-1)}$ for the $t$-th coordinate update.

\paragraph{Notation.}
Let $\mathcal R$ consist of the powers of two from $1$ up to the first power that is at least $2r_+$, so $|\mathcal R|=O(\log(2r_+))$.
We use dyadic lengths $2^j$ for $0\leq j<\ell$, where $\ell=O(\log(2m))$ is the number of powers of two at most $m$.
Let $\rho_t\in\mathcal R\cup\{0\}$ denote the scale chosen at time $t$ by the algorithm below.
When its norm estimator succeeds, $\rho_t$ is a constant-factor approximation to $\|x^{(t)}\|_1$, with $\rho_t=0$ when $x^{(t)}=0$.
We say that update $t$ is assigned to $\rho\in\mathcal R$ if $\rho_t=\rho$; an update with $\rho_t=0$ is unassigned.
For $\rho\in\mathcal R$, $0\leq j<\ell$, and a multiple $a\geq0$ of $2^j$, define a block $B=(\rho,j,a)$ to be the stream interval starting just after the $a$-th update assigned to $\rho$ and ending with the $(a+2^j)$-th such update.
Blocks with $a=0$ start before the first stream update.
The length $2^j$ counts the updates assigned to $\rho$ in the block.
The block contains every stream update between its endpoints, including updates assigned to other scales and unassigned updates.

\paragraph{Algorithm: Initialization.}
We use a sketch $\mathcal S_{\mathrm{norm}}$ to robustly estimate $\|x^{(t)}\|_1$ throughout the stream.
The robust $\ell_1$-norm estimator of~\cite{focs/GLWYZ26} applies to a vector on $U$ undergoing at most $m$ coordinate updates of bounded magnitude.
In our case, $\|x^{(t)}\|_1\leq r_+$ and
$\|u_t\|_1\leq\|x^{(t)}\|_1+\|x^{(t-1)}\|_1\leq2r_+$.
Initialize $\mathcal S_{\mathrm{norm}}$ on the zero vector using this estimator, on $U$ for at most $m$ updates, each of magnitude at most $2r_+$, with a sufficiently small constant accuracy parameter and failure probability $\delta/2$.
The estimator uses $\poly(\log(m|U|r_+/\delta))$ space, with the failure parameter obtained by universe padding.

For each $\rho\in\mathcal R$, initialize a counter $J_\rho=0$ to count the updates assigned to $\rho$.
For each $\rho\in\mathcal R$ and $0\leq j<\ell$, initialize an empty sketch $\mathcal S_{\rho,j}$ using \Cref{lemma:inner_prod_non} for the first block $(\rho,j,0)$.
Each instance is initialized on $U$ for at most $m$ updates with norm upper bound $2r_+$, and
\[
  \eta:=\frac{\varepsilon}{8\ell},
  \qquad \delta':=\frac{\delta}{4m|\mathcal R||\mathcal G|}
\]
as its accuracy and failure probability.
All these sketches use independent randomness, also independent of $\mathcal S_{\mathrm{norm}}$.
For each pair $(\rho,j)$, initialize $\widehat I_{\rho,j}=\bot$, indicating that no output mapping has been stored yet.

\paragraph{Algorithm: Updates.}
On update $u_t$, first feed $u_t$ to $\mathcal S_{\mathrm{norm}}$ and to every $\mathcal S_{\rho,j}$.
{From the estimate returned by $\mathcal S_{\mathrm{norm}}$ after this update, obtain a nonnegative estimate $R_t$ such that, with probability at least $1-\delta/2$,}
\[
  \|x^{(t)}\|_1\leq R_t\leq2\|x^{(t)}\|_1
  \qquad\text{for all }t\in[m].
\]
If $R_t=0$, set $\rho_t=0$ and end the update procedure without changing any $J_\rho$.
Otherwise, compute
\[
  \rho_t:=\min\{\rho\in\mathcal R:\rho\geq\min\{R_t,2r_+\}\},
\]
and increment $J_{\rho_t}$ by one.
The largest scale is at least $2r_+$, so this choice is defined even if the norm estimate is inaccurate.

Set $\rho=\rho_t$.
For each $j$ such that $2^j$ divides the new value of $J_\rho$, the block
$B=(\rho,j,J_\rho-2^j)$ has just completed.
Let $\widehat I_B$ denote the output mapping of $\mathcal S_{\rho,j}$ at this moment.
Store its sample and norm representation by setting $\widehat I_{\rho,j}\leftarrow\widehat I_B$.
Then replace $\mathcal S_{\rho,j}$ by a new instance with the same parameters and fresh independent randomness.
The new instance starts with update $u_{t+1}$.
Thus $\widehat I_{\rho,j}$ stores the output mapping for the most recently completed block of length $2^j$ at scale $\rho$.
This mapping receives no updates and is replaced only when the next block of this length and scale completes.
No output uses the state of $\mathcal S_{\rho,j}$ before its block completes.

\paragraph{Algorithm: Answering Queries.}
Consider a query $g\in\mathcal G$ after update $t$.
If $\rho_t=0$, return $\widehat I(g)=0$.
Otherwise, let $\rho=\rho_t$ and let $J$ be the value of $J_\rho$ after this update.
Write the binary expansion of $J$ as
\[
  J=\sum_{i=1}^q2^{j_i},
  \qquad j_1>\cdots>j_q\geq0,
  \qquad q\leq\ell.
\]
Define $b_0=0$, $b_i=\sum_{h=1}^i2^{j_h}$, and $B_i=(\rho,j_i,b_{i-1})$.
The blocks $B_1,\ldots,B_q$ are consecutive intervals covering the stream up to the $J$-th update assigned to $\rho$.
For each $i$, the binary expansion gives
\[
  0\leq J-b_i<2^{j_i},
  \qquad
  2^{j_i}\left\lfloor J/2^{j_i}\right\rfloor=b_i.
\]
Thus the last completed block of length $2^{j_i}$ ends at assignment $b_i$, and the update rule gives
$\widehat I_{\rho,j_i}=\widehat I_{B_i}$.
Return
\[
  \widehat I(g):=\sum_{i=1}^q\widehat I_{\rho,j_i}(g).
\]
We prove the linearity of $\widehat I$.
By \Cref{lemma:inner_prod_non}, each stored mapping $\widehat I_{\rho,j_i}$ is linear in query $g\in\R^U$.
The scale $\rho$ and the selected blocks are determined by the stored state independently of the current query $g$, so their sum is also linear in $g$.
When $\rho_t=0$, the returned mapping is the zero functional.
Thus linearity holds for each update, regardless of whether the accuracy guarantees hold.

\paragraph{Analysis: Robustness.}
The norm sketch $\mathcal S_{\mathrm{norm}}$ already tolerates adaptively chosen updates.
The block sketches and the adversary together form an adaptive strategy whose randomness is independent of $\mathcal S_{\mathrm{norm}}$ and whose updates depend on this sketch only through its earlier estimates.
Thus the robustness guarantee of $\mathcal S_{\mathrm{norm}}$ applies to the entire execution.
We need to justify using the non-adaptive algorithm of \Cref{lemma:inner_prod_non} for the block sketches.
The key is that an unfinished block's sketch contributes nothing to the reported answers.
Its output is used only after its input sequence is complete.
We show that this prevents the adversary from choosing that input sequence based on the sketch's randomness.

For this argument, we may view the independent randomness for every possible block as sampled in advance; this is equivalent to sampling it when the block starts.
Consider a possible block $B=(\rho,j,a)$ and fix the randomness of the adversary, $\mathcal S_{\mathrm{norm}}$, and every other block sketch.
Before $B$ starts, no computation uses its randomness.
While $B$ is unfinished, every reported answer uses only the norm estimates, the counters, and the mappings from completed blocks.
Thus, inductively, all answers reported before $B$ completes, and all updates received by its sketch through completion, are independent of the randomness for $B$.
The same holds for the starting and completion times of $B$, since these are determined by the norm estimates and the counters.
Under this fixing of the other randomness, the input to the sketch for $B$ is therefore an oblivious sequence, as required by \Cref{lemma:inner_prod_non}.
Once a block completes, its input and output mapping remain fixed, even when later updates and queries depend on that mapping.

\paragraph{Analysis: Accuracy and Failure Probability.}
We first bound the errors of all completed block estimates, and then use these bounds to control the error in each reported inner product.
To apply \Cref{lemma:inner_prod_non}, we first verify its input requirements.
At any time while processing a block, its sketch represents the difference between the current stream vector and the stream vector just before the block started.
The triangle inequality bounds this difference by $2r_+$ in $\ell_1$ norm, and the sketch receives at most $m$ updates.
These are the norm and stream-length bounds used to initialize the sketch.

Let $v_B$ be the sum of the updates in a completed block $B$.
Together with the independence established in the robustness analysis, the preceding input bounds allow us to apply \Cref{lemma:inner_prod_non}.
For each possible block $B$ and fixed $g\in\mathcal G$, the probability that $B$ completes but violates
\begin{equation}
  \label{eqn:inner_prod_block_error}
  |\widehat I_B(g)-\langle v_B,g\rangle|
  \leq\eta\|v_B\|_1
\end{equation}
is at most $\delta'$.
This bound holds after fixing the other randomness as above, and hence also after averaging over it.
The input bounds hold for any sequence of scale assignments, including those produced by inaccurate norm estimates.
Thus this step establishes accuracy of the block mappings; accuracy of the final answer will additionally require accurate norm estimates.

Let $\mathcal E_{\mathrm{ip}}$ denote the event that \eqref{eqn:inner_prod_block_error} holds for every completed block $B$ and every $g\in\mathcal G$.
For each scale, the number of possible block labels that can complete is at most
\[
  \sum_{j=0}^{\ell-1}\left\lfloor\frac{m}{2^j}\right\rfloor<2m.
\]
A union bound over these blocks, all $|\mathcal R|$ scales, and all $g\in\mathcal G$ gives
\[
  \Pr[\mathcal E_{\mathrm{ip}}]\geq1-2m|\mathcal R||\mathcal G|\delta'
  =1-\delta/2.
\]
Finally, let $\mathcal E_{\mathrm{norm}}$ denote the event that
$\|x^{(t)}\|_1\leq R_t\leq2\|x^{(t)}\|_1$ for every $t\in[m]$.
By the robustness of $\mathcal S_{\mathrm{norm}}$ established above,
$\Pr[\mathcal E_{\mathrm{norm}}]\geq1-\delta/2$.
Both probability bounds hold for the original adaptive execution, so a union bound gives
$\Pr[\mathcal E_{\mathrm{norm}}\cap\mathcal E_{\mathrm{ip}}]\geq1-\delta$.

We now assume $\mathcal E_{\mathrm{norm}}\cap\mathcal E_{\mathrm{ip}}$ and bound the error in the reported inner product using \eqref{eqn:inner_prod_block_error}.
Fix a time $t$ and a query $g\in\mathcal G$.
If $x^{(t)}=0$, then $R_t=\rho_t=0$ and the query procedure returns zero.
Otherwise, $R_t\geq\|x^{(t)}\|_1\geq1$ and $R_t\leq2r_+$.
The definition of $\rho_t$ therefore gives
$R_t\leq\rho_t\leq2R_t$, and hence
\begin{equation}
  \label{eqn:inner_prod_scale}
  \|x^{(t)}\|_1\leq\rho_t\leq4\|x^{(t)}\|_1.
\end{equation}

For each $\rho$, let $\tau_{\rho,a}$ denote the index in the original stream of the $a$-th update assigned to $\rho$, and set $\tau_{\rho,0}=0$.
We use these indices only in the analysis to express the block vectors in terms of their endpoints.
For a completed block $B=(\rho,j,a)$, the definition of $v_B$ gives
\[
  v_B=\sum_{h=\tau_{\rho,a}+1}^{\tau_{\rho,a+2^j}}u_h
      =x^{(\tau_{\rho,a+2^j})}-x^{(\tau_{\rho,a})}.
\]
Let $\rho=\rho_t$, and use $J$, $b_i$, and $B_i$ from the query procedure at time $t$.
By definition of $J$, the $J$-th assignment to $\rho$ occurs at time $t$, so $\tau_{\rho,J}=t$.
Consequently,
\[
  \sum_{i=1}^q v_{B_i}
  =\sum_{i=1}^q
    \left(x^{(\tau_{\rho,b_i})}-x^{(\tau_{\rho,b_{i-1}})}\right)
  =x^{(t)}.
\]
To express the final error relative to the current $\|x^{(t)}\|_1$, we sharpen the global block norm bound using the current scale $\rho$.
Each endpoint of a selected block is either time zero or an update assigned to $\rho$.
At every such assignment, \eqref{eqn:inner_prod_scale} bounds the norm of the stream vector by $\rho$.
Both endpoints of each block vector $v_{B_i}$ therefore have norm at most $\rho$, giving $\|v_{B_i}\|_1\leq2\rho$.
Using $\widehat I_{\rho,j_i}=\widehat I_{B_i}$ and applying \eqref{eqn:inner_prod_block_error} to these blocks, we obtain
\begin{align*}
  |\widehat I(g)-\langle x^{(t)},g\rangle|
  &\leq\sum_{i=1}^q
       |\widehat I_{B_i}(g)-\langle v_{B_i},g\rangle|\\
  &\leq\eta\sum_{i=1}^q\|v_{B_i}\|_1
   \leq2\eta\rho\ell\\
  &\leq8\eta\ell\|x^{(t)}\|_1
   =\varepsilon\|x^{(t)}\|_1.
\end{align*}
Since $\mathcal E_{\mathrm{norm}}\cap\mathcal E_{\mathrm{ip}}$ gives these bounds for all blocks and queries, \eqref{eqn:inner_prod_error} holds simultaneously at every time, including for queries chosen adaptively from $\mathcal G$.

\paragraph{Analysis: Space.}
The algorithm stores $|\mathcal R|$ counters $J_\rho$, $|\mathcal R|\ell$ sketches $\mathcal S_{\rho,j}$, and at most $|\mathcal R|\ell$ mappings $\widehat I_{\rho,j}$.
Each counter uses $O(\log m)$ bits.
By \Cref{lemma:inner_prod_non}, each sketch and its output mapping use
\[
  O\!\left(\eta^{-2}\log\frac{m}{\delta'}\right)
  \cdot\polylog\!\left(\frac{m|U|r_+}{\eta}\right)
\]
space.
Substituting $\eta=\varepsilon/(8\ell)$ and $\delta'=\delta/(4m|\mathcal R||\mathcal G|)$, their total space is
\[
  O\!\left(\frac{|\mathcal R|\ell^3}{\varepsilon^2}
       \log\frac{m^2|\mathcal R||\mathcal G|}{\delta}\right)
  \cdot\polylog\!\left(\frac{m|U|r_+\ell}{\varepsilon}\right).
\]
Adding the space of $\mathcal S_{\mathrm{norm}}$ and the counters gives the claimed bound for $r_-=1$, since $|\mathcal R|=O(\log(2r_+))$ and $\ell=O(\log(2m))$.
The dependence on $|\mathcal G|$ is logarithmic because it enters only through $\log(1/\delta')$.
For the original stream, dividing the updates by $r_-$ replaces the upper bound $r_+$ by $r_+/r_-$, as required.
\end{proof}

\section{Main Theorem: Robust Streaming Algorithm}
\label{sec:combine}

\begin{theorem}
\label{thm:main}
For \kzC and given $0 < \varepsilon < 1$,
there is an adversarially robust streaming algorithm that
takes as input an adaptive sequence of $m$ point insertion and deletions from $[\Delta]^d$,
reports for all $t \in [m]$ a function $\widehat{\cost}_t$ after the $t$-th update,
such that with high constant probability,
$\widehat{\cost}_t$ is an $\varepsilon$-oracle for $P_t$ simultaneously for all $t \in [m]$,
where $P_t$ is the dataset after the $t$-th update.
The algorithm uses $\tO(kd^{z}\varepsilon^{-O(z)})$ space for general $z$,
and $\tO(kd \varepsilon^{-O(z)})$ for $z = 1, 2$.
\end{theorem}

Let $\mu_1, \ldots, \mu_m \in \mathbb{N}^{\cX}$ be the frequency vectors after each update.
We start with some general transformation,
such that our cost oracle will be randomized and the plan is to devise an oracle
that works for all $k$-point subset $C$ that come only from a candidate set $X_\varepsilon$
(defined below in \eqref{eqn:Xeps}) with high probability.
Then in \Cref{lem:finite-queries} we show that after some cheap postprocessing one can turn this into a true $\varepsilon$-oracle.
In \Cref{sec:alg_ingredient} we give several subroutines/ingredients of the algorithm,
and we present the proof of \Cref{thm:main} in \Cref{sec:proof_main}.

\paragraph{Universal Domain $X_\varepsilon$.}
Let $D := \sqrt{d}\Delta$ be an upper bound on the diameter of the
input domain $[\Delta]^d$, and choose an enlarged radius
$R := \Theta(zD/\varepsilon)$.
Let $u := O(\varepsilon/(mz))$ be the spacing parameter, and choose a fixed
dyadic grid $\mathcal N$ covering $B(0,R)$, with coordinate
spacing $u/\sqrt d$, so that rounding any point in this ball moves it
by at most $u$.
Throughout the construction below, we use the following discrete finite
domain in place of $\mathbb R^d$:
\begin{equation}
  \label{eqn:Xeps}
\xfar := R e_1, \quad
X_\varepsilon
:=
\mathcal N\cup\cX
\cup\bigcup_{q=1}^m\left(q^{-1}\mathbb Z^d\cap[1,\Delta]^d\right)
\cup\{\xfar\}.
\end{equation}
Here $\mathcal N$ allows us to round query centers, $\cX$ contains the
stream's input points, the union over $q$ contains all possible cluster
means, and $\xfar$ handles queries whose centers lie outside $B(0,R)$.
Indeed, a cluster of integer mass $q\leq m$ has its mean in
$q^{-1}\mathbb Z^d\cap[1,\Delta]^d$.
The added mean points lie in $[1,\Delta]^d$ and number at most
$m(m\Delta+1)^d$, so $\log|X_\varepsilon|=\tO(d)$.

We first record a lower bound for nonzero clustering costs.
\begin{fact}\label{fact:positive_cost}
For every $\mu\in\mathbb Z_{\geq0}^{\cX}$, if $|\supp(\mu)|\leq k$,
then $\opt(\mu)=0$; otherwise, $\opt(\mu)\geq2^{1-z}$.
\end{fact}
\begin{proof}
If $|\supp(\mu)|\leq k$, placing a center at each support point gives
cost zero. Otherwise, for any nonempty center set $Q$ of size at most
$k$, two distinct points $x,y\in\supp(\mu)$ have the same nearest
center, after breaking ties arbitrarily. Since their weights are
positive integers and $\|x-y\|_2\geq1$,
\[
  \cost(\mu,Q)
  \geq\dist(x,Q)^z+\dist(y,Q)^z
  \geq2^{1-z}\|x-y\|_2^z
  \geq2^{1-z}.
\]
Taking the minimum over $Q$ proves the claim.
\end{proof}

\begin{lemma}\label{lem:finite-queries}
  For every nonnegative vector \(\mu\in\mathbb{Z}^{\cX}\), with \(|\supp(\mu)|>k\) and \(\|\mu\|_1\le m\), suppose a function \(\widehat{\cost}\) satisfies
\[
\widehat{\cost}(C)\in(1\pm\varepsilon/10)\cost(\mu,C)
\]
for every nonempty \(C\subseteq X_\varepsilon\) with \(|C|\le k\).
Then deterministic post-processing turns \(\widehat{\cost}\) into an \(\varepsilon\)-oracle, using \(\widetilde O(1)\) additional working bits.
\end{lemma}

\begin{proof}
We construct an estimate $\widehat{\cost}'(C)$ for every nonempty
$C\subseteq\mathbb R^d$ with $|C|\le k$.
Given such a query, let
\[
a\coloneqq\dist(0,C),
\qquad
C_{\mathrm{near}}\coloneqq C\cap B(0,R).
\]
When $C_{\mathrm{near}}\ne\varnothing$, let $C'$ be obtained by
rounding each center in $C_{\mathrm{near}}$ coordinatewise to
$\mathcal N$, using a fixed rule to break ties and removing
repetitions. Return
\[
\widehat{\cost}'(C)\coloneqq
\begin{cases}
\widehat{\cost}(C'), & C_{\mathrm{near}}\ne\varnothing,\\[2mm]
\displaystyle
\left(\frac aR\right)^z \widehat{\cost}(\{\xfar\}),
& C_{\mathrm{near}}=\varnothing.
\end{cases}
\]
Every evaluation of $\widehat{\cost}$ uses a nonempty subset of $X_\varepsilon$
of size at most $k$.

The intuition is that, if all centers are far from all data points ($C_{\mathrm{near}} = \emptyset$),
then simply assigning all points to the nearest center point changes the cost by only a little (and even rounding it to $\xfar$ is fine, subject to scaling).
In the other case, we can discard the far centers, and for near centers we can
round and use the proxy in $X_\varepsilon$.
This rounding cost is bounded because we choose the spacing $u = O(1 / m)$ which is small enough.

We now prove the approximation guarantee.
For the analysis, let
\[
n\coloneqq\|\mu\|_1,
\qquad
\beta\coloneqq\frac{R+D}{R-D}.
\]
Every input point has norm at most $D$, and our choice of $R$ gives
\[
\beta^z
\le \exp\left(\frac{3zD}{R}\right)
\le 1+\frac{\varepsilon}{10}.
\]

First suppose that $C_{\mathrm{near}}=\varnothing$, so $a>R$.
The triangle inequality gives
\[
n(a-D)^z
\le \cost(\mu,C)
\le n(a+D)^z,
\]
and
\[
n(R-D)^z
\le \cost(\mu,\{\xfar\})
\le n(R+D)^z.
\]
Combining these inequalities with $a>R$ and the guarantee for
$\widehat{\cost}(\{\xfar\})$ yields
\[
(1-\varepsilon/10)\beta^{-z}
\le
\frac{\widehat{\cost}'(C)}{\cost(\mu,C)}
\le
(1+\varepsilon/10)\beta^z.
\]
Both bounds lie within $1\pm\varepsilon$.

Now suppose that $C_{\mathrm{near}}\ne\varnothing$.
We first bound the effect of discarding the other centers.
For every $x\in[\Delta]^d$,
\[
\dist(x,C)
\le \dist(x,C_{\mathrm{near}})
\le \beta\,\dist(x,C).
\]
Indeed, if a nearest center belongs to $C_{\mathrm{near}}$, the
two distances are equal. Otherwise, $\dist(x,C)>R-D$, whereas
any retained center has distance at most $R+D$ from $x$.
Consequently,
\[
\cost(\mu,C)
\le \cost(\mu,C_{\mathrm{near}})
\le (1+\varepsilon/10)\cost(\mu,C).
\]

To control rounding, \Cref{fact:positive_cost} and the assumption
$|\supp(\mu)|>k$ give, for every nonempty center set $Q$ of size at most $k$,
\[
\cost(\mu,Q)^{1/z}\ge\frac12.
\]

Each retained center moves by at most $u$, so
\[
\bigl|\dist(x,C')-\dist(x,C_{\mathrm{near}})\bigr|
\le u
\qquad\text{for every }x\in[\Delta]^d.
\]
The triangle inequality for the weighted $\ell_z$ norm therefore
implies
\[
\left|
\cost(\mu,C')^{1/z}
-
\cost(\mu,C_{\mathrm{near}})^{1/z}
\right|
\le n^{1/z}u
\le mu
=\frac{\varepsilon}{100z}.
\]
Using $\cost(\mu,C_{\mathrm{near}})^{1/z}\ge1/2$, we obtain
\[
\left(1-\frac{\varepsilon}{50z}\right)^z
\le
\frac{\cost(\mu,C')}{\cost(\mu,C_{\mathrm{near}})}
\le
\left(1+\frac{\varepsilon}{50z}\right)^z.
\]
These bounds lie within $1\pm\varepsilon/10$.
Combining the rounding bound, the bound for discarding centers,
and the accuracy of $\widehat{\cost}(C')$ gives
\[
(1-\varepsilon/10)^2\cost(\mu,C)
\le \widehat{\cost}'(C)
\le (1+\varepsilon/10)^3\cost(\mu,C).
\]
For $0<\varepsilon<1$, this implies
\[
\widehat{\cost}'(C)\in(1\pm\varepsilon)\cost(\mu,C).
\]
Since $C$ was arbitrary, the guarantee holds simultaneously for
every valid query.

Finally, consider the additional working space under the
read-only query-coordinate access convention.
The grid $\mathcal N$ is represented by its spacing and coordinate
bounds. Coordinates of $C'$ are generated on demand.
Neither $\mathcal N$ nor $C'$ needs to be stored explicitly.
This uses a constant number of
$O(\log(kmdz\Delta/\varepsilon))$-bit indices and counters, together
with the $O(1)$ real-arithmetic registers allowed by the query
model. The additional working space, beyond evaluating $\widehat{\cost}$, is
therefore $\widetilde O(1)$ bits.
\end{proof}

\subsection{Ingredients of Streaming Algorithm}
\label{sec:alg_ingredient}

\paragraph{Dimension Reduction and Condition on Bounded Distortion.}
We apply \Cref{lemma:robust:projection}
with the ground set $X_\varepsilon$ as in \eqref{eqn:Xeps},
obtain $\tilde{d} = \poly(z\varepsilon^{-1}\log k)$, $n_g = (z \varepsilon^{-1}kd)^{O(z)}$ and (low-space representation of)
$G_1, \ldots G_{n_g}$.
Now, use the $G_j$'s to define a proxy of the cost
\begin{equation}
  \label{eqn:cost_hat}
  \widetilde{\cost}(\mu, C) := \frac{1}{n_g}\sum_{j \in [n_g]} \cost((G_j)_\#\mu, G_j C),
\end{equation}
where $((G_j)_\#\mu)_y\coloneqq\sum_{x:G_j(x)=y}\mu_x$ maps the frequency to the projected domain.
Let $\mathcal E_{\mathrm{dis}}$ be the event that the distance-preservation
guarantee \eqref{eq:projection-guarantee} of \Cref{lemma:robust:projection} holds.
In the remaining arguments we condition on $\mathcal E_{\mathrm{dis}}$.
Then \Cref{lemma:robust:projection} gives
\begin{equation}
  \label{eqn:distortion}
  \forall C \subseteq X_\varepsilon, 1 \leq |C| \leq k, \quad
  \widetilde{\cost}(\mu, C) \in (1 \pm \varepsilon) \cost(\mu, C).
\end{equation}
Moreover, by \Cref{lem:finite-queries}, after rescaling $\varepsilon$ by an absolute constant,
it suffices to maintain a function $\widehat{\cost} : \{C\subseteq X_\varepsilon : 1\leq |C|\leq k\} \to \mathbb{R}$ in adaptive stream,
such that with high constant probability,
\begin{equation}
  \label{eqn:finite_query}
  \forall C \subseteq X_\varepsilon, 1 \leq |C| \leq k, \qquad \widehat{\cost}(C) \in (1 \pm \varepsilon) \widetilde{\cost}(\mu, C).
\end{equation}

\paragraph{Defining $A, g_C$ via HLA.}
Next, we define $A$ and $g_C$
whose inner product approximates $\widetilde{\cost}(\mu,C)$,
with a remainder controlled below.
To this end, we need to obtain an HLA using \Cref{lemma:HLA}
in the \emph{low-dimensional target space} $\mathbb{R}^{\tilde d}$.
On $\mathcal E_{\mathrm{dis}}$, applying
\eqref{eq:projection-guarantee} to singleton center sets and using
nonnegativity gives, for every $j\in[n_g]$ and $x,y\in X_\varepsilon$,
\[
  \|G_jx-G_jy\|_2
  \leq \bigl((1+\varepsilon)n_g\bigr)^{1/z}\|x-y\|_2.
\]
Choose $s_1$ to be the smallest dyadic scale at least
$\max\{1,((1+\varepsilon)n_g)^{1/z}\diam(X_\varepsilon)\}$, and let
$s_\ell=2^{1-\ell}s_1$.
Since $\diam(X_\varepsilon)=O(z\sqrt d\Delta/\varepsilon)$,
we have $s_1\leq(zmkd\Delta/\varepsilon)^{O(1)}$.
Write $K=O(z\sqrt{\tilde d})$, $K\geq1$, for the locality bound of the HLA:
\[
  F_\ell(a,y)>0\ \Longrightarrow\ \|a-y\|_2\leq Ks_\ell,
  \qquad
  U_\ell(b,a)>0\ \Longrightarrow\ \|a-b\|_2\leq Ks_\ell.
\]
Choose $L$ so that
\begin{equation}
  \label{eqn:finest_scale}
  Ks_L=\frac{\varepsilon}{m(zmkd\Delta/\varepsilon)^{O(z)}}.
\end{equation}
The number of levels is
$L=1+\log_2(s_1/s_L)=\tO(z)$.
Apply \Cref{lemma:HLA} with dimension $\tilde d$, order $z$, and
$L$ levels at scales $s_1,\ldots,s_L$.
Let $W_\ell\subseteq\mathbb R^{\tilde d}$, $F_\ell$ for $\ell\in[L]$,
and $U_\ell$ for $2\leq\ell\leq L$ be the resulting HLA.
Its representation and entry evaluation use $\tO(\tilde d)$ space.

The rows of $A$, and the coordinates of $A\mu$ and $g_C$, are indexed by $(\ell,a,j)$,
where $\ell \in [L]$ is a level, $a \in W_\ell$ is a representative,
and $j \in [n_g]$ indexes a projection matrix $G_j$.
For each $x\in X_\varepsilon$, define column $x$ of $A$ by
\begin{equation}
  \label{eqn:def_A}
  A_{(\ell, a, j), x} := \frac{s_\ell^z}{n_g} F_\ell(a, G_j x)
\end{equation}
so $(A\mu)_{\ell, a, j} = \frac{s_\ell^z}{n_g} \sum_{x \in X_\varepsilon} F_\ell(a, G_j x)\mu_x$.
Input vectors on $\cX$ are extended by zero to $X_\varepsilon$.
We use the same cost and pushforward definitions for vectors on this domain.
\begin{align}
  \label{eqn:def_gC}
  (g_C)_{\ell, a, j} :=
  \begin{cases}
    \frac{1}{s_\ell^z} (\dist^z(a, G_j C) -  \sum_{b \in W_{\ell - 1}} U_\ell(b, a) \dist^z(b, G_j C)) & 2\leq\ell\leq L \\
    \frac{1}{s_\ell^z}\dist^z(a, G_j C) & \ell = 1
  \end{cases}
\end{align}
\begin{lemma}
  \label{lemma:cost_tilde_equal}
  For every $\mu\in\mathbb{R}^{X_\varepsilon}$ and every
  $C\subseteq X_\varepsilon$ with $1\leq|C|\leq k$,
  \begin{equation}
    \label{eqn:cost_terminal_identity}
    \widetilde{\cost}(\mu,C)=\langle A\mu,g_C\rangle+E_C(\mu),
  \end{equation}
  where $E_C$ is linear in $\mu$ and satisfies
  \[
    |E_C(\mu)|\leq\frac{\varepsilon2^{-z}}{m}\|\mu\|_1.
  \]
\end{lemma}
\begin{proof}
  We directly verify the definition.
  Fix $j\in[n_g]$. For $y\in\mathbb R^{\tilde d}$ and $\ell\in[L]$, write
  \[
    h_{\ell,j}(y):=\sum_{a\in W_\ell}F_\ell(a,y)\dist^z(a,G_jC).
  \]
  For $2\leq\ell\leq L$, refinement gives
  \begin{align*}
    &\sum_{a\in W_\ell}F_\ell(a,y)
      \sum_{b\in W_{\ell-1}}U_\ell(b,a)\dist^z(b,G_jC)\\
    &\qquad=\sum_{b\in W_{\ell-1}}\dist^z(b,G_jC)
      \sum_{a\in W_\ell}U_\ell(b,a)F_\ell(a,y)\\
    &\qquad=\sum_{b\in W_{\ell-1}}F_{\ell-1}(b,y)\dist^z(b,G_jC)
      =h_{\ell-1,j}(y).
  \end{align*}
  All these sums have finite support. Substituting
  \eqref{eqn:def_A} and \eqref{eqn:def_gC}, the factors $s_\ell^z$
  cancel and the levels telescope:
  \begin{align*}
    \langle A\mu,g_C\rangle
    &=\frac{1}{n_g}\sum_{j\in[n_g]}\sum_{x\in X_\varepsilon}\mu_x
      \left(h_{1,j}(G_jx)+\sum_{\ell=2}^{L}
        \bigl(h_{\ell,j}(G_jx)-h_{\ell-1,j}(G_jx)\bigr)\right)\\
    &=\frac{1}{n_g}\sum_{j\in[n_g]}\sum_{x\in X_\varepsilon}\mu_x h_{L,j}(G_jx).
  \end{align*}
  The difference from $\widetilde{\cost}(\mu,C)$ defines the linear
  remainder in \eqref{eqn:cost_terminal_identity}.
  If $F_L(a,G_jx)>0$, locality and the $1$-Lipschitz property of distance give
  \[
    |\dist^z(a,G_jC)-\dist^z(G_jx,G_jC)|
    \leq zKs_L(s_1+Ks_L)^{z-1}
    \leq 2^{z-1}zKs_Ls_1^{z-1}
    \leq\frac{\varepsilon2^{-z}}{m}.
  \]
  Here $\dist(G_jx,G_jC)\leq s_1$, and \eqref{eqn:finest_scale}
  also gives $Ks_L\leq s_1$.
  Column normalization gives the same bound for
  $|h_{L,j}(G_jx)-\dist^z(G_jx,G_jC)|$.
  Multiplying by $|\mu_x|$, summing over $x$, and averaging over $j$
  proves the claim.
\end{proof}

The algorithm will estimate only the inner product. We will bound the
omitted remainder in the accuracy analysis using \Cref{lemma:cost_tilde_equal}.
For the inner product, our plan is to apply the estimation subroutine in \Cref{lemma:linear_query}.
This essentially requires two conditions: a bounded $\|g_C\|_\infty$
and a bounded $\|A\mu\|_1$.
However, neither holds directly, and we need additional steps.

\paragraph{Clipping Error of $g_C$.}
The first step is to clip $g_C$ to obtain a bounded $\ell_\infty$ norm
while controlling the change in the inner product.
For a clipping parameter $r>0$, define
$\gamma^{[r]} : \mathbb{R} \to \mathbb{R}$ by
\begin{align*}
  \gamma^{[r]}(x) := \begin{cases}
    -r & x < -r \\
    r & x > r\\
    x & \text{otherwise}
  \end{cases}.
\end{align*}
Write $\gamma^{[r]}(g_C)$ for its coordinatewise application to $g_C$.
Clearly, $\|\gamma^{[r]}(g_C)\|_\infty \leq r$.
We give the following error bound.
\begin{lemma}
  \label{lemma:clipping_error}
  There is a choice $r=\poly(z\tilde d)^z\varepsilon^{-O(z)}$ such that every
  $\mu\in\mathbb{R}_{\geq0}^{\cX}$ and every
  $C\subseteq X_\varepsilon$ with $1\leq|C|\leq k$ satisfy
  \[
    |\langle A\mu,g_C-\gamma^{[r]}(g_C)\rangle|
    \leq\varepsilon\widetilde{\cost}(\mu,C).
  \]
\end{lemma}
\begin{proof}
  We use locality and column normalization.
  Fix $j\in[n_g]$ and $x\in\cX$.

  First consider level $1$. The choice of $s_1$ gives
  $\dist(G_jx,G_jC)\leq s_1$,
  so every representative receiving mass from $G_jx$ satisfies
  \[
    0\leq(g_C)_{1,a,j}
    \leq\left(\frac{\dist(G_jx,G_jC)}{s_1}+K\right)^z
    \leq(1+K)^z\leq r.
  \]
  Hence level $1$ contributes no clipping error.

  Now fix $2\leq\ell\leq L$ and $a$ with $F_\ell(a,G_jx)>0$.
  If $U_\ell(b,a)>0$, locality gives
  \[
    \dist(a,G_jC),\dist(b,G_jC)\leq\dist(G_jx,G_jC)+2Ks_\ell.
  \]
  Since distance to a set is $1$-Lipschitz,
  \[
    |\dist^z(a,G_jC)-\dist^z(b,G_jC)|
    \leq zKs_\ell\bigl(\dist(G_jx,G_jC)+2Ks_\ell\bigr)^{z-1}.
  \]
  Averaging with the weights $U_\ell(b,a)$, which sum to one, yields
  \begin{equation}
    \label{eqn:clipping_increment}
    s_\ell^z|(g_C)_{\ell,a,j}|
      \leq zKs_\ell\bigl(\dist(G_jx,G_jC)+2Ks_\ell\bigr)^{z-1}.
  \end{equation}
  For $z=1$, this gives $|(g_C)_{\ell,a,j}|\leq K\leq r$,
  proving that the clipping error is zero.

  Suppose $z>1$. The clipping function changes a coordinate only when
  $|(g_C)_{\ell,a,j}|>r$; coordinates of magnitude at most $r$ are unchanged.
  For a coordinate receiving mass from $G_jx$,
  \eqref{eqn:clipping_increment} and the choice of $r$ then imply
  \[
    \frac{\dist(G_jx,G_jC)}{s_\ell}+2K>\frac{8zK}{\varepsilon},
    \qquad\text{and hence}\qquad
    s_\ell<\frac{\varepsilon\dist(G_jx,G_jC)}{4zK}.
  \]
  At such a level, $2Ks_\ell<\varepsilon\dist(G_jx,G_jC)/(2z)$, so
  \begin{align*}
    \bigl(\dist(G_jx,G_jC)+2Ks_\ell\bigr)^{z-1}
      &\leq\left(1+\frac{\varepsilon}{2z}\right)^{z-1}
        \dist^{z-1}(G_jx,G_jC)\\
      &\leq2\dist^{z-1}(G_jx,G_jC).
  \end{align*}
  For any real $v$, the magnitude of this change is
  \[
    |v-\gamma^{[r]}(v)|=\max\{|v|-r,0\}=(|v|-r)_+.
  \]
  Thus the subscript $+$ denotes the positive part.
  Using \eqref{eqn:clipping_increment} and
  $\sum_{a\in W_\ell}F_\ell(a,G_jx)=1$, each such level contributes
  at most $2zKs_\ell\dist^{z-1}(G_jx,G_jC)$ to the error for $G_jx$.
  All other levels contribute zero. Since the scales halve at each level,
  the sum of scales below any threshold is at most twice that threshold.
  Consequently,
  \begin{align}
    &\sum_{\ell=1}^{L}\sum_{a\in W_\ell}
      s_\ell^zF_\ell(a,G_jx)(|(g_C)_{\ell,a,j}|-r)_+\notag\\
    &\qquad\leq2zK\dist^{z-1}(G_jx,G_jC)
      \sum_{\substack{\ell\in[L]:\\
        s_\ell<\varepsilon\dist(G_jx,G_jC)/(4zK)}}s_\ell\notag\\
    \label{eqn:clipping_pointwise}
    &\qquad\leq2zK\dist^{z-1}(G_jx,G_jC)
      \frac{\varepsilon\dist(G_jx,G_jC)}{2zK}
      =\varepsilon\dist^z(G_jx,G_jC).
  \end{align}

  Finally, $A\mu$ is nonnegative because $\mu$ and $F_\ell$ are nonnegative.
  Expanding the inner product and then substituting \eqref{eqn:def_A} gives
  \begin{align*}
    &|\langle A\mu,g_C-\gamma^{[r]}(g_C)\rangle|\\
    &\quad=\left|\sum_{\ell,a,j}(A\mu)_{\ell,a,j}
      \bigl((g_C)_{\ell,a,j}-\gamma^{[r]}((g_C)_{\ell,a,j})\bigr)\right|\\
    &\quad\leq\sum_{\ell,a,j}(A\mu)_{\ell,a,j}(|(g_C)_{\ell,a,j}|-r)_+\\
    &\quad=\frac{1}{n_g}\sum_{j\in[n_g]}\sum_{x\in\cX}\mu_x
      \sum_{\ell=1}^{L}\sum_{a\in W_\ell}
      s_\ell^zF_\ell(a,G_jx)(|(g_C)_{\ell,a,j}|-r)_+\\
    &\quad\leq\frac{\varepsilon}{n_g}\sum_{j\in[n_g]}\sum_{x\in\cX}
      \mu_x\dist^z(G_jx,G_jC)
      =\varepsilon\widetilde{\cost}(\mu,C).
  \end{align*}
  The first inequality is the triangle inequality and the formula for
  $|v-\gamma^{[r]}(v)|$ above. The last inequality applies
  \eqref{eqn:clipping_pointwise} to each $(j,x)$, and the last equality
  is \eqref{eqn:cost_hat}.
\end{proof}

\paragraph{Sparse Reference.}
We introduce a \emph{sparse reference} vector $\nu$ on $X_\varepsilon$
such that $\|A(\mu-\nu)\|_1$ is bounded by the optimum.
Its sparse description lets us evaluate the reference contribution
$\langle A\nu,\gamma^{[r]}(g_C)\rangle$ exactly.
The algorithm will find such a reference using the batched queries below.

\begin{lemma}
  \label{lemma:sparse_reference}
  Fix a precision
  \[
    \rho:=\left(\frac{\varepsilon}{zmkd\Delta}\right)^{O(z)}.
  \]
  For every $\mu\in\mathbb Z_{\geq0}^{\cX}$ with $\|\mu\|_1\leq m$,
  there exists $\nu\in(\rho\mathbb Z_{\geq0})^{X_\varepsilon}$ such that
  $\|\nu\|_1=\|\mu\|_1$,
  \[
    |\supp(\nu)|\leq
    \begin{cases}
      k,&z=1,2,\\
      k(d+1)^{z-1},&z>2,
    \end{cases}
    \qquad
    \|A(\mu-\nu)\|_1\leq O(4^zL\tilde d^{z/2})\opt(\mu).
  \]
\end{lemma}
\begin{proof}
  Let $P_i^*$ partition $\supp(\mu)$ into the clusters of an optimal solution
  with centers $c_i^*$, and let $\mu_i$ be the restriction of $\mu$ to $P_i^*$.
  We extend $\mu_i$ by zero to $X_\varepsilon$ and define $\nu:=\sum_i w_i$.
  For each cluster, we will choose an optimal solution
  $w_i\in\mathbb R_{\geq0}^{X_\varepsilon}$ of the linear program
  \begin{align*}
    \text{minimize}\quad
      &\sum_{x\in X_\varepsilon}(w_i)_x\|x-c_i^*\|_2^z\\
    \text{subject to}\quad
      &\sum_{x\in X_\varepsilon}(w_i)_x x^\alpha
        =\sum_{x\in P_i^*}(\mu_i)_x x^\alpha
      \quad\text{for }\alpha\in\mathbb{Z}_{\geq0}^d,\ \|\alpha\|_1\leq z-1,
  \end{align*}
  where $x^\alpha:=\prod_{t=1}^d x_t^{\alpha_t}$.
  The vector $\mu_i$ is feasible, and the constraint for $\alpha=0$ fixes
  the total mass. Since $X_\varepsilon$ is finite, an optimal solution exists.

  For $z=1$, mass is the only constraint. Assigning the entire cluster
  mass to a point of $X_\varepsilon$ nearest to $c_i^*$ is therefore optimal.
  For $z=2$, write
  \[
    \bar x_i:=\frac{1}{\|\mu_i\|_1}
      \sum_{x\in P_i^*}(\mu_i)_x x.
  \]
  The cluster mass is an integer at most $m$, so $\bar x_i\in X_\varepsilon$
  by \eqref{eqn:Xeps}. Take $w_i:=\|\mu_i\|_1e_{\bar x_i}$.
  This matches the mass and every first moment of $\mu_i$.
  The optimal squared-distance center of the cluster is $\bar x_i$,
  so this solution has objective zero and is optimal.
  Both constructions use one point with an integer weight.

  For $z>2$, choose an optimal extreme point. There is one equality
  constraint for each monomial of total degree at most $z-1$, namely
  \[
    \binom{d+z-1}{d}
    \leq(d+1)^{z-1}.
  \]
  An extreme point has at most as many positive coordinates as there are
  equality constraints, giving $|\supp(w_i)|\leq(d+1)^{z-1}$.
  In every case, feasibility and optimality give
  \[
    \|w_i\|_1=\|\mu_i\|_1,
    \qquad
    \cost(w_i,\{c_i^*\})\leq\cost(\mu_i,\{c_i^*\}).
  \]
  Summing over the at most $k$ clusters gives the support and mass claims for $\nu$.

  Write $\lambda_i:=\mu_i-w_i$. The LP constraints say that $\mu_i$ and $w_i$
  have equal weighted sums for each monomial of degree at most $z-1$.
  Since every polynomial $q:\mathbb{R}^d\to\mathbb{R}$ of degree at most
  $z-1$ is a linear combination of these monomials, we have
  \[
    \sum_{x\in X_\varepsilon}\lambda_{i,x}q(x)
    =\sum_{x\in P_i^*}(\mu_i)_xq(x)
      -\sum_{x\in X_\varepsilon}(w_i)_xq(x)=0.
  \]
  Fix $\ell\leq L$, $j\in[n_g]$, and $y\in P_i^*$.
  Apply the order-$z$ Lipschitz approximation property of the HLA at $G_jy$.
  By \eqref{eqn:lip}, there exist polynomials
  $\{\pi_{\ell,a,G_jy}\}_{a\in W_\ell}$ of degree at most $z-1$ such that,
  for every $x\in\mathbb{R}^d$,
  \[
    \sum_{a\in W_\ell}
      \left|F_\ell(a,G_jx)-F_\ell(a,G_jy)-\pi_{\ell,a,G_jy}(G_jx)\right|
    \leq O(2^z)\left(\frac{\|G_j(x-y)\|_1}{s_\ell}\right)^z.
  \]
  Fix such a family of polynomials.
  Since $G_j$ is linear, the function
  $x\mapsto F_\ell(a,G_jy)+\pi_{\ell,a,G_jy}(G_jx)$ also has degree at most
  $z-1$. Its weighted sum against $\lambda_i$ is therefore zero, and
  \begin{align*}
    &s_\ell^z\sum_{a\in W_\ell}
      \left|\sum_{x\in X_\varepsilon}\lambda_{i,x}F_\ell(a,G_jx)\right|\\
    &\quad=s_\ell^z\sum_{a\in W_\ell}
      \left|\sum_{x\in X_\varepsilon}\lambda_{i,x}
        \bigl(F_\ell(a,G_jx)-F_\ell(a,G_jy)
          -\pi_{\ell,a,G_jy}(G_jx)\bigr)\right|\\
    &\quad\leq s_\ell^z\sum_{x\in X_\varepsilon}|\lambda_{i,x}|
      \sum_{a\in W_\ell}
        \left|F_\ell(a,G_jx)-F_\ell(a,G_jy)
          -\pi_{\ell,a,G_jy}(G_jx)\right|\\
    &\quad\leq O(2^z)\sum_{x\in X_\varepsilon}|\lambda_{i,x}|
      \|G_j(x-y)\|_1^z\\
    &\quad\leq O(2^z\tilde d^{z/2})
      \sum_{x\in X_\varepsilon}(\mu_i+w_i)_x\|G_j(x-y)\|_2^z.
  \end{align*}
  The first inequality is the triangle inequality, the second uses the HLA
  approximation bound \eqref{eqn:lip}, and the last uses
  $|\lambda_{i,x}|\leq(\mu_i+w_i)_x$ and
  $\|v\|_1\leq\sqrt{\tilde d}\|v\|_2$ in the projected space.
  Average over $j$ using \eqref{eq:projection-guarantee}, which applies
  since $x,y\in X_\varepsilon$, and then over $y\in P_i^*$
  with weights $(\mu_i)_y/\|\mu_i\|_1$.
  The resulting bound is at most $O(2^z\tilde d^{z/2})$ times
  \begin{align*}
    &\frac{1}{\|\mu_i\|_1}\sum_{x\in X_\varepsilon}\sum_{y\in P_i^*}
      (\mu_i+w_i)_x(\mu_i)_y\|x-y\|_2^z\\
    &\qquad\leq 2^{z-1}
      \bigl(\cost(w_i,\{c_i^*\})+3\cost(\mu_i,\{c_i^*\})\bigr)
      \leq 2^{z+1}\cost(\mu_i,\{c_i^*\}),
  \end{align*}
  by the triangle inequality and the mass and cost bounds above.
  Using \eqref{eqn:def_A} and summing over clusters and levels yields
  \[
    \|A(\mu-\nu)\|_1
    \leq O(4^z L\tilde d^{z/2})\opt(\mu).
  \]

  \paragraph{Rounding the Reference Weights.}
  Choose $\rho^{-1}\in\mathbb N$ at the stated order.
  The weights are already integers for $z=1,2$, and also when
  $\opt(\mu)=0$, since a zero LP objective forces $w_i=\mu_i$.
  In these cases the precision claim holds.
  Otherwise, keep the support points fixed and round each weight of $\nu$
  down to a multiple of $\rho$. Put the remaining mass on one of these
  points, and call the resulting vector $\nu'$.
  Since the total mass is an integer, this remaining mass is also a
  multiple of $\rho$. Thus $\nu'$ has the required weight precision,
  the same mass, and support contained in $\supp(\nu)$.
  The total change in weights satisfies
  \[
    \|\nu'-\nu\|_1\leq2\rho k(d+1)^{z-1}.
  \]
  Column normalization gives $\|Ae_x\|_1=\sum_{\ell=1}^L s_\ell^z
  \leq2s_1^z$. Consequently,
  \[
    \|A(\nu'-\nu)\|_1
    \leq4\rho k(d+1)^{z-1}s_1^z
    \leq O(2^{-z})\leq O(\opt(\mu)).
  \]
  Here we use $s_1\leq(zmkd\Delta/\varepsilon)^{O(1)}$ and the choice of $\rho$.
  The last inequality uses $\opt(\mu)\geq2^{1-z}$ when $\opt(\mu)>0$,
  by \Cref{fact:positive_cost}.
  The triangle inequality now gives the claimed residual bound for $\nu'$.
\end{proof}

\paragraph{Robust Batched $\ell_1$ Queries.}

To find a reference, we enumerate a fixed finite family $\mathcal V$
of candidate vectors on $X_\varepsilon$, with the support bound and weight
precision from \Cref{lemma:sparse_reference}.
The main theorem proof below defines this family explicitly.
After each stream update, we query the residual norms $\|A(\mu_t-\nu)\|_1$ for all $\nu\in\mathcal V$ and retain a candidate with the smallest estimate among those of the required mass. 
These queries form one batch, so there are $m$ batches. 
We use the following batched variant of the estimator-corrector-learner construction of~\cite{focs/GLWYZ26}, whose space depends logarithmically on $|\mathcal V|$.

\begin{restatable}{lemma}{lemmabatchquery}
\label{lemma:batched-subtraction}
\cite{focs/GLWYZ26}
Let $U$ be a finite domain and $\mathcal V\subseteq\mathbb R^U$ a fixed finite query family. 
Given $\varepsilon,\delta\in(0,1)$ and $0<r_-\leq r_+$, there exists an adversarially robust streaming algorithm that processes $m$ coordinate updates to a frequency vector $\mu\in\mathbb R^U$, where $r_-\leq\|\mu-\nu\|_1\leq r_+$ for every $\nu\in\mathcal V$ after every update. 
The algorithm has coordinate access to the query vectors. 
After each update, the algorithm reports a mapping $\widehat r:\mathcal V\to\mathbb R_{\geq0}$ such that, with probability at least $1-\delta$,
\[\forall\nu\in\mathcal V,\qquad\left|\widehat r(\nu)-\|\mu-\nu\|_1\right|\leq\varepsilon\|\mu-\nu\|_1\]
holds simultaneously after all updates.
The algorithm uses space
\[\poly\left(\frac1\varepsilon\log\left(2m|U|\frac{r_+}{r_-}\right)\right)\cdot\left(\log|\mathcal V|+\log\frac1\delta\right).
\]
\end{restatable}
Here we briefly describe how to achieve \Cref{lemma:batched-subtraction} using the framework of \cite{focs/GLWYZ26}, which maintains an estimator, a corrector that flags inaccurate estimates, and a learner that updates the estimator using flagged queries. 
The framework recursively partitions the stream into blocks, so that within each recursive block, the number of learning updates is at most $n_{\mathrm{corr}}=\poly(\varepsilon^{-1},\log(m|U|r_+/r_-))$, independent of the number of queries tested per stream update. 
Algorithmically, we process each batch with the learner state fixed and advance the recursive counter once per batch. 
If all queries pass the corrector's tests, we return the estimator's answers.
Otherwise, we select one flagged query for the usual learner update and answer the entire batch using one previously unused independent sketch copy, maintained alongside the original sketches. 
Since the batch is fixed before this copy is revealed, its non-adaptive guarantee applies simultaneously to $\cV$ by a union bound.
We retire the copy after releasing its answers, using standard sketch switching~\cite{jacm/Ben-EliezerJWY22}.
Each corrected batch therefore consumes one copy and answers one learning update, so $n_{\mathrm{corr}}$ copies suffice per block, while the recursive counter remains upper bounded by $m$.

Finally, we summarize the bounded computation-path argument of \cite{focs/GLWYZ26}. 
Fix the randomness of the copies used to answer corrected batches, which is independent of the checking and learning sketches. 
Their answers are then determined by the stream prefix and the copy used.
At each correction, the path need only additionally specify which flagged query $\nu\in\cV$ the learner used. 
Over $b$ corrections this gives at most $|\cV|^b$ witness sequences. 
The failure probabilities of each sketch is then set to allow a union bound over these sequences and over all queries in each batch. 

\subsection{Proof of \texorpdfstring{\Cref{thm:main}}{Theorem}}
\label{sec:proof_main}
We now combine these ingredients and track the resulting approximation
and space bounds.
  We start with describing the algorithm.
  Use $X_\varepsilon$, the projections, and the scales defined earlier,
  and choose $r$ as guaranteed by \Cref{lemma:clipping_error}.
  We first obtain an $O(\varepsilon)$-oracle;
  rescaling $\varepsilon$ gives the required accuracy.

  \paragraph{A Finite Reference Family.}
  Let $\rho$ be the weight precision from \Cref{lemma:sparse_reference},
  and write
  \[
    k':=\begin{cases}
      k,&z=1,2,\\
      k(d+1)^{z-1},&z>2,
    \end{cases}
  \]
  for its support bound.
  {Define the fixed family of candidate references by}
  \[
    \mathcal V:=\left\{\nu\in(\rho\mathbb Z_{\geq0})^{X_\varepsilon}:
      \|\nu\|_1\leq m,\quad |\supp(\nu)|\leq k'\right\}.
  \]

  {
  \paragraph{Parameters for the Sketches.}
  Set the following parameter as an upper bound on the norms of all residual vectors
  $A(\mu_t-\nu)$, with $\nu\in\mathcal V$:
  \begin{equation}
    \label{eqn:main_residual_bound}
    M:=4ms_1^z.
  \end{equation}
  Indeed, column normalization and the geometric scales give
  $\|Ae_x\|_1=\sum_{\ell=1}^L s_\ell^z\leq2s_1^z$.
  Since $\|\mu_t\|_1\leq m$ and $\|\nu\|_1\leq m$ for every $\nu\in\mathcal V$,
  \[
    \|A(\mu_t-\nu)\|_1
    \leq\sum_{x\in X_\varepsilon}|(\mu_t-\nu)_x|\,\|Ae_x\|_1
    \leq2s_1^z(\|\mu_t\|_1+\|\nu\|_1)
    \leq M.
  \]

  The sketches use as their coordinate universe the rows of $A$ that can receive mass
  from a point in $X_\varepsilon$. Define
  \begin{equation}
    \label{eqn:main_feature_universe}
    \mathcal U:=\bigcup_{x\in X_\varepsilon}\supp(Ae_x),
    \qquad N:=|\mathcal U|.
  \end{equation}
  Locality gives finite support for each column, so $\mathcal U$ is finite.
  Every row outside $\mathcal U$ is identically zero on $X_\varepsilon$.
  Thus we regard $A\mu$ and $A\nu$ as vectors on $\mathcal U$ and evaluate
  inner product query vectors at these same coordinates, preserving all residual norms
  and inner products.

  We give a general transformation that satisfies the norm lower bound $r_-$ in
  \Cref{lemma:linear_query,lemma:batched-subtraction}, so that the remainder of the proof
  can ignore this requirement.
  For any $\tau>0$, append $\tau$ to the maintained vector $v$ and $0$ to each
  reference $w$ or inner product query $h$. Then
  \[
    \begin{aligned}
      \|(v,\tau)\|_1&=\|v\|_1+\tau\geq\tau,\\
      \|(v,\tau)-(w,0)\|_1&=\|v-w\|_1+\tau\geq\tau,\\
      \langle(v,\tau),(h,0)\rangle&=\langle v,h\rangle.
    \end{aligned}
  \]
  Thus both lemmas satisfy their lower norm requirement with $r_-=\tau$,
  even when the original vector or residual is zero.
  The added coordinate is initialized once and never updated, so the transformation
  adds one coordinate and one update, and preserves the query-family sizes.

  Here take $\tau=2^{-z}$. By \Cref{fact:positive_cost}, positive OPT is at least
  $2\tau$, so the added norm contribution is at most $\opt(\mu_t)/2$;
  zero-OPT states will be handled exactly by sparse recovery.
  The upper norm bound $2M$ verified below becomes $r_+=2M+\tau$, giving
  \begin{equation}
    \label{eqn:main_norm_ratio}
    \log\frac{r_+}{r_-}
    =\log\left(1+\frac{2M}{\tau}\right)
    =O(\log(2M)+z).
  \end{equation}
  This handles the $r_-$ requirements. Henceforth we apply both lemmas as if these
  requirements were absent, with the transformation understood, and omit the extra
  coordinate and $\tau$ from the notation.
  }

  \paragraph{Algorithm: Initialization.}
  {Write an input update as $\sigma e_x$, where $\sigma=1$ for an insertion
  and $\sigma=-1$ for a deletion.}
  The algorithm initializes the following structures before the stream starts.
  \begin{itemize}
    \item The representations of $G_1,\ldots,G_{n_g}$ from
      \Cref{lemma:robust:projection} and of the HLA from \Cref{lemma:HLA}.
      These give coordinate access to $A$ and $g_C$.
    \item A sparse recovery structure $\mathcal S_{\mathrm{rec}}$ on
      $\cX$~\cite{ciss/Cormode006}, using $\tO(kd)$ space and returning
      $\mu_t$ for $|\supp(\mu_t)|\leq k$. Otherwise, it returns the
      symbol $\bot$, indicating that the support has more than $k$ points.
      Successful outputs are determined by the input and reveal no private
      randomness, so a failure bound of $1/(100m)$ per recovery gives
      failure probability at most $1/100$ over the adaptive stream.
    \item The batched $\ell_1$-estimation structure $\mathcal S_{\mathrm{norm}}$ of
      {\Cref{lemma:batched-subtraction}, maintaining $A\mu_t$ on $\mathcal U$
      with query family $\{A\nu:\nu\in\mathcal V\}$ and relative accuracy $1/2$.
      Allow at most $mN+1$ coordinate updates,
      including initialization. Write $\widehat r_t(\nu)$ for twice the estimate
      returned for reference $A\nu$ after original update $t$.}
    \item The inner product estimation structure $\mathcal S_{\mathrm{ip}}$
      of \Cref{lemma:linear_query}.
      {Initialize this structure on $\mathcal U$ for at most
      $mN+1$ coordinate updates, including initialization.
      Set its accuracy to
      \begin{equation}
        \label{eqn:main_inner_product_accuracy}
        \eta:=\frac{\varepsilon}{r4^zL\tilde d^{z/2}}.
      \end{equation}
      Use the fixed query family
      \begin{equation}
        \label{eqn:main_inner_product_family}
        \mathcal H:=\left\{r^{-1}\gamma^{[r]}(g_C):
          C\subseteq X_\varepsilon,\quad 1\leq|C|\leq k\right\}.
      \end{equation}
      }
  \end{itemize}
  Give $\mathcal S_{\mathrm{rec}}$, $\mathcal S_{\mathrm{norm}}$, and $\mathcal S_{\mathrm{ip}}$
  failure probability $1/100$ each, and use independent private randomness
  for them, independent of the projections.
  Also maintain the total mass $\|\mu_t\|_1$ and the current reference,
  {starting with $\nu_0=0$.}

  \paragraph{Algorithm: Updates.}
  {On an update $\sigma e_x$, first update the mass counter and
  $\mathcal S_{\mathrm{rec}}$. Scan the coordinates of $\sigma Ae_x$ once,
  feeding each coordinate increment to $\mathcal S_{\mathrm{norm}}$.
  After the complete scan, enumerate the members $\nu\in\mathcal V$ with
  $\|\nu\|_1=\|\mu_t\|_1$. Query $\mathcal S_{\mathrm{norm}}$ at $A\nu$
  as each candidate is enumerated, and choose a minimizer $\nu_t$ of the scaled
  estimates $\widehat r_t(\nu)$, with a fixed rule for ties.
  References of the required mass exist by \Cref{lemma:sparse_reference}.}
Each candidate is enumerated as a list of at most $k'$ points and their
  positive weights. The search keeps the current list and the best list,
  and stores the minimizing list as $\nu_t$.
  {Next process the residual change}
  \[
    A(\mu_t-\nu_t)-A(\mu_{t-1}-\nu_{t-1})
      =\sigma Ae_x-A(\nu_t-\nu_{t-1})
  \]
  {by scanning its coordinates once and feeding each coordinate increment to $\mathcal S_{\mathrm{ip}}$.
  Its residual vector after the scan is $A(\mu_t-\nu_t)$.
  Use its output mapping only after the scan is complete, and write $\operatorname{Ans}(h,t)$
  for its answer at $h\in\mathcal H$.
  For the error analysis, define}
  \[
    {B_t:=\widehat r_t(\nu_t).}
  \]
  {Whenever both sketches succeed, the inner product error is at most
  $\eta B_t$, as verified in the accuracy analysis below.}
  After this update, discard $\nu_{t-1}$.

  \paragraph{Algorithm: Answering Queries.}
  If $\mathcal S_{\mathrm{rec}}$ recovers $\mu_t$ with
  $|\supp(\mu_t)|\leq k$, the algorithm reports the exact oracle
  {$\widehat{\cost}_t:C\mapsto\cost(\mu_t,C)$} from its recovered points and weights.
  This case is needed because $\opt(\mu_t)=0$, and the reduction in
  \Cref{lem:finite-queries} requires $|\supp(\mu_t)|>k$.
  Otherwise, $\mathcal S_{\mathrm{rec}}$ returns $\bot$, which implies
  $|\supp(\mu_t)|>k$ on its success event. The algorithm then evaluates an oracle $H_t$ on
  candidate center sets $C\subseteq X_\varepsilon$, $1\leq|C|\leq k$.
  It queries $\mathcal S_{\mathrm{ip}}$ at {$r^{-1}\gamma^{[r]}(g_C)$},
  adds the exact reference contribution
  $\langle A\nu_t,\gamma^{[r]}(g_C)\rangle$, and returns
  \begin{equation}
    \label{eqn:main_oracle}
    H_t(C):=
      \langle A\nu_t,\gamma^{[r]}(g_C)\rangle
      +r\operatorname{Ans}\bigl(r^{-1}\gamma^{[r]}(g_C),t\bigr).
  \end{equation}
  Here $\operatorname{Ans}(h,t)$ is the answer returned by $\mathcal S_{\mathrm{ip}}$.
  {The algorithm reports $\widehat{\cost}_t$, obtained by applying the
  post-processing of \Cref{lem:finite-queries} to $H_t$.}
  Since $\nu_t$ is stored explicitly, the reference contribution is
  evaluated exactly as
  \[
    \langle A\nu_t,\gamma^{[r]}(g_C)\rangle
    =\frac{1}{n_g}\sum_{x\in\supp(\nu_t)}(\nu_t)_x
      \sum_{j=1}^{n_g}\sum_{\ell=1}^L s_\ell^z
      \sum_{a\in W_\ell}F_\ell(a,G_jx)
        \gamma^{[r]}((g_C)_{\ell,a,j}).
  \]
  The sums over representatives have finite support by locality.
  This evaluation uses $\tO(|\supp(\nu_t)|)$ additional space beyond the stored reference
  and the coordinate-evaluation workspace for $A$ and $g_C$.

  \paragraph{Accuracy and Robustness.}
  We verify the remaining input requirements of \Cref{lemma:batched-subtraction,lemma:linear_query}.
  Both need a finite universe, fixed query families,
  a bound on the number of coordinate updates, and upper bounds on the relevant norms after each update.
  In addition, the batched estimator requires coordinate access to its references,
  and the inner product estimator requires queries of infinity norm at most one.
  The query families $\{A\nu:\nu\in\mathcal V\}$ and $\mathcal H$ consist of vectors on
  the finite universe $\mathcal U$. They are finite and, after fixing the projections,
  independent of the sketches' private randomness.
  The representations of the projections and the HLA give coordinate access to $Ae_x$;
  a coordinate of $A\nu$ is computed by summing the contributions of the points
  in the sparse description of $\nu$.
  For each sketch, initialization uses one coordinate update, and each original update
  gives at most $N$ more, for a total of at most $mN+1$.

  During the scan for original update $t$, let $y$ be the current vector on
  $\mathcal U$ maintained by $\mathcal S_{\mathrm{norm}}$.
  Each coordinate of $y$ equals the corresponding coordinate of either $A\mu_{t-1}$
  or $A\mu_t$. Consequently, for every $\nu\in\mathcal V$,
  \[
    \|y-A\nu\|_1
    \leq\|A(\mu_{t-1}-\nu)\|_1+\|A(\mu_t-\nu)\|_1
    \leq2M.
  \]
  During the scan for $\mathcal S_{\mathrm{ip}}$, each coordinate of the residual
  $y'$ on $\mathcal U$ equals the corresponding coordinate of either $A(\mu_{t-1}-\nu_{t-1})$
  or $A(\mu_t-\nu_t)$. Hence
  \[
    \|y'\|_1
    \leq\|A(\mu_{t-1}-\nu_{t-1})\|_1
      +\|A(\mu_t-\nu_t)\|_1
    \leq2M.
  \]
  Initially, the batched residuals on $\mathcal U$ have norm $\|A\nu\|_1\leq M$,
  and the inner product residual is zero. Thus the upper bound $2M$ holds
  at initialization and throughout every scan.
  Finally, clipping gives, for every $h=r^{-1}\gamma^{[r]}(g_C)\in\mathcal H$,
  \[
    \|h\|_\infty=\frac1r\|\gamma^{[r]}(g_C)\|_\infty\leq1.
  \]
  This verifies all requirements of the two lemmas.

  Define the following success events for the three data structures.
  \begin{itemize}
    \item $\mathcal E_{\mathrm{rec}}$: at every time $t\in[m]$,
      $\mathcal S_{\mathrm{rec}}$ returns $\mu_t$ if $|\supp(\mu_t)|\leq k$,
      and $\bot$ otherwise.
    \item {$\mathcal E_{\mathrm{norm}}$: the simultaneous success event of
      \Cref{lemma:batched-subtraction} for $\mathcal S_{\mathrm{norm}}$,
      over the entire coordinate stream and its fixed query family.}
    \item {$\mathcal E_{\mathrm{ip}}$: the simultaneous success event of
      \Cref{lemma:linear_query} for $\mathcal S_{\mathrm{ip}}$, with accuracy $\eta$
      from \eqref{eqn:main_inner_product_accuracy},
      over the entire coordinate stream and its fixed query family.}
  \end{itemize}
  Condition on $\mathcal E_{\mathrm{dis}}\cap\mathcal E_{\mathrm{rec}}
    \cap\mathcal E_{\mathrm{norm}}\cap\mathcal E_{\mathrm{ip}}$.
  Sparse recovery handles $|\supp(\mu_t)|\leq k$ exactly;
  all remaining states have $|\supp(\mu_t)|>k$, as required by
  \Cref{lem:finite-queries}.
  {Each $\widehat r_t(\nu)$ is twice a relative $1/2$-estimate, so it lies
  between one and three times the norm estimated by $\mathcal S_{\mathrm{norm}}$.
  The parameter setup bounds that norm by
  $\|A(\mu_t-\nu)\|_1+\opt(\mu_t)/2$ in these states, giving
  \begin{equation}
    \label{eqn:main_reference_estimate}
    \widehat r_t(\nu)
    \leq3\|A(\mu_t-\nu)\|_1+\frac32\opt(\mu_t).
  \end{equation}
  By \Cref{lemma:sparse_reference}, $\mathcal V$ contains a reference of the required
  mass and residual norm. Applying this bound and the choice of $\nu_t$ gives
  \begin{align*}
    B_t&=\min_{\nu\in\mathcal V:\,\|\nu\|_1=\|\mu_t\|_1}
        \widehat r_t(\nu)\\
      &\leq3\min_{\nu\in\mathcal V:\,\|\nu\|_1=\|\mu_t\|_1}
        \|A(\mu_t-\nu)\|_1+\frac32\opt(\mu_t)\\
      &\leq O(4^zL\tilde d^{z/2})\opt(\mu_t).
  \end{align*}}
  {We use this bound on $B_t$ to control the additive error from \Cref{lemma:linear_query}.}
  Fix a time $t$ with $|\supp(\mu_t)|>k$ and a query
  $C\subseteq X_\varepsilon$ with $1\leq|C|\leq k$, and let
  $h_C:=r^{-1}\gamma^{[r]}(g_C)\in\mathcal H$.
  {On $\mathcal E_{\mathrm{norm}}\cap\mathcal E_{\mathrm{ip}}$,
  $B_t=\widehat r_t(\nu_t)$ upper bounds the norm estimated at reference $A\nu_t$.
  This is also the norm used in the inner product sketch's error guarantee, which therefore gives}
  \[
    \left|\operatorname{Ans}(h_C,t)
      -\langle A(\mu_t-\nu_t),h_C\rangle\right|
    {\leq\eta B_t.}
  \]
  Multiplying by $r$ and using $rh_C=\gamma^{[r]}(g_C)$ gives
  \[
    \left|r\operatorname{Ans}(h_C,t)
      -\langle A(\mu_t-\nu_t),\gamma^{[r]}(g_C)\rangle\right|
    \leq r\eta B_t.
  \]
  By the definition of $H_t$ in \eqref{eqn:main_oracle} and the linearity of $A$,
  the exact reference contribution cancels:
  \begin{align*}
    &H_t(C)-\langle A\mu_t,\gamma^{[r]}(g_C)\rangle\\
    &\quad=\langle A\nu_t,\gamma^{[r]}(g_C)\rangle
      +r\operatorname{Ans}(h_C,t)
      -\langle A\mu_t,\gamma^{[r]}(g_C)\rangle\\
    &\quad=r\operatorname{Ans}(h_C,t)
      -\langle A(\mu_t-\nu_t),\gamma^{[r]}(g_C)\rangle.
  \end{align*}
  Therefore, substituting the choice of $\eta$ and the bound on $B_t$, we obtain
  \[
    \left|H_t(C)-\langle A\mu_t,\gamma^{[r]}(g_C)\rangle\right|
    \leq r\eta B_t
    =\frac{\varepsilon B_t}{4^zL\tilde d^{z/2}}
    \leq O(\varepsilon)\opt(\mu_t).
  \]

  Since $\|\mu_t\|_1\leq m$,
  \Cref{lemma:cost_tilde_equal} bounds the omitted remainder by
  \[
    |E_C(\mu_t)|
    \leq\frac{\varepsilon2^{-z}}{m}\|\mu_t\|_1
    \leq\varepsilon2^{-z}
    \leq\varepsilon\opt(\mu_t).
  \]
  Combining the remainder bound with \Cref{lemma:clipping_error},
  the inner product estimate, and \eqref{eqn:distortion} gives
  \begin{align*}
    |H_t(C)-\cost(\mu_t,C)|
      &\leq O(\varepsilon)\opt(\mu_t)
        +\varepsilon\widetilde{\cost}(\mu_t,C)
        +\varepsilon\cost(\mu_t,C)\\
      &\leq O(\varepsilon)\cost(\mu_t,C).
  \end{align*}
  After rescaling $\varepsilon$, \Cref{lem:finite-queries} gives the claimed
  oracle accuracy.

  To bound the failure probability, fix an arbitrary adaptive adversary
  and any realization of $G_1,\ldots,G_{n_g}$ for which
  $\mathcal E_{\mathrm{dis}}$ holds.
  Then $A$, $\mathcal V$, and {$\mathcal H$} are fixed independently of
  the private randomness of the three data structures.
  Each of $\mathcal E_{\mathrm{rec}}$, $\mathcal E_{\mathrm{norm}}$, and
  $\mathcal E_{\mathrm{ip}}$ fails with probability at most $1/100$,
  even when updates depend on previous outputs.
  {For either sketch, the other structures and their independent random seeds
  can be viewed as part of its adaptive adversary. In particular, the batched estimate
  used to choose $\nu_t$ is available before the corresponding coordinate updates to
  $\mathcal S_{\mathrm{ip}}$, so this choice is allowed by \Cref{lemma:linear_query}.}
  A union bound gives failure probability at most $3/100$ for these
  three events, for every such realization of the projections.
  Together with the success probability of $\mathcal E_{\mathrm{dis}}$,
  this shows that the four success events hold with high constant
  probability against this adversary. On their intersection, the preceding
  analysis guarantees that $\widehat{\cost}_t$ is an $\varepsilon$-oracle
  simultaneously for every $t\in[m]$, even when subsequent updates and
  queries depend on all previously reported oracles.

  \paragraph{Space Accounting.}
  {We now bound the sizes of the universe and query families defined above,
  and substitute the parameters into \Cref{lemma:batched-subtraction,lemma:linear_query}.
  First, fix a level $\ell$ and a projection $j$.
  For any two rows $(\ell,a,j),(\ell,b,j)\in\mathcal U$,
  the definition of $\mathcal U$ gives points $x,y\in X_\varepsilon$ with
  $F_\ell(a,G_jx)>0$ and $F_\ell(b,G_jy)>0$.
  On $\mathcal E_{\mathrm{dis}}$, locality and the choice of $s_1$ give
  \[
    \|a-b\|_2
    \leq\|a-G_jx\|_2+\|G_jx-G_jy\|_2+\|G_jy-b\|_2
    \leq s_1+2Ks_\ell.
  \]
  Thus the range of each coordinate of these representatives has length at most
  $s_1+2Ks_\ell$. By \Cref{lemma:HLA}, $W_\ell=s_\ell\mathbb Z^{\tilde d}$,
  so there are at most $1+s_1/s_\ell+2K$ choices in each coordinate.
  Consequently,
  \[
    N=|\mathcal U|
    \leq n_g\sum_{\ell=1}^L
      (1+s_1/s_\ell+2K)^{\tilde d}
    \leq n_gL\bigl[O(s_1/s_L+K)\bigr]^{\tilde d}.
  \]
  We enumerate $\mathcal U$ by scanning the column supports in a fixed order,
  skipping a row if it occurs in an earlier column, which can be checked by rescanning
  the earlier columns. This stores only the current row, the current points of
  $X_\varepsilon$, and loop counters.
  The enumeration reports failure if its count exceeds the explicit bound above,
  which cannot occur on $\mathcal E_{\mathrm{dis}}$.

  The magnitude bound satisfies
  \[
    \log M=\log(4m)+z\log s_1
    =O\left(z\log\frac{2zmkd\Delta}{\varepsilon}\right).
  \]
  Using the bounds on $s_1,s_L,n_g$, and $K$, we obtain
  \begin{align*}
    \log N
    &\leq O\left(\log(n_gL)
      +\tilde d\log(s_1/s_L+K)\right)\\
    &=\tO(z\tilde d+z^2).
  \end{align*}
  }

  The choice of $\rho$ gives $\log(1/\rho)=\tO(z)$, and
  \[
    \log|\mathcal V|
      \leq k'\log\bigl(1+|X_\varepsilon|(m/\rho+1)\bigr)
      =\tO(k'(d+z)),
    \qquad
    \log|\mathcal H|=\tO(kd).
  \]

  We can now substitute these parameters into the two space bounds.
  {For $\mathcal S_{\mathrm{norm}}$, apply \Cref{lemma:batched-subtraction}
  with universe size $N+1$, at most $mN+1$ coordinate updates,
  relative accuracy $1/2$, and failure probability $1/100$.
  Its query family has size at most $|\mathcal V|$.
  Using \eqref{eqn:main_norm_ratio} and
  $\log(2(mN+1)(N+1))=O(\log(2mN))$, its space is at most
  \[
    \poly\!\left(\log(2mN)+\log(2M)+z\right)
    \bigl(1+\log|\mathcal V|\bigr).
  \]}
  {For $\mathcal S_{\mathrm{ip}}$, substitute the same universe size,
  update bound, and norm ratio, together with accuracy $\eta$ and failure probability $1/100$,
  into \Cref{lemma:linear_query}.
  Its query-family size is $|\mathcal H|$, so the resulting space is
  \[
    \log|\mathcal H|\cdot
    \poly\!\left(\eta^{-1}\bigl(\log(2M)+z\bigr)\log(2mN)\right).
  \]
  }

  An explicit reference uses the same order of space as the bound above on
  $\log|\mathcal V|$, in particular $\tO(kd)$ for $z=1,2$.
  Each mean point is represented by its integer denominator at most $m$
  and its $d$ integer coordinate numerators at most $m\Delta$.
  The sparse recovery structure uses another $\tO(kd)$ space.
  Coordinate evaluation uses the stored reference, the projection seed,
  and the HLA workspace. Projected coordinates and grid anchors are
  recomputed as needed in the allowed real registers.
  The projection representation uses $\tO(zkd)$ space, and the remaining
  coordinate workspace is bounded by $\tO(k'(d+z)+z\tilde d+z^2)$.
  Adding these bounds and substituting
  {$\tilde d=\poly(z\varepsilon^{-1}\log k)$, $r$, and $\eta$} gives total space
  \[
    \begin{cases}
      \tO(kd\varepsilon^{-O(1)}),&z=1,2,\\
      \tO\bigl(kd(d+1)^{z-1}(z/\varepsilon)^{O(z)}\bigr),&z>2.
    \end{cases}
  \]
  Since $z$ is fixed and $(d+1)^{z-1}\leq2^{z-1}d^{z-1}$,
  these bounds give $\tO(kd^z\varepsilon^{-O(z)})$ space for general $z$,
  and $\tO(kd\varepsilon^{-O(z)})$ for $z=1,2$, as claimed.

\section{A Polynomial-Time Constant Approximation for \texorpdfstring{$k$}{k}-Median}
\label{poly:sec:polynomial}
In this section, we describe our polynomial-time algorithm for \kMedian with constant-approximation and linear space in both $k$ and $d$, which is stated in \Cref{poly:thm:centers}.
Here we start with a proof overview, and the formal proof can be found in \Cref{poly:sec:refinement}.

\begin{theorem}[Polynomial-time constant approximation]
\label{poly:thm:centers}
For \kMedian and every $0<\delta<1/2$, there is an adversarially robust streaming algorithm that takes as input an adaptive sequence of $m$ point insertions and deletions from $[\Delta]^d$, and reports after the $t$-th update a set $C_t$ of at most $k$ centers, such that with probability at least $1-\delta$, simultaneously for all $t\in[m]$ $\cost(\mu_t,C_t)\leq O(1)\opt(\mu_t)$.
The algorithm uses $\widetilde O(kd)$ bits of space and $\poly(k,d,\log(m\Delta/\delta))$ time per update.
\end{theorem}

For a signed vector $\lambda\in \mathbb{R}^{\cX}$ with total mass zero, we define its transportation cost $\mathcal K(\lambda)$ to be the minimum cost of moving the positive part of $\lambda$ onto its negative part, at cost $\|x-y\|_2$ per unit moved from $x$ to $y$ (see the formal definition below). 
Observe that for $|\nu|=|\mu|$, we have
\[ \cost(\mu, \supp \nu) \le\mathcal K(\mu-\nu).\]
Hence, we aim to construct a weight vector $\nu$ with support size at most $k$ with $\mathcal K(\mu-\nu)=O(\opt(\mu))$, so that $\supp \nu$ gives a constant-approximate clustering solution.

\begin{definition}[Transportation cost]
\label{def:transportation}
For a finitely supported signed vector $\lambda$ of total mass zero, let $\mathcal K(\lambda)$ be the minimum cost of transporting its positive part onto its negative part, at cost $\|x-y\|_2$ per unit moved from $x$ to $y$:
\[\mathcal K(\lambda):=\min_\pi\Bigl\{\sum_{x,y}\pi_{x,y}\|x-y\|_2:\ \pi\geq0,\ \sum_y\pi_{x,y}=\lambda^+_x,\ \sum_x\pi_{x,y}=\lambda^-_y\Bigr\}.\]
\end{definition}

\paragraph{Iterative Reconstruction.}
Instead of enumerating a family of exponentially many candidates, we start from an arbitrary candidate set $C$ of at most $k$ centers and repeatedly improve it.
Let $\nu$ be the weight vector defined by moving the weight of $\mu$ to the nearest center in $C$, so $|\nu|=|\mu|$ and $|\supp \nu|\le k$. 
We devise a reconstruction subroutine (which is our key lemma and we shall describe in more detail shortly), which given $\nu$ and a budget $S\ge\mathcal K(\mu-\nu)$, computes $\sigma\ge0$ on $\widetilde O(k/\eta)$ points with $\mathcal K(\mu-\sigma)\le O(\opt(\mu))+\eta S$ for a sufficiently small constant $\eta$. 
Given the output $\sigma$ from this subroutine, we reduce its support to $k$ points by applying the single-swap local search algorithm for \kMedian \cite{siamcomp/AryaGKMMP04} to $\sigma$ and moving each weight to its nearest center. 
This computes a new vector $\tau$ with $|\supp \tau|\le k$ and $\mathcal K(\mu-\tau)\le O(\opt(\mu))+S/10$. 
We set $\nu_{\text{new}}:=\tau$ and $S_{\text{new}}:=O(\opt(\mu))+S_\text{old}/10$ and repeat this procedure as long as $S_{\text{new}}$ reduces by half, otherwise, when $S_\text{new} \ge S_\text{old}/2$, we have $S_\text{new} \le O(\opt)+S_\text{new}/5$.
Thus $S=O(\opt(\mu))$ after $O(\log(md\Delta))$ rounds, and $\supp \nu$ at the last step gives us the desired constant approximation.

\paragraph{Sparse Candidate Locations.}
We now describe the reconstruction subroutine used above to find $\sigma$. 
We first select the candidate locations for $\sigma$, without considering their weights. 
Fix the input candidate vector $\nu$ and the budget $S$. 
At a high level, we retain all points in $\supp\nu$ as ``anchors'' and sample additional points from the residual $\mu-\nu$ to cover the points far from the anchors. 
Specifically, we randomly hash the weighted residual points from $\mu-\nu$ to $\widetilde O(k/\eta)$ buckets.
We discard every bucket that contains an anchor and $\ell_1$ sample a candidate point from each remaining bucket. 
Consider a cluster $K$ in the optimal solution and suppose that most of its weight is far from the anchors. 
Intuitively, our bucket construction ensures that points close to $K$ has total sampling weight $\Omega(\eta S/k)$ in the remaining buckets. 
On the other hand, the budget $\mathcal K(\mu-\nu)\le S$ upper bounds the expected total sampling weight by $\widetilde O(S)$, and thus, hashing into $\widetilde O(k/\eta)$ buckets limits the sampling weight from points far from $K$ in each bucket to $O(\eta S/k)$. 
Hence, with constant probability, at least one sampler associated with a bucket containing points of $K$ returns a candidate near its optimal center. 
By the triangle inequality, moving the cluster’s points to the sampled candidate costs at most their transportation cost to the optimal center plus the cluster’s mass times the distance from that center to the candidate. 
The latter term is at most the clustering cost of $K$ plus $\eta S/k$, so summing over all clusters gives total transportation cost $O(\opt(\mu))+\eta S$. 
Repeating this sampler polylogarithmic times returns a good set of candidate locations for all clusters with high probability. 
For adaptive streams, we apply this construction in the dyadic decomposition framework in \Cref{sec:inner_prod}, i.e., we divide the stream into dyadic blocks and conduct the procedure above privately in each block to achieve robustness.

\paragraph{Sparse Candidate Scores.}
Given the recovered candidate locations $\supp\sigma$, we next need to attribute the weights to these points such that $\mathcal K(\mu-\sigma)\le O(\opt(\mu))+\eta S$. 
From the discussion above, if we could send the weight of each input point to the nearest location in $\supp\sigma$, then the resulting weighted vector would satisfy this requirement. 
However, we cannot store $\mu$ to compute these weights exactly. 
Instead, we apply the HLA to estimate them with small space. 
Fix a candidate $p\in\supp\sigma$ and define a score function $\xi:\mathcal X\to[0,1]$ so that $\xi(x)$ is the fraction of the input weight $\mu_x$ attributed to $p$. 
The total weight $\sigma_p$ is therefore $\sum_{x\in X}\mu_x \xi(x)$. 
Recall that HLA induces a linear operator $A$ that defines a fractional assignment of each input point among fixed representatives at each scale. 
The score $\xi(x)$ can then be decomposed by telescoping across these scales, with a query vector $g_\xi$ encoding the differences in $\xi$ between adjacent levels divided by their geometric scales. 
This gives $\sum_x\mu_x \xi(x)=\langle A\mu, g_\xi \rangle$, and the weight $\sigma_p$ attributed to point $p$ is then
\[\langle A\mu, g_\xi \rangle = \langle A\nu, g_\xi \rangle + \langle A(\mu-\nu),g_\xi\rangle,\]
where the first term is computed exactly, and the second term is computed by our inner-product estimator in \Cref{lemma:linear_query}. 
Note that this formulation is analogous to \eqref{eqn:inner_prod}, with the pointwise cost $\dist(x,C)$ replaced by the score $\xi(x)$.

To apply \Cref{lemma:linear_query}, we need $\|g_\xi\|_\infty \le 1$. 
Unfortunately, if we define $\xi$ by the nearest-candidate attribution, the value of $\xi(x)$ could jump under small movements of the location of $x$. 
Even when the representatives of two adjacent scales are only $2^{-\ell}$ apart, they may choose different candidates, so their difference in $\xi$ is $1$ and $\|g_\xi\|_\infty$ could be as large as $2^\ell$, which is prohibitively large at fine scales. 
We therefore use a smooth score function $\xi$: each point spreads its weight among nearby candidates in $\supp \sigma$, with the scores varying gradually as the point moves.
This smoothness controls the differences between adjacent levels by the geometric scale, and hence bounds $\|g_\xi\|_\infty$ while preserving the transportation cost bound.
In addition, the Lipschitz property of the HLA gives $\|A(\mu-\nu)\|_1\le \widetilde O(d)\mathcal K(\mu-\nu)\le \widetilde O(dS)$, bounding the residual norm that governs the inner-product estimation error.
Together, these two guarantees allow us to apply \Cref{lemma:linear_query} and obtain a good candidate vector $\widehat \sigma$. 
Finally, some entries of $\widehat \sigma$ may be negative due to estimation error, so $\widehat \sigma$ is not a valid input for the clustering algorithm used to reduce the support to $k$ points. 
Thus we implement a minimum-cost-flow algorithm \cite{GoldbergTarjan1990,KiralyKovacs2012} to cancel the negative entries of $\widehat \sigma$ and obtain $\sigma$, preserving the bound $\mathcal K(\mu-\sigma)\le O(\opt(\mu))+\eta S$.

\paragraph{Time-efficient HLA.}
Finally, we need to use a more efficient HLA construction than what we have in \Cref{sec:general-dimension} which works for general $z$.
In particular, we need the support of $F_\ell(\cdot, x)$ be bounded by $\poly(d)$ and efficiently enumerated.
Specifically, the construction in \Cref{sec:general-dimension}
takes the tensor-products of the one-dimensional fractional assignments, so we need to compute $\exp(d)$ values for each input point, incurring exponential time. 
For $z=1$, we avoid this by sharing the one-dimensional fractional assignments across all dimensions, giving the time-efficient HLA of \Cref{poly:lem:diagonal}.
This efficient implementation of HLA preserves the same Refinement, Locality, and Lipschitz properties in \Cref{def:HLA}.

\subsection{Proof of \texorpdfstring{\Cref{poly:thm:centers}}{Theorem}}
\label{poly:sec:refinement}
In this section, we present the reconstruction lemma, whose proof is delayed to \Cref{poly:sec:locations}, and prove the main result. 
We use the following standard properties.

\begin{lemma}[{see, e.g.,~\cite[Theorem~1.14]{books/Villani03} and~\cite[Theorem~4]{isr/GibbsS02}}]
\label{poly:lem:transport}
Let $\lambda,\lambda'$ be finitely supported signed vectors of zero total mass.
\begin{enumerate}[(i)]
\item $\mathcal K(-\lambda)=\mathcal K(\lambda)$ and $\mathcal K(\lambda+\lambda')\leq\mathcal K(\lambda)+\mathcal K(\lambda')$. Hence $\mathcal K(\alpha-\alpha'')\leq\mathcal K(\alpha-\alpha')+\mathcal K(\alpha'-\alpha'')$ for equal-mass $\alpha,\alpha',\alpha''$.
\item For equal-mass $\alpha,\beta$ and nonempty $C\subseteq\R^d$, $|\cost(\alpha,C)-\cost(\beta,C)|\leq\mathcal K(\alpha-\beta)$.
\item If $\lambda$ is supported on $[\Delta]^d$, then $\|\lambda\|_\infty\leq\|\lambda\|_1/2\leq\mathcal K(\lambda)$.
\item If $\mu'\geq0$ and $\nu'$ is signed with $|\nu'|=|\mu'|$, then $\cost(\mu',\supp\nu')\leq\mathcal K(\mu'-\nu')$, where the left side is zero if $\mu'=0$.
\end{enumerate}
\end{lemma}
\begin{proof}
Items (i)--(iii) are standard consequences of the Kantorovich--Rubinstein formula and of the fact that distinct points of $[\Delta]^d$ are at distance at least one.
For (iv), we may assume $\mu'\neq0$, so that $\nu'\neq0$ since $|\nu'|=|\mu'|>0$.
Let $\lambda:=\mu'-\nu'$ and fix an optimal plan, which moves the positive part of $\lambda$ onto its negative part.
Mass is moved only into points $y$ with $\lambda_y<0$, i.e., $\nu'_y>\mu'_y\ge0$, so it is moved only into $\supp\nu'$.
A point $x\notin\supp\nu'$ has $\lambda_x=\mu'_x\ge0$, so the plan moves all of $\mu'_x$ out of $x$ and into $\supp\nu'$, at cost at least $\dist(x,\supp\nu')$ per unit.
Since points of $\supp\nu'$ have $\dist(x,\supp\nu')=0$, summing over $x$ gives
\[
 \cost(\mu',\supp\nu')=\sum_{x\notin\supp\nu'}\mu'_x\dist(x,\supp\nu')\le\mathcal K(\lambda).
\]
\end{proof}

\paragraph{Reconstruction Lemma.}
Given vector $\nu$ and budget $S$, reconstruction computes $\sigma$ with support $\widetilde{O}(k)$ that reduces the cost bound by an $\eta$-multiplicative factor, for a sufficiently small constant $\eta\leq1/300$.
In the following, we restrict the weights to a fixed dyadic grid $\gamma\Z$ with $\log(1/\gamma)=\polylog(2mkd\Delta/\delta)$ and $\gamma\le1/(100(m+1))$.
Let $S_{\max}:=10^6(m+1)d(\Delta+1)$ be the upper bound on the budget $S$.

\begin{lemma}[Reconstruction]
\label{poly:lem:reconstruction}
Given a signed vector $\nu$ and a budget $S$ satisfying
\begin{equation}
\label{poly:eq:premises}
 |\supp\nu|\leq k+1,\qquad |\nu|=|\mu|,\qquad
 \|\nu\|_1\leq m+2,\qquad
 \mathcal K(\mu-\nu)\leq S\in[1,S_{\max}].
\end{equation}
There is an algorithm that returns $\sigma\geq0$ on $\widetilde O(k)$ points of $[\Delta]^d$ with $|\sigma|=|\mu|$, and a rational certificate $U$ such that
\begin{equation}
\label{poly:eq:reconstruction}
 \mathcal K(\mu-\sigma)\leq U\leq112\opt(\mu)+\eta S.
\end{equation}
Suppose we call this algorithm $O(\log S_{\text{max}})$ times per update, with probability at least $1-\delta$, \eqref{poly:eq:reconstruction} holds at every call for which \eqref{poly:eq:premises} has held at all previous calls so far, even under adaptive updates.
The algorithm stores a sketch of residual $\mu-\nu$ of size $\widetilde O(kd)$, and each update and reconstruction call takes $\poly(k,d,\log(m\Delta/\delta))$ time.
\end{lemma}

\paragraph{Reducing the Support.}
We next cluster $\sigma$ and move its weights to $k$ points to obtain $\tau$.
The next lemma shows that this clustering step yields a valid candidate solution.

\begin{lemma}
\label{poly:lem:replacement}
Given $\sigma$ and $U$ satisfying \eqref{poly:eq:reconstruction}, there is a replacement step that computes vector $\tau$ and certificate $Q$, such that $\tau\geq0$ with $|\tau|=|\mu|$ has at most $k$ locations, and
\begin{equation}
\label{poly:eq:refinement}
 \mathcal K(\mu-\tau),\ \cost(\mu,\supp\tau)
 \leq Q\leq O(1) \,\opt(\mu)+S/10.
\end{equation}
The replacement uses $\widetilde{O}(kd)$ bits and $\poly(k,d,\log(m\Delta/\delta))$ time.
\end{lemma}
\begin{proof}
If $|\mu|=0$, return $\tau=0$ and $Q=0$.
We use the weighted single-swap algorithm of~\cite[Section~2 and Theorem~3.2]{siamcomp/AryaGKMMP04}: starting from any $k$ centers in $\supp \sigma$, repeatedly take a swap that decreases the cost by a factor at least $1/(4k)$, stopping at zero cost or when no such swap remains.
Let $F$ be the final cost and let $\mathrm{OPT}$ be the cost of the best $k$ centers in $\supp\sigma$.
At termination, every possible swap from the final center set saves at most $F/(4k)$.
In contrast, \cite{siamcomp/AryaGKMMP04}'s analysis identifies at most $k$ swaps from the final center set whose savings sum to at least $F-5\mathrm{OPT}$.
Hence
\[
 F-5\mathrm{OPT}
 \le \text{total saving of these swaps}
 \le k\cdot\frac{F}{4k}
 =\frac F4,
\]
which gives $F\le\frac{20}{3}\mathrm{OPT}$.
Thus this returns $C\subseteq\supp\sigma$ with $|C|\leq k$ and $\cost(\sigma,C) \le\frac{20}{3}\mathrm{OPT}$.
Moving each optimal center to the nearest point of its cluster at most doubles the cost of that cluster, so $\cost(\sigma,C)\le\frac{20}{3}\mathrm{OPT}\leq\frac{40}{3}\opt(\sigma)\leq14\opt(\sigma)$.
Each possible swap is examined by rescanning $\sigma$, storing only the current centers, indices, and $O(1)$ active distances and cost values.
For $|\supp\sigma|>k$, clustering costs lie in $[\gamma,m\sqrt d\,\Delta]$, so there are $O(k\log(2m\sqrt d\,\Delta/\gamma))$ accepted swaps and at most $k|\supp\sigma|$ possible swaps per round.
Since $|\supp\sigma|=\widetilde O(k)$, this uses $\widetilde O(kd)$ bits and $\poly(k,d,\log(m\Delta/\delta))$ time.
Compute an upper approximation $Q\in\gamma\Z$ to $U+\cost(\sigma,C)$ with excess less than $1/20$.

Moving the weights from $\supp \sigma$ to $C$ preserves their grid and mass, and for the transportation cost of $\tau$, by \Cref{poly:lem:transport}(i) and \eqref{poly:eq:reconstruction}, we have
\[
 \mathcal K(\mu-\tau)\leq\mathcal K(\mu-\sigma)+\mathcal K(\sigma-\tau)\leq U+\cost(\sigma,C)\leq Q.
\]
Here, $\mathcal K(\sigma-\tau)\leq \cost(\sigma,C)$ because moving each $\sigma_x$ with $x\notin C$ to its nearest center in $C$ is a transportation plan for $\sigma-\tau$ of cost $\sum_x\sigma_x\dist(x,C)=\cost(\sigma,C)$.
Since $\tau\geq0$ and $|\tau|=|\mu|$, \Cref{poly:lem:transport}(iv) gives $\cost(\mu,\supp\tau)\leq\mathcal K(\mu-\tau)\leq Q$.
Let $C^\circ$ be an optimal center set for $\mu$.
Since $\sigma$ and $\mu$ have equal mass, \Cref{poly:lem:transport}(ii) gives
\[
 \opt(\sigma)\leq\cost(\sigma,C^\circ)\leq\cost(\mu,C^\circ)+\mathcal K(\mu-\sigma)\leq\opt(\mu)+U.
\]
Combining this with the clustering bound and the excess of $Q$,
\[
 Q\leq U+\cost(\sigma,C)+\frac1{20}\leq U+14\opt(\sigma)+\frac1{20}\leq15U+14\opt(\mu)+\frac1{20}.
\]
Substituting $U\leq112\opt(\mu)+\eta S$ from \eqref{poly:eq:reconstruction} gives
\[
 Q\leq(15\cdot112+14)\opt(\mu)+15\eta S+\frac1{20}=1694\opt(\mu)+15\eta S+\frac1{20}.
\]
Finally, $\eta\leq1/300$ and $S\geq1$ give $Q\leq1700\opt(\mu)+S/10$.
\end{proof}

\paragraph{Main Algorithm: Iterative Reconstruction.}
We repeat the reconstruction and clustering procedure as long as the budget halves.
Note that \Cref{poly:lem:reconstruction} works for any input vector as long as \eqref{poly:eq:premises} is satisfied.
Here, we use the weight vector $\nu$ obtained after the last stream update as a convenient initialization, since adding the input update to both $\mu$ and $\nu$ preserves the residual and its budget. 
Remark that the update could delete a point that contributes a large amount to the previous clustering cost, so that $\opt(\mu)$ is subsequently much smaller.
Thus $\supp\nu$ could no longer give a good estimate even if the residual is preserved, and we still need to operate the reconstruction lemma. 
Whenever $\nu$ is replaced while $\mu$ is fixed, the algorithm sends $\nu_{\mathrm{old}}-\nu_{\mathrm{new}}$ to every active residual sketch.
Adding an input update to both $\mu$ and $\nu$ leaves these sketches unchanged.
The algorithm proceeds as follows.
\begin{enumerate}[(1)]
\item Initialize the reconstruction data structure and an exact mass counter, and set $\nu:=0$ and $S:=1$.
\item On an input update $\varsigma e_x$, update the mass counter and set $\nu:=\nu+\varsigma e_x$.
If the input mass is zero, replace $\nu$ by $0$, set $S:=1$, and stop.
Otherwise, repeat:
\begin{enumerate}[(a)]
\item Set $S_{\mathrm{old}}:=S$, reconstruct $(\sigma,U)$ from $(\nu,S)$, compute $(\tau,Q)$ by \Cref{poly:lem:replacement}, and replace $\nu$ by $\tau$.
\item If $Q<1/2$, round $\nu$ coordinatewise to the nearest integer, set $S:=1$, and stop.
\item Set $S:=\max\{1,Q\}$.
Stop if $Q\geq S_{\mathrm{old}}/2$; otherwise, return to Step~(2a).
\end{enumerate}
\item Report the centers $\supp\nu$.
\end{enumerate}

\begin{proof}[Proof of \Cref{poly:thm:centers}]

We first show the premises of \Cref{poly:lem:reconstruction} hold at every call by induction.
Adding an input update to both $\mu$ and $\nu$ leaves the residual and the budget unchanged, and adds at most one point and one unit of $\ell_1$ mass to $\nu$.
Moreover, after a replacement, \Cref{poly:lem:replacement} gives a nonnegative vector $\tau$ on at most $k$ points with the input mass and $\mathcal K(\mu-\tau)\leq Q\leq\max\{1,Q\}$, and $Q\leq1700m\sqrt d\,\Delta+S_{\max}/10\leq S_{\max}$ since $\opt(\mu)\leq m\sqrt d\,\Delta$.
So every call satisfies \eqref{poly:eq:premises}.

We now prove the algorithm stops with a good budget.
If the input mass is zero, then $\nu=\mu=0$.
If $Q<1/2$, then $\|\mu-\tau\|_\infty\leq\mathcal K(\mu-\tau)<1/2$ by \Cref{poly:lem:transport}(iii), so rounding recovers $\nu=\mu$.
Otherwise, the loop stops once $Q\geq S_{\mathrm{old}}/2$, and \eqref{poly:eq:refinement} with the budget $S_{\mathrm{old}}$ of that call gives $Q\leq C\opt(\mu)+Q/5$, i.e., $Q\leq5C/4\opt(\mu)$, where $C$ is the constant factor in \eqref{poly:eq:refinement}.
This and \Cref{poly:lem:replacement} prove 
\[\cost(\mu,\supp\nu)\leq Q\leq O(1)\opt(\mu).\]
While the loop continues, $Q<S_{\mathrm{old}}/2$, so the budget halves as long as $Q\geq1$; once $Q<1$, it becomes one and the next iteration stops.
Hence each update makes at most $O(\log S_{\max})=O(\log(m,d,\Delta))$ calls, and by \Cref{poly:lem:reconstruction,poly:lem:replacement}, each call and replacement takes polynomial time and $\widetilde O(kd)$ bits.
This proves the complexity bounds of \Cref{poly:thm:centers}.
\end{proof}

\subsection{Time-Efficient Hierarchical Lipschitz Assignment}
\label{sec:HLA_z_1}
In this section, we introduce our time-efficient construction of HLA to facilitate time-efficient \kMedian algorithms.
Fix the number of levels $L$ and the dimension $d$.
We obtain the same scalar shifts $S_\ell$ and $K_\ell$ for $\ell \in [L]$ from the one-dimensional construction in \Cref{sec:one-dimensional} with $z=1$, where $S_\ell$ is uniform and $K_\ell$ is binomial.
We use this single $S_\ell$ in every dimension at level $\ell$, and this single $K_\ell$ in every dimension of the parent map.
This differs from \Cref{sec:general-dimension}, which takes independent copies of the scalar hierarchy for different dimensions.
Let $h_\ell^{(1)}$ and $p_\ell^{(1)}$ denote the scalar maps defined as \eqref{eq:hash-parent}.
Set $W_\ell:=s_\ell\Z^{d}$ and define the vector maps, for $x\in\R^{d}$, $v\in\Z^{d}$, and $i\in[d]$, by
\[(h_\ell(x))_i :=h_\ell^{(1)}(x_i)
 =s_\ell\left\lfloor\frac{x_i}{s_\ell}-S_\ell\right\rfloor,\qquad (p_\ell(s_\ell v))_i :=p_\ell^{(1)}(s_\ell v_i)
 =s_{\ell-1}\left\lfloor\frac{v_i-K_\ell}{2}\right\rfloor.\]
In particular, evaluating another coordinate does not draw new randomness.
Define the probability matrices
\[
 F_\ell(a,x):=\Pr[h_\ell(x)=a],
 \qquad
 U_\ell(b,a):=\Pr[p_\ell(a)=b],
 \qquad a\in W_\ell,\ b\in W_{\ell-1}.
\]
This gives the fractional assignment of the time-efficient HLA, and we have the following lemma:

\begin{lemma}
\label{poly:lem:diagonal}
Let $L,d,s_\ell,W_\ell$ be defined as in \Cref{def:HLA}, there exists an order-$1$ HLA ($z=1$) that satisfies refinement, locality with radius $2\sqrt{d}\,s_\ell$, and
\[
 \|F_\ell(\cdot,x)-F_\ell(\cdot,y)\|_1
 \le \frac{2\|x-y\|_1}{s_\ell}.
\]
The HLA can be described using $\widetilde O(d)$ space, and evaluating an entry of $F$ or $U$ takes $\widetilde O(d)$ space and $O(dL)$ time.
Moreover, each column of $F_\ell$ has at most $d+1$ nonzero entries and each column of $U_\ell$ at most two.
These entries can be enumerated one at a time in $O(d^2)$ time and $\widetilde O(d)$ space.
\end{lemma}
\begin{proof}
We first prove that our construction satisfies \Cref{def:HLA}.
Column normalization follows immediately, and one-dimensional locality in \Cref{sec:one-dimensional} gives locality of the time-efficient HLA with radius $2\sqrt{d}\,s_\ell$.
For refinement, \Cref{sec:one-dimensional} also gives $h_{\ell-1}=p_\ell\circ h_\ell$.
Since $K_\ell$ is independent of $S_\ell$, the parent map is independent of $h_\ell(x)$, and hence $F_{\ell-1}=U_\ell F_\ell$.
For the Lipschitz bound, similar as \Cref{sec:one-dimensional}, the uniform shift gives
$\Pr[h_\ell^{(1)}(x_i)\ne h_\ell^{(1)}(y_i)]
\le |x_i-y_i|/s_\ell$ for every coordinate $i$.
A union bound therefore yields
\[
 \|F_\ell(\cdot,x)-F_\ell(\cdot,y)\|_1
 \le 2\Pr[h_\ell(x)\ne h_\ell(y)]
 \le \frac{2\|x-y\|_1}{s_\ell}.
\]
This verifies the Lipschitz bound of \Cref{def:HLA}, taking all approximation polynomials to be zero.

To evaluate $F_\ell(a,x)$, note that $h_\ell(x)=a$ holds when $S_\ell$ falls into the following interval
\[
 [0,1)\cap\bigcap_{i=1}^{d}
 \left(\frac{x_i-a_i}{s_\ell}-1,\frac{x_i-a_i}{s_\ell}\right].
\]
We compute in $O(d)$ time by scanning the endpoints.
To evaluate $U_\ell(b,a)$, check whether $p_\ell(a)=b$ for each of $K_\ell=0$ and $K_\ell=1$.
Since each bit value has probability $1/2$, the entry equals half the number of successful checks.
Each check scans $d$ coordinates, so evaluation takes $O(d)$ time.

For a fixed $x$, each coordinate of $h_\ell(x)$ changes at most once as the common shift $S_\ell$ ranges over $[0,1)$.
Thus $h_\ell(x)$ takes at most $d+1$ values with positive probability, so each column of $F_\ell$ has at most $d+1$ nonzero entries.
Similarly, for a fixed $a$, the parent $p_\ell(a)$ depends only on the single bit $K_\ell\in\{0,1\}$, so each column of $U_\ell$ has at most two nonzero entries.
To identify the nonzero entries, for $F_\ell(\cdot,x)$, sort the fractional parts of $x_i/s_\ell$ together with $0$ and $1$.
On each interval $(u,v)$ between consecutive distinct values, $h_\ell(x)$ is constant, so evaluating it with $S_\ell=(u+v)/2$ gives a nonzero row index $a$ with entry $F_\ell(a,x)=v-u$.
These intervals cover all shifts except finitely many endpoints of probability zero, so this enumerates every nonzero entry.
For $U_\ell(\cdot,a)$, evaluate $p_\ell(a)$ with $K_\ell=0$ and $K_\ell=1$, assigning weight $1/2$ to each resulting parent and combining coincident parents.
To sum up, columns of $F$ and $U$ have at most $d+1$ and $2$ nonzero entries, enumerable in $O(d^2)$ and $O(d)$ time, respectively, using $\widetilde O(d)$ space as we compute one entry at a time.
\end{proof}

\subsection{The Reconstruction Lemma}
\label{poly:sec:locations}

In this section, we prove \Cref{poly:lem:reconstruction}.
We first recover the locations of points in $\supp \sigma$ in \Cref{poly:lem:locations}, and then we assign weights to those recovered locations in \Cref{poly:lem:masses}.
We begin with the Gaussian projection family used in this section, which is nearly the same as in \Cref{lemma:robust:projection}, except that we obtain constant distortion and target dimension $\tilde d=\Theta\left(\log\frac{2mkd\Delta}{\delta}\right)$.
We use part (i) in location recovery and part (ii) in weight assignment.

\begin{lemma}
\label{poly:lem:jl}
Let $\tilde d=\Theta(\log(2mkd\Delta/\delta))$, with a sufficiently large constant.
The following hold.
\begin{enumerate}[(i)]
\item
Given nonempty $X\subseteq[\Delta]^d$ and $0<\delta'<1$ that satisfies $\frac{|X|^2}{\delta'} \le \left(\frac{2mkd\Delta}{\delta}\right)^4$, there is a random projection $G\in\R^{\tilde d\times d}$ such that, with probability at least $1-\delta'$,
\[
 \forall x,y\in X,\qquad
 \tfrac12\|x-y\|_2
 \le \|G(x-y)\|_2
 \le 2\|x-y\|_2,
 \qquad \|G\|_{\mathrm F}\le(2mkd\Delta/\delta)^4.
\]

\item
Let $\left(\frac{2mkd\Delta}{\delta}\right)^{-4}\le\zeta\le\tfrac12$ and set
$n_g=O(\zeta^{-1}d\log(2\Delta/\delta))$
for a sufficiently large constant.
There is a random family $G_1,\ldots,G_{n_g}\in\R^{\tilde d\times d}$ such that, with probability at least $1-\delta/4$, every nonzero difference $v=x-y$ with $x,y\in[\Delta]^d$ satisfies
\begin{equation}
\label{poly:eq:jl}
 \Pr_j\!\left[
   \|G_jv\|_2^2\notin(1\pm\tfrac14)\|v\|_2^2
 \right]\le\zeta,
 \qquad
 \frac{1}{n_g}\sum^{n_g}_{j=1}\|G_jv\|_2^2\le\tfrac32\|v\|_2^2,
\end{equation}
\end{enumerate}
Both the random matrix in (i) and the shared seed describing the family in (ii) use $\widetilde O(d)$ bits.
\end{lemma}

\begin{proof}
This follows the proof of \Cref{lemma:robust:projection}, with constant distortion, target dimension $\tilde d$, and the stated $n_g$, applied to the at most $(2\Delta+1)^d$ nonzero grid differences.
For a fixed difference, Gaussian norm concentration makes a single map distort it with probability at most $\delta'/(2|X|^2)$, so a union bound over $X-X$ gives the distance guarantee in (i), while the Chernoff and chi-squared bounds of the proof of \Cref{lemma:robust:projection} and a union bound over the grid differences give (ii).
We use the same derandomization procedure as in  \Cref{lemma:robust:projection}, with error parameter $\delta'/(16(|X|^2+1))$ per test in (i), including the Frobenius-norm check, and $\delta/[16(2\Delta+1)^d]$ per test in (ii). Each test reads the random tape once using $\widetilde O(d)$ workspace, so the resulting space bound remains $\widetilde O(d)$.
\end{proof}

\subsubsection*{Recovering the Candidate Locations}
We use the robust framework in the proof of \Cref{lemma:linear_query}, which decomposes the stream into blocks of dyadic length and performs the estimators privately.
Here, since each reconstruction call is followed by a residual update $\nu_{\mathrm{old}} - \nu_{\mathrm{new}}$, we treat reconstruction calls as the time steps of this decomposition.
In particular, a call with budget $S$ is assigned to the dyadic scale $\rho$ with $\rho \le S < 2\rho$, just as \Cref{lemma:linear_query} assigns an update to the scale of its norm estimate.
Then for each scale $\rho$, we maintain blocks of dyadic length, where the length counts the reconstruction calls assigned to $\rho$.
Each block maintains its own sketches with fresh private randomness and receives every residual change between its endpoints, including changes produced by calls assigned to other scales.
As in the proof of \Cref{lemma:linear_query}, each block is closed before its sketches are revealed to the adversary, and the resulting replacement $\nu_{\mathrm{old}}-\nu_{\mathrm{new}}$ is fed only to open, hidden blocks.
Thus the same argument shows that each block's input is independent of its private randomness.
There are $O(m\log(2S_{\max}))$
calls and $O(m\log^2(2S_{\max}))$ possible blocks across the budget scales,
so a union bound gives correctness throughout the adaptive stream.

\paragraph{Anchor Set and Recovery.}
Fix a reconstruction call with input $\mu$, candidate $\nu$, budget $S\ge\mathcal K(\mu-\nu)$, and corresponding scale $\rho$ with $\rho\le S<2\rho$.
As in the proof of \Cref{lemma:linear_query}, the most recent completed blocks of scale $\rho$ form a set $\mathcal B$ with $|\mathcal B|=O(\log(m\log(S_{\max})))$ and
\[
 \mu-\nu=\sum_{b\in\mathcal B}\lambda_b,
 \qquad
 \lambda_b:=(\mu_{b,1}-\nu_{b,1})-(\mu_{b,0}-\nu_{b,0}),
\]
where $(\mu_{b,0},\nu_{b,0})$ and $(\mu_{b,1},\nu_{b,1})$ are the input and candidate at the start and end of $b$.
The \emph{anchors} of block $b$ are defined as $\mathcal A_b:=\supp\nu_{b,0}\cup\supp\nu_{b,1}$.
Due to our premises, we have $|\mathcal A_b|\le2(k+1)$ and $\cost(\mu_{b,e},\mathcal A_b)\le2\rho$ for $e\in\{0,1\}$, and moreover, $\supp\nu\subseteq\bigcup_{b\in\mathcal B}\mathcal A_b$.

Within a block $b$, to recover clusters whose mass is costly to move to the anchors, we group points into cells of randomly shifted grids and hash these cells into buckets.
We then discard all buckets containing an anchor and sample from the remaining ones.
Note that expensive clusters include light clusters far from the anchors and heavy clusters close to the anchors, so a single cell diameter does not work well for all clusters.
Thus, we use dyadic cell diameters $\ell_j=2^j\ell_{\min}$ and multiply each residual weight by $\ell_j$ at scale $j$.
Then in each bucket, we create a label $(j,x)$ with weight $\ell_j\lambda_b(x)$ if the cell of point $x$ at scale $j$ is hashed to this bucket, and sample from the resulted weight vector. 
Ideally, we would $\ell_1$ sample from the original locations $x\in[\Delta]^d$, but a sampler over a universe of size $N$ uses $\log^2N$ space~\cite{siamcomp/JayaramW21}, incurring a $d^2$ overhead.
We therefore draw a fresh private projection $G:\R^d\to\R^{\tilde d}$ from \Cref{poly:lem:jl}(i), preserving the distances up to a factor $2$.
This reduces the $d^2$ overhead to a $O(\tilde d^2)=\widetilde O(1)$ overhead.
We then sample from projected cells and recover the sample's original location with the decoding sketches discussed later.
Note that the decoder is based on CountSketch, which uses $\log N$ bits instead, translating to only $\widetilde{O}(d)$ bits.

\paragraph{Algorithm: Recovering Locations.}
Fix a block $b$ and write $\lambda=\lambda_b$ and $\mathcal A=\mathcal A_b$.
It maintains $O(\log((m+k)J\log(2S_{\max})/\delta))$ independent repetitions that share $G$, each as follows.
\begin{enumerate}
\item \textbf{Hash to cells and buckets:} Use the scales $\ell_j:=2^j\ell_{\min}$ for $0\le j<J$, where $\ell_{\min}$ is the largest power of two at most $\eta/(1024m)$ and $J$ is least with $\ell_{J-1}\ge\sqrt d\Delta$, and perform:
\begin{enumerate}
    \item At each scale $j$, let $w_j:=\ell_j/\sqrt{\tilde d}$, and draw a shift $\theta_j\in[0,w_j)^{\tilde d}$ whose coordinates are independent and uniform over $[0,w_j)$.
The \emph{cell} of $x$ at scale $j$ is
\[
 Q_j(x):=\Bigl(\Bigl\lfloor\frac{(Gx)_i-\theta_{j,i}}{w_j}\Bigr\rfloor\Bigr)_{i\in[\tilde d]}\in\Z^{\tilde d},
\]
that is, the index of the cube $\theta_j+w_j\bigl(Q_j(x)+[0,1)^{\tilde d}\bigr)$ containing $Gx$.
We use $(j,Q_0(x))$ as the encoded label of $(j,x)$ as we cannot store the location of $x$.
    \item Hash the cells $(j,Q)$ of all scales into $K=\Theta(\tilde dJk\log(2m\log(2S_{\max}))/\eta)$ buckets by a pairwise independent hash $h$ and run an independent $\ell_1$ sampler on each bucket.
    The sampler in bucket $q\in[K]$ receives the vector
\[
 \sum_{0\le j<J}\ \sum_{x:\,h(j,Q_j(x))=q}\ell_j\,\lambda(x)\,e_{(j,Q_0(x))}.
\]
\end{enumerate}
\item \textbf{Maintain residual updates:} On a residual update $\varsigma e_x$, add $\ell_j\varsigma$ to the label $(j,Q_0(x))$ in bucket $h(j,Q_j(x))$ for every scale $j$, and forward it, with location $x$, to the decoding sketches of each sampler trial.
\item \textbf{Output:} Discard the buckets $h(j,Q_j(u))$ for all anchors $u\in\mathcal A$ and scales $j$, and run $\ell_1$ sampler on each remaining bucket.
Decode the original address $\widehat x$ of each sampled label, discarding addresses outside $[\Delta]^d$.
Let $P_b$ be the points retained by block $b$, return $P:=\bigcup_{b\in\mathcal B}(\mathcal A_b\cup P_b)$.
\end{enumerate}

\paragraph{Sampling and Decoding.}
To run in polynomial time, each bucket uses the approximate $\ell_1$ sampler of~\cite{siamcomp/JayaramW21} with fast update time.

\begin{claim}
\label{clm:fast-sampler} \cite{siamcomp/JayaramW21}
For every constant $c>0$, there is an $\ell_1$ sampler such that on every nonzero input vector $v$ over a universe of size $N$ induced by a turnstile stream, it fails with probability at most $1/4$, and otherwise returns each coordinate $i$ with probability $(1\pm\frac1{10})\frac{|v_i|}{\|v\|_1}\pm N^{-c}$.
It uses $\widetilde O(\log^2 N)$ bits of space and $\polylog(N)$ update and query time.
\end{claim}
The sampler in \cite[Sections~5--6]{siamcomp/JayaramW21} works by duplicating every coordinate of $v$ polynomially many times and scaling each copy by the inverse of an independent exponential variable.
Then it returns a maximum coordinate $r^*$ of the scaled vector $z$, and if the sample succeeds, every returned coordinate satisfies $|z_{r^*}|\ge\|z\|_2/100$, which is useful for constructing our decoder.
The sampler outputs a coordinate of $v$ that is a label $(j,Q_0(x))$, but we still need to decode the location $x$ since we cannot store the original dataset.
Let $b_1(x),\ldots,b_H(x)$ be the bits of the binary address of $x\in[\Delta]^d$, where $H:=d\lceil\log_2(\Delta+1)\rceil$, let $x(r)$ be the point inherited by coordinate $r$ of $z$, and let
\[
 z^{(q)}_r:=b_q(x(r))\,z_r
\]
be the \emph{masked vector} that restricts to all bit-$1$ entries, maintained by forwarding each update $w$ of $z_r$ as $b_q(x(r))\,w$.
The \emph{decoder} keeps CountSketches~\cite[Theorem~1]{siamcomp/JayaramW21} of $z$ and every $z^{(q)}$, with accuracy $1/1000$ and failure probability $\delta_{\mathrm{dec}}$, using the auxiliary-sketch construction of \cite[Section~6]{siamcomp/JayaramW21}.
For the sampled $r^*$, it outputs $\widehat b_q:=\mathbf1\{2|\widehat z^{(q)}_{r^*}|>|\widehat z_{r^*}|\}$ for $q\in[H]$.
At a heavy coordinate $r$, the estimate of $z^{(q)}_r$ has absolute value close to $|z_r|$ if bit $q$ of $x(r)$ is one and close to zero otherwise, so the threshold reads off the bit.
Hence we reconstruct the original location by the binary representation.
We next state the formal guarantee of the decoder:

\begin{lemma}
\label{lem:heavy-address}
Given $0<\delta_{\mathrm{dec}}<1$, the probability that a sampler trial returns a coordinate $r^*$ but $\widehat b_q\ne b_q(x(r^*))$ for some $q\in[H]$ is at most $(H+1)\delta_{\mathrm{dec}}$ plus the sampler's failure probability.
The decoder uses $O(H\log(2/\delta_{\mathrm{dec}}))$ counters and $O(H\,\polylog(N/\delta_{\mathrm{dec}}))$ time per update to $v$ and per query.
\end{lemma}
\begin{proof}
Outside the sampler's failure event, the returned coordinate $r^*$ satisfies $|z_{r^*}|\ge\|z\|_2/100$ by~\cite[Corollary~1 and Lemma~9]{siamcomp/JayaramW21}.
The coordinate $r^*$ and the vector $z$ are determined by the input and the randomness of the sampler, independently of the added CountSketch seeds, so we condition on them.
By~\cite[Theorem~1]{siamcomp/JayaramW21}, a CountSketch with $O(\log(2/\delta_{\mathrm{dec}}))$ counters estimates a coordinate within $1/1000$ times the $\ell_2$ norm of its vector, except with probability $\delta_{\mathrm{dec}}$.
Since $\|z^{(q)}\|_2\le\|z\|_2\le100|z_{r^*}|$, a union bound makes all $H+1$ estimates at $r^*$ accurate within $|z_{r^*}|/10$, except with probability $(H+1)\delta_{\mathrm{dec}}$.
Then $|\widehat z_{r^*}|\in[\frac9{10}|z_{r^*}|,\frac{11}{10}|z_{r^*}|]$, while $|\widehat z^{(q)}_{r^*}|$ is at most $\frac1{10}|z_{r^*}|$ if $b_q(x(r^*))=0$ and at least $\frac9{10}|z_{r^*}|$ otherwise.
Since $2\cdot\frac9{10}>\frac{11}{10}$ and $2\cdot\frac1{10}<\frac9{10}$, the threshold reads every bit correctly.
The auxiliary-sketch construction of \cite[Section~6]{siamcomp/JayaramW21} incurs only $\polylog(N/\delta_{\mathrm{dec}})$ additional time per update to $v$ and per query for each sketch, giving the stated time bound.
\end{proof}

We now show that the returned set serves the input well: moving all input mass to its nearest returned point costs a constant times the optimum, plus the permitted additive error.
Importantly, the input threshold $S$ is reduced to $\eta S$.

\begin{lemma}[Candidate locations]
\label{poly:lem:locations}
For $0<\eta\le1$, the algorithm returns a set $P\subseteq[\Delta]^d$ of size $\widetilde O(k/\eta)$ such that, with probability at least $1-\delta/2$, simultaneously at all reconstruction calls,
\[
 \cost(\mu,P)\le7\opt(\mu)+\eta S+4m\ell_{\min},
\]
where $\mu$, $\nu$, and $S$ denote the input, candidate, and budget at that call.
It uses $\widetilde O(kd/\eta)$ bits and polynomial update and readout time.
\end{lemma}
\begin{proof}
We first show that each block finds a point near every ball carrying enough residual mass far from its anchors, and then use these points to cover the optimal clusters.

Fix a block and drop the subscript $b$ from $\lambda$, $\mathcal A$, $\mathcal S$, and $G$.
Here $\lambda$ is the residual change over the block, $\mathcal A$ consists of its endpoint candidate supports, $\mathcal S:=\supp\lambda\cup\mathcal A$ with $|\mathcal S|=O(m+k)$, and $\rho$ is its budget scale.
Condition on the randomness external to this block.
Its input and anchors are then fixed independently of its private randomness, since nothing depending on that randomness is revealed before closure.
Recall that there are $O(m\log^2(2S_{\max}))$ possible blocks.
By \Cref{poly:lem:jl}(i) with $X=\mathcal S$ and $\delta'=\delta/(C m\log^2(2S_{\max}))$ for a sufficiently large absolute constant $C$, with probability at least $1-\delta'$,
\begin{equation}
\label{eq:anchor-jl}
 \tfrac12\|x-y\|_2\le\|G(x-y)\|_2\le2\|x-y\|_2
 \qquad\text{for all }x,y\in\mathcal S.
\end{equation}
We condition on this event and the norm bound in \Cref{poly:lem:jl}(i).
Distinct points of $\mathcal S\subseteq[\Delta]^d$ have projected distance at least $1/2$, whereas a finest cell has diameter $\ell_{\min}<1/2$.
Thus $Q_0$ is injective on $\mathcal S$ for every shift, and the label $(j,Q_0(x))$ determines $(j,x)$ there.
Intermediate collisions are harmless: by linearity, every source outside $\mathcal S$ cancels by closure, including in the masked sketches.
We formally define the \emph{mass} of a set of labels as their total weight, where the label $(j,Q_0(x))$ has weight $\ell_j|\lambda(x)|$.

\textbf{Sampling and Decoding.}
Fix a repetition, together with its shifts and hash.
For the decoder, use \Cref{lem:heavy-address} with a local failure parameter $\delta_{\mathrm{dec}}\le c_0/(K(H+1))$, for a sufficiently small absolute constant $c_0$.
We aim to prove a grouped $\ell_1$ sampling guarantee for any subset of entries.
Note that the sampler runs on a label universe supported on at most $O(J(m+k))$ entries, and each of them could have additive error $1/N^{-c}$.
Pad the label universe to $N=\poly(m,k,\Delta)$ entries so that the total additive sampling error is at most $c_0/K$ for any subset.
This preserves $\log N=\widetilde O(\tilde d)$.
The sampler guarantee and a union bound over the $H+1$ sketches then give the following.
For every bucket with nonzero vector $v$ and every set $F$ of its labels, both fixed independently of the samplers and the decoders, there is an universal constant $c_{\mathrm{samp}}$ such that
\begin{equation}
\label{eq:anchor-sample}
 \Pr[\text{the bucket returns a correctly decoded label in $F$}]
 \ge c_{\mathrm{samp}}\Bigl(\frac{\|v_F\|_1}{\|v\|_1}-\frac1{100K}\Bigr).
\end{equation}

\textbf{Background Mass after Deletion.}
We first bound the mass of cells not covered by the anchors.
A cell is \emph{anchor-free} if it contains no $Gu$ with $u\in\mathcal A$.
Let $Y$ be the total mass of the labels in anchor-free cells, over all scales.
A point $x$ lies in an anchor-free cell at scale $j$ only if the shift separates $Gx$ from $Gu$ for a nearest anchor $u$.
By \eqref{eq:anchor-jl}, this happens with probability at most
\[
 \frac{\|G(x-u)\|_1}{\ell_j/\sqrt{\tilde d}}\le\frac{\tilde d\|G(x-u)\|_2}{\ell_j}\le\frac{2\tilde d\dist(x,\mathcal A)}{\ell_j},
\]
using the continuous uniform shifts specified above.
In anchor-free cells the candidate terms $\nu_{b,0}(x)$ and $\nu_{b,1}(x)$ vanish, so $|\lambda(x)|\le\mu_{b,0}(x)+\mu_{b,1}(x)$ there.
Therefore,
\[
 \E Y\le\sum_{j<J}\sum_x\ell_j\bigl(\mu_{b,0}(x)+\mu_{b,1}(x)\bigr)\frac{2\tilde d\dist(x,\mathcal A)}{\ell_j}
 =2\tilde d J\bigl(\cost(\mu_{b,0},\mathcal A)+\cost(\mu_{b,1},\mathcal A)\bigr)\le8\tilde d J\rho,
\]
and $\Pr[Y\le32\tilde d J\rho]\ge3/4$ by Markov's inequality.
Conversely, a projected cell at scale $j$ has diameter $\ell_j$, so by \eqref{eq:anchor-jl}, a point $x$ with $\dist(x,\mathcal A)>2\ell_j$ lies in an anchor-free cell for every shift, and two points in a common cell at scale $j$ are within distance $2\ell_j$.

\textbf{Isolating a Ball.}
We next show that the $\ell_1$ samples give good estimates of balls not covered by the anchor points.
Intuitively, a ball carrying enough mass far from the anchors lies in anchor-free cells that survive deletion, and since there are many buckets relative to the background mass, most of this mass shares its buckets with little else, so some sampler returns one of its points.
Fix a ball $B=B(u,\|u-v\|_2)$ with $u,v\in\mathcal S$ and a scale $j$.
Call $x\in\mathcal S\cap B$ a \emph{far point} if $\dist(x,\mathcal A)>2\ell_j$, and assume that the far points carry mass
\begin{equation}
\label{eq:anchor-far}
 T:=\ell_j\sum_{x\in\mathcal S\cap B:\,\dist(x,\mathcal A)>2\ell_j}|\lambda(x)|
 \ge\frac{\eta\rho}{16k|\mathcal B|}.
\end{equation}
Mark the cells at scale $j$ that contain a far point.
Every point in a marked cell lies within $2\ell_j$ of $B$.
We show that one repetition decodes such a point with constant probability.

Assume $Y\le32\tilde d J\rho$.
Then the choice of $K$ gives $TK\ge64\tilde d J\rho\ge2Y$.
Call a marked cell \emph{good} if its bucket contains no anchor cell, and receives mass at most $8Y/K$ from other anchor-free cells.
There are at most $2(k+1)J\le K/8$ anchor cells, and the expected mass from other anchor-free cells is at most $Y/K$.
Then by pairwise independence and Markov's inequality, each marked cell is bad with probability at most $1/4$.
Hence the far mass in bad cells has expectation at most $T/4$, and with probability at least $1/2$, the good marked cells carry far mass at least $T/2$.

Fix such a hash.
For a bucket $\kappa \in [K]$, let $g_\kappa$ be the mass of its good marked cells and $\bar g_\kappa$ its total mass.
Every bucket with $g_\kappa>0$ survives deletion and satisfies $\bar g_\kappa\le g_\kappa+8Y/K$.
The function $t\mapsto t/(t+8Y/K)$ is concave and vanishes at zero, hence is subadditive, so $\sum_\kappa g_\kappa\ge T/2$ gives
\[
 \sum_{\kappa:\,g_\kappa>0}\frac{g_\kappa}{\bar g_\kappa}
 \ge\frac{TK}{TK+16Y}
 \ge\frac1{17}.
\]
Let $p_\kappa$ be the probability that bucket $\kappa$ returns a correctly decoded label from its good marked cells.
Applying \eqref{eq:anchor-sample} and summing over these buckets gives
\[
 \sum_{\kappa:\,g_\kappa>0}p_\kappa
 \ge c_{\mathrm{samp}}\left(\sum_{\kappa:\,g_\kappa>0}\frac{g_\kappa}{\bar g_\kappa}-\frac1{100}\right)
 \ge c_{\mathrm{samp}}/25.
\]
The samplers and decoders of distinct buckets are independent, so their joint failure probability is at most $\prod_\kappa(1-p_\kappa)\le e^{-c_{\mathrm{samp}}/25}$.
Hence one repetition decodes a point within $2\ell_j$ of $B$ with probability at least $\frac38(1-e^{-c_{\mathrm{samp}}/25})=\Omega(1)$.
Recall that $|\mathcal S|=O(m+k)$, and $J$ is the number of cell scales.
The $O(\log((m+k)J\log(2S_{\max})/\delta))$ repetitions therefore achieve this for all of the at most $O(J(m+k)^2)$ pairs $(B,j)$ satisfying \eqref{eq:anchor-far}, except with probability $\delta'$.
Together with the projection guarantee, this shows that, except with probability $2\delta'$, the recovered set $P_b$ contains a point within distance $2\ell_j$ of $B$ simultaneously for every pair $(B,j)$ satisfying \eqref{eq:anchor-far}.

\textbf{Serving the Optimal Clusters.}
We now bound $\cost(\mu,P)$.
We show that each optimal cluster either has a nearby anchor, or has enough residual mass near its center in some block for the sampling guarantee to recover a nearby candidate.
Recall that $\mathcal B$ is the current collection of completed blocks, with $|\mathcal B|=O(\log(2m\log(2S_{\max})))$, and let $P_0:=\bigcup_{b\in\mathcal B}\mathcal A_b$ be their anchors.
Since $\mu-\nu=\sum_b\lambda_b$ and $\supp\nu\subseteq P_0$, every $x\notin P_0$ satisfies $\mu_x\le\sum_b|\lambda_b(x)|$.

Fix an optimal cluster with mass $M>0$, center $c$, and cost $M\bar r$.
Since its average distance to $c$ is $\bar r$, its core $B(c,2\bar r)$ contains at least $M/2$ of the cluster's mass.
If $\ell_{J-1}<\eta S/(8kM)$, sending the entire cluster to any point of $\supp\nu\subseteq P_0$ costs at most $M\sqrt d\Delta\le M\ell_{J-1}<\eta S/(8k)$, and such a point exists because $|\nu|=|\mu|>0$.
Otherwise, choose the least scale $\ell_j$ with $\ell_j\ge\eta S/(8kM)$.
This choice gives $4M\ell_j\le\eta S/k+4M\ell_{\min}$.

If an anchor in $P_0$ lies within distance $6\bar r+4\ell_j$ of $c$, sending the cluster to this anchor costs at most $M\bar r+M(6\bar r+4\ell_j)=7M\bar r+4M\ell_j$.
Otherwise, every point of $B(c,6\bar r)$ is farther than $4\ell_j$ from $P_0$.
In particular, the core contains no anchor, so summing the residual bound over the core and averaging over the blocks gives some $b\in\mathcal B$ with $\sum_{x\in B(c,2\bar r)}|\lambda_b(x)|\ge\frac{M}{2|\mathcal B|}$.
Choose $u\in\supp\lambda_b\cap B(c,2\bar r)$ and let $v$ be a point in this set farthest from $u$, allowing $u=v$.
The ball $B(u,\|u-v\|_2)$ contains every point of $\supp\lambda_b\cap B(c,2\bar r)$ and is contained in $B(c,6\bar r)$.
Every point of this ball is therefore farther than $4\ell_j>2\ell_j$ from $\mathcal A_b$.
Its scaled residual mass at scale $j$ is at least
\[
\frac{\ell_j M}{2|\mathcal B|}
\ge\frac{\eta S}{16k|\mathcal B|}
\ge\frac{\eta\rho}{16k|\mathcal B|},
\]
so \eqref{eq:anchor-far} holds.
The preceding sampling guarantee therefore gives a point of $P_b$ within distance $2\ell_j$ of this ball, and hence within distance $6\bar r+2\ell_j$ of $c$.
Sending the cluster to this candidate costs at most $7M\bar r+2M\ell_j$.

In all cases, the cluster contributes at most $7M\bar r+\eta S/k+4M\ell_{\min}$ to $\cost(\mu,P)$.
Summing over the at most $k$ clusters, whose total mass is at most $m$, gives
\[
\cost(\mu,P)\le7\opt(\mu)+\eta S+4m\ell_{\min}.
\]

\textbf{Union Bound and Complexities.}
Each block fails with probability at most $\delta'$ for \eqref{eq:anchor-jl} and $\delta'$ for the sampler.
Since each block is analyzed with private randomness, as we discussed earlier in this section, a union bound over the $O(m\log^2(2S_{\max}))$ possible blocks gives failure probability at most $\delta/2$, even under adaptive updates.

For space, at each of the polylogarithmically many scales, $O(\log(2m\log(2S_{\max})))$ blocks are open or kept, and there are at most $1+|\mathcal B|(2(k+1)+\widetilde O(K))=\widetilde O(k/\eta)$ output points.
Note that by the norm bound in \Cref{poly:lem:jl}(i), each coordinate of the cell index $Q_0(x)$ has magnitude polynomial in $2mkd\Delta/\delta$ and $\ell_{\min}^{-1}$ and $Gx \in \mathbb{R}^{\tilde d}$, so the labels $(j,Q_0(x))$ lie in a universe of size $N$ with $\log N=\widetilde O(\tilde d)$.
Thus each sampler ran on the projected weight vector uses $\widetilde O(\tilde d^2)=\widetilde O(1)$ bits of space by \Cref{clm:fast-sampler}, and all CountSketches in one decoder use $\widetilde O(d)$ bits of space by \Cref{lem:heavy-address}, so the total space is $\widetilde O(kd/\eta)$ bits.

For the update time, each residual change $\nu_{\mathrm{old}}-\nu_{\mathrm{new}}$ can be processed as $O(k)$ coordinate updates $\varsigma e_x$.
For each such update and each open block, we compute $Gx$ once using $O(d\tilde d)$ arithmetic operations.
For each repetition and scale $j$, we compute $Q_j(x)$ and $Q_0(x)$ in $O(\tilde d)$ operations, identify the bucket $h(j,Q_j(x))$, and update its sampler at label $(j,Q_0(x))$ by $\ell_j\varsigma$.
Updating the sampler and its decoding sketches takes $O(H\polylog(N/\delta_{\mathrm{dec}}))=\widetilde O(d)$ time by \Cref{clm:fast-sampler,lem:heavy-address}.
At closure, for each repetition, a block evaluates the hash at the $O(kJ)$ anchor cells to identify the discarded buckets and queries each remaining sampler.
There are at most $K$ samplers per repetition, and each successful sample is decoded using $H+1$ point queries, taking $\widetilde O(d)$ time in total.
Since $K=\widetilde O(k/\eta)$ and the numbers of blocks, repetitions, and scales are polylogarithmic, updates and reports take $\poly(k,d,\log(m\Delta/\delta))$ time for fixed $\eta$.
\end{proof}

\subsubsection*{Assigning Weights to the Candidate Locations}
Let $P=\{p_1,\ldots,p_n\}\subseteq[\Delta]^d$ be the candidate locations recovered by \Cref{poly:lem:locations} with $n=\widetilde O(k/\eta)$ points, and let $M=|\mu|$ be the current total weight.
Our goal is to obtain a nonnegative weight vector $\sigma$ on $P$ of total weight $M$ that is close to $\mu$ in transportation cost.

\paragraph{Success Events.}
We use the public projections $G_1,\ldots,G_{n_g}$ of \Cref{poly:lem:jl}(ii), drawn before the stream with a sufficiently small $\zeta=\widetilde\Theta(\eta^2/(dk)^2)$, and the inner-product estimator of \Cref{lemma:linear_query} on the residual features $A(\mu-\nu)$
Define the following success events.
\begin{itemize}
\item $\mathcal E_{\mathrm{proj}}$: the guarantee \eqref{poly:eq:jl} holds for $G_1,\ldots,G_{n_g}$.
\item $\mathcal E_{\mathrm{query}}$: $\widehat I$ answers every query in the proof of \Cref{wt:lem:direct} within the accuracy of \Cref{lemma:linear_query}.
\end{itemize}
Each event holds with probability at least $1-\delta/4$, uniformly over all inputs and candidate pools, and hence also under adaptive updates.
In the rest of this subsection, we condition on $\mathcal E_{\mathrm{proj}}\cap\mathcal E_{\mathrm{query}}$.

Let $\mathcal R\subseteq[n_g]$ be the set of projections that preserve every pair in $P$ within squared distortion $1\pm1/4$, which we call \emph{pair preservation}.
For a function $h$, define
\[
 \phi:=\frac{|\mathcal R|}{n_g},
 \qquad
 \E^*h(G):=\frac{1}{|\mathcal R|}\sum_{j\in\mathcal R}h(G_j).
\]
Choose the constant in $\zeta$ so that, on $\mathcal E_{\mathrm{proj}}$, a union bound gives $\phi\ge1-\binom n2\zeta\ge0.99$ for every candidate set of $\widetilde O(k/\eta)$ points.

\paragraph{Smooth Weight Attribution.}
Intuitively, each input point attributes its weight among the candidates by their distance relative to its nearest-candidate distance: those at relative distance at most $1$ receive full weight, those at relative distance in $(1,2)$ receive a weight that decreases smoothly to zero, and those farther away receive none.
Let $r_0$ be the largest power of two at most $\eta/(100m)$.
Put $\omega:=\lceil\log_2(2n_{\max})\rceil$, where $n_{\max}=\widetilde O(k/\eta)$ is the upper bound of the candidate set size $n$, and for a retained projection $G$ and $y\in\R^{\tilde d}$, define
\[
 \begin{split}
 \psi(u)&:=\bigl(\max\{0,\min\{1,2-u\}\}\bigr)^{4\omega},
 \qquad
 t(y):=\max\Bigl\{r_0,\min_i\|y-Gp_i\|_2\Bigr\},\\
 \xi_i(y)&:=
 \frac{\psi(\|y-Gp_i\|_2/t(y))}
      {\sum_h\psi(\|y-Gp_h\|_2/t(y))},
 \end{split}
\]
where $\xi_i$ is the normalized smooth score function for candidate $p_i$.
Define the ideal weights and the projected nearest-candidate cost by
\[
 \sigma_i^*:=\sum_x\mu_x\E^*\xi_i(Gx),
 \qquad
 D_P:=\sum_x\mu_x\E^*\min_i\|Gx-Gp_i\|_2.
\]

The next claim bound the transportation cost and the gradient of the ideal assignment.
\begin{claim}
\label{wt:clm:smooth}
The ideal weights and the projected nearest-candidate cost satisfy
\[
 \mathcal K(\mu-\sigma^*)\le8D_P+8mr_0,
 \qquad
 D_P\le2\cost(\mu,P),
 \qquad
 \sum_i\|\nabla \xi_i(y)\|_2\le\frac{24e\omega}{t(y)},
\]
where the gradient bound holds at every point of differentiability.
\end{claim}
\begin{proof}
We first prove the cost bounds, and then bound the gradient.

\textbf{Cost Bounds.}
Fix $x$, a nearest candidate $p$ of $x$ in the original space, and $r:=\|x-p\|_2$.
Since the retained projections form at least a $0.99$ fraction, \eqref{poly:eq:jl} and Cauchy--Schwarz give
\[
 \E^*\|G(x-p)\|_2\le\bigl(\E^*\|G(x-p)\|_2^2\bigr)^{1/2}\le\Bigl(\frac{3}{2\cdot0.99}\Bigr)^{1/2}r\le\frac54r.
\]
Moreover, all but an $n\zeta/\phi\le0.01$ fraction of the retained projections preserve every difference $x-p_i$ within $1\pm1/4$ in squared norm, so $M_x:=\E^*\min_i\|Gx-Gp_i\|_2\ge0.99\sqrt{3/4}\,r\ge r/2$.
Since also $M_x\le\E^*\|G(x-p)\|_2\le5r/4$, summing over $x$ with weights $\mu_x$ gives $D_P\le2\cost(\mu,P)$.
Sending the weight $\mu_x$ of each point $x$ to the candidates in proportions $\E^*\xi_i(Gx)$ is a transportation plan from $\mu$ to $\sigma^*$.
Fix a retained $G$ and put $y:=Gx$.
If $\xi_i(y)>0$, then $\|y-Gp_i\|_2\le2t(y)$, and
\[
 \|x-p_i\|_2\le r+\|p-p_i\|_2\le r+\tfrac65\|G(p-p_i)\|_2\le r+\tfrac65\bigl(2t(y)+\|G(x-p)\|_2\bigr).
\]
The first inequality is the triangle inequality, the second uses pair preservation, and the third uses $\|Gp-Gp_i\|_2\le\|Gp-y\|_2+\|y-Gp_i\|_2$.
Averaging with the scores $\xi_i(y)$ and over the retained projections, and using $\E^*t(Gx)\le M_x+r_0$ and $r\le2M_x$, we get
\[
 \E^*\sum_i\xi_i(Gx)\|x-p_i\|_2\le\tfrac52r+\tfrac{12}5(M_x+r_0)\le8M_x+8r_0.
\]
Multiplying by $\mu_x$ and summing over $x$ gives $\mathcal K(\mu-\sigma^*)\le8D_P+8Mr_0\le8D_P+8mr_0$.

\textbf{Gradient Bound.}
Write $t:=t(y)$, $d_i:=\|y-Gp_i\|_2$, and $\psi_i:=\psi(d_i/t)$.
Wherever $\psi_i$ has a nonzero derivative, $d_i\le2t$.
Since $d_i$ and $t$ are $1$-Lipschitz, $\|\nabla(d_i/t)\|_2\le1/t+d_i/t^2\le3/t$.
As $|\psi'(u)|\le4\omega\psi(u)^{1-1/(4\omega)}$, this gives $\|\nabla\psi_i\|_2\le(12\omega/t)\psi_i^{1-1/(4\omega)}$.
The quotient rule gives $\sum_i\|\nabla \xi_i\|_2\le2\sum_i\|\nabla\psi_i\|_2/\sum_h\psi_h$, and H\"older's inequality, with $\sum_h\psi_h\ge1$, gives
\[
 \sum_i\psi_i^{1-1/(4\omega)}\le n^{1/(4\omega)}\Bigl(\sum_h\psi_h\Bigr)^{1-1/(4\omega)}\le e\sum_h\psi_h.
\]
Combining the last three bounds gives $\sum_i\|\nabla \xi_i\|_2\le24e\omega/t$, which proves the claim.
\end{proof}

\paragraph{The Operator $A$.}
We use the order-$1$ HLA of \Cref{sec:alg_ingredient}, with levels $1,\ldots,L$, implemented by the time-efficient HLA of \Cref{poly:lem:diagonal}.
By \eqref{poly:eq:jl} applied to the unit coordinate differences, $\|G_je_i\|_2\le\sqrt{3n_g/2}$ for every $i,j$.
Choose the smallest power of two $\widetilde\Delta\ge d\Delta\sqrt{3n_g/2}$, so all projected points lie in $[-\widetilde\Delta,\widetilde\Delta]^{\tilde d}$, and put $s_0:=4\widetilde\Delta$ and $s_\ell:=2^{-\ell}s_0$.
For a prescribed accuracy $0<\beta<1$, choose the least $L$ before the stream so that
\[
 s_L\le\tfrac14,
 \qquad
 C(m+1)\omega\sqrt{\tilde d}\,s_L\le\beta/8
\]
where $C$ is a sufficiently large absolute constant.
Thus the number of levels is polylogarithmic and the feature map is fixed independently of the recovered candidate locations.
Replace the level weight $s_\ell$ by the power of two $w_\ell:=2^{\lceil\log_2(40\sqrt{\tilde d}\,s_\ell)\rceil}$, which dominates the transport displacement $3\sqrt{\tilde d}\,s_\ell$, and define
\[
 A_{(\ell,a,j),x}:=\frac{w_\ell}{n_g}F_\ell(a,G_jx).
\]
We introduce a single level-$0$ representative at the origin, setting
$F_0(0,y)=1$ for every $y$ and $U_1(0,a)=1$ for every level-$1$
representative $a$.
Thus every level-$1$ representative has the origin as its only parent.
The feature coordinates are triples $(\ell,a,j)$ with $1\le\ell\le L$
such that $F_\ell(a,G_jx)>0$ for some $x\in[\Delta]^d$.
As in \Cref{sec:alg_ingredient}, we maintain sketches of $A$ without
explicitly storing the matrix: for an update at $x$, we enumerate
the nonzero entries of $F_\ell(\cdot,G_jx)$ for each $\ell,j$
and apply the corresponding weighted updates.
We now bound the residual norm by transportation cost.

\begin{claim}
\label{poly:clm:feature-bound}
Every $\lambda$ on $[\Delta]^d$ of zero total weight satisfies $\|A\lambda\|_1\le O(\tilde d(L+1))\mathcal K(\lambda)$.
\end{claim}
\begin{proof}
We bound the features of a single transported unit.
At each level $1\le\ell\le L$, the Lipschitz bound of \Cref{poly:lem:diagonal} and $w_\ell\le80\sqrt{\tilde d}\,s_\ell$ give, for projected points $u,v$,
\[
 w_\ell\|F_\ell(\cdot,u)-F_\ell(\cdot,v)\|_1\le80\sqrt{\tilde d}\,s_\ell\cdot\frac{2\|u-v\|_1}{s_\ell}\le160\tilde d\|u-v\|_2.
\]
Summing over the levels and averaging over the projections, Cauchy--Schwarz and \eqref{poly:eq:jl} give $\|A(e_x-e_y)\|_1\le O(\tilde d(L+1))\|x-y\|_2$.
Finally, let $\pi$ be an optimal transportation plan for $\lambda$, so that $\sum_y\pi_{x,y}=\lambda^+_x$, $\sum_x\pi_{x,y}=\lambda^-_y$, and $\mathcal K(\lambda)=\sum_{x,y}\pi_{x,y}\|x-y\|_2$.
Then $\lambda=\lambda^+-\lambda^-=\sum_{x,y}\pi_{x,y}(e_x-e_y)$, so the linearity of $A$, the triangle inequality, and the bound above give
\[
 \|A\lambda\|_1\le\sum_{x,y}\pi_{x,y}\|A(e_x-e_y)\|_1\le O(\tilde d(L+1))\sum_{x,y}\pi_{x,y}\|x-y\|_2=O(\tilde d(L+1))\,\mathcal K(\lambda).
\]
\end{proof}

\paragraph{Direct Weight Estimates.}
Let $\xi_i^{G_j}$ denote the score function $\xi_i$ with $G=G_j$ and, following \Cref{eqn:def_A,eqn:def_gC}, define
\begin{equation}
\label{wt:eq:g}
 (g_i)_{\ell,a,j}
 :=
 \frac{\xi_i^{G_j}(a)-\sum_bU_\ell(b,a)\xi_i^{G_j}(b)}
      {\phi\,w_\ell}
\end{equation}
for retained $j$ and $\ell\ge1$, and set the other coordinates to zero.
At level $1$, the subtracted term is the value at the root.
Write $\lambda:=\mu-\nu$.
Telescoping as in \Cref{sec:alg_ingredient} cancels the root contribution because $|\mu-\nu|=0$ and leaves the terminal error
\[
 E_i(\lambda):=\sum_x\lambda_x\E^*\Bigl[\xi_i(Gx)-\sum_aF_L(a,Gx)\xi_i(a)\Bigr].
\]
Using the linear output map $\widehat I$ from \Cref{lemma:linear_query}, we obtain
\begin{equation}
\label{wt:eq:estimate}
 \sigma_i^*
 =
 \sum_x\nu_x\E^*\xi_i(Gx)
 +\ip{A(\mu-\nu)}{g_i}+E_i(\lambda),
 \qquad
 \widehat\sigma_i:=
 \sum_x\nu_x\E^*\xi_i(Gx)
 +\widehat I(g_i),
\end{equation}
Similarly, define $g_P$ by replacing the $\xi_i^{G_j}(\cdot)$ in \eqref{wt:eq:g} by $\min_i\|\cdot-G_jp_i\|_2$ and define $\widehat D_P$ analogously from $g_P$.
Since $\sum_i \xi_i\equiv1$ and the columns of $F_L$ sum to one, $\sum_iE_i(\lambda)=0$.
We next bound the estimation error.

\begin{lemma}
\label{wt:lem:direct}
For $0<\beta<1$, simultaneously for all candidate pools and valid calls,
\[
 |\widehat\sigma|=M,
 \qquad
 \mathcal K(\widehat\sigma-\sigma^*)+|\widehat D_P-D_P|\le\beta S.
\]
The estimates use $\widetilde O((kd/\eta)\beta^{-O(1)})$ bits and polynomial time.
\end{lemma}
\begin{proof}
We reduce $\mathcal K(\widehat\sigma-\sigma^*)$ to the errors of finitely many bounded linear queries by the Kantorovich--Rubinstein formula.
We first analyze the unrounded estimates.

\textbf{Reduction to Inner Product Estimation.}
Both $\widehat\sigma$ and $\sigma^*$ are supported on $P$ and have total weight $M$.
For $\sigma^*$, this holds because the scores sum to one.
For $\widehat\sigma$, the same fact makes the telescoping coefficients of $\sum_i\xi_i\equiv1$ vanish, so $\sum_ig_i=0$, and the linearity of $\widehat I$ gives $|\widehat\sigma|=M$.
Let
\[
 \Pi:=\bigl\{\pi\in\R^P:\ \pi_1=0,\ |\pi_i-\pi_h|\le\|p_i-p_h\|_2\text{ for all }i,h\bigr\}
\]
be the $1$-Lipschitz potentials on $P$, normalized at $p_1$. The normalization loses nothing, since $\widehat\sigma-\sigma^*$ has zero mass.
By the Kantorovich--Rubinstein formula, and by \eqref{wt:eq:estimate} and the linearity of $\widehat I$,
\[
 \mathcal K(\widehat\sigma-\sigma^*)=\max_{\pi\in\Pi}\sum_i\pi_i(\widehat\sigma_i-\sigma_i^*)
 =\max_{\pi\in\Pi}\Bigl(\widehat I(g_\pi)-\ip{A(\mu-\nu)}{g_\pi}-\sum_i\pi_iE_i(\lambda)\Bigr),
 \qquad
 g_\pi:=\sum_i\pi_ig_i.
\]
Thus it suffices to bound the query error and the terminal remainder uniformly over $\pi\in\Pi$.

\textbf{Query Errors.}
The vector $g_\pi$ is the coefficient vector of $\xi_\pi:=\sum_i\pi_i\xi_i$.
Fix $y$ and a candidate $p_j$ minimizing $\|y-Gp_j\|_2$.
Since $\sum_i\nabla \xi_i=0$,
\[
 \nabla \xi_\pi(y)=\sum_i(\pi_i-\pi_j)\nabla \xi_i(y).
\]
Whenever $\nabla \xi_i(y)\neq0$, we have $\|y-Gp_i\|_2\le2t(y)$ and $\|y-Gp_j\|_2\le t(y)$, so pair preservation gives $|\pi_i-\pi_j|\le\|p_i-p_j\|_2\le4t(y)$.
With \Cref{wt:clm:smooth}, this gives $\|\nabla \xi_\pi(y)\|_2\le4t(y)\cdot24e\omega/t(y)=O(\omega)$ at every point of differentiability, so $\xi_\pi$ is $O(\omega)$-Lipschitz.
The terminal-remainder argument of \Cref{sec:alg_ingredient}, locality, and $\|\lambda\|_1\le\|\mu\|_1+\|\nu\|_1\le2m+2$ therefore give
\[
 \sup_{\pi\in\Pi}\Bigl|\sum_i\pi_iE_i(\lambda)\Bigr|
 \le O\bigl((m+1)\omega\sqrt{\tilde d}\,s_L\bigr)\le\beta S/8,
 \qquad
 |E_P(\lambda)|\le O\bigl((m+1)\sqrt{\tilde d}\,s_L\bigr)\le\beta S/8.
\]
These bounds hold for every candidate locations set and potential without an additional failure event.
Every transport of $U_\ell$ moves by at most $3\sqrt{\tilde d}\,s_\ell\le w_\ell$, including the root transition by the choice of $s_0$, so \eqref{wt:eq:g} and $\phi\ge0.99$ give
\[
 \|g_\pi\|_\infty=O(\omega),
 \qquad
 \|g_P\|_\infty=O(1),
\]
where the second bound holds because $\min_i\|\cdot-Gp_i\|_2$ is $1$-Lipschitz.
By \Cref{poly:clm:feature-bound} and $\mathcal K(\mu-\nu)\le S$,
\[
 \|A(\mu-\nu)\|_1\le O(\tilde d(L+1))K(\mu-\nu)\le O(\tilde d(L+1))S.
\]
As in \Cref{sec:combine}, we append a constant coordinate $1$ with zero coefficient in every query to meet the estimator's lower norm requirement. Since $S\ge1$, its norm contribution is absorbed in the preceding bound.
We normalize the queries by a common $O(\omega)$ bound and apply \Cref{lemma:linear_query} with accuracy a sufficiently small multiple of $\beta/(\omega\tilde d(L+1))$.
After rescaling, every query error is at most $\beta S/8$, so the terminal-error bounds give $\mathcal K(\widehat\sigma-\sigma^*)\le\beta S/4$ and $|\widehat D_P-D_P|\le\beta S/4$ conditioned on $\mathcal E_{\mathrm{query}}$.
The event $\mathcal E_{\mathrm{query}}$ covers a fixed finite family of queries.
Both the query error and the terminal remainder are linear in $\pi$, so their maxima over the polytope $\Pi$ are attained at vertices.
A vertex is determined by $n-1$ tight constraints among at most $2n^2$, so $\Pi$ has at most $(2n^2)^{n-1}$ vertices.
We therefore include, for every pool of $\widetilde O(k/\eta)$ grid points, the queries $g_\pi$ at these vertices together with $g_P$.
The logarithm of the size of this family is
\[
 \widetilde O\bigl((k/\eta)\,d\log(2\Delta)\bigr).
\]
Conditional on the public projections, the family is fixed independently of the estimator, so the guarantee covers pools chosen adaptively via a union bound.

\textbf{Rounding and Complexities.}
Before the stream, in addition to its earlier requirements, we choose the dyadic grid $\gamma\Z$ so that $n\gamma\sqrt d\,\Delta\le\beta/16$ for every pool, which is possible since $n=\widetilde O(k/\eta)$.
Round $\widehat\sigma_2,\ldots,\widehat\sigma_{n}$ to $\gamma\Z$, and set $\widehat\sigma_1:=M-\sum_{i=2}^{n}\widehat\sigma_i$.
This preserves the total weight exactly, changes each $\widehat\sigma_i$ with $i\ge2$ by at most $\gamma$, and hence changes $\widehat\sigma_1$ by at most $n\gamma$.
Since all candidates lie in $[\Delta]^d$, the transportation error grows by at most $2n\gamma\sqrt d\,\Delta\le\beta S/8$.
Round $\widehat D_P$ to $\gamma\Z$ with additive error at most $\beta/8$.
Together with the previous paragraphs, this proves the error bound.
With the accuracy and family size above, \Cref{lemma:linear_query} uses $\widetilde O((kd/\eta)\beta^{-O(1)})$ bits, and the readout is evaluated in polynomial time.
\end{proof}

\paragraph{A Sparse Graph for the Correction.}
The estimate $\widehat\sigma$ may have negative entries, which we cancel by moving positive weight onto them.
Note that we may lose $k^2$ if we run directly the minimum-cost-flow algorithm on the graph constructed by the candidate set $P=\{p_1,\ldots,p_n\}\subseteq[\Delta]^d$ of $\widetilde O(k/\eta)$ points.
We thus construct a sparse graph on $P$ using the greedy spanner of Alth\"ofer et al.~\cite{Althofer1993}.
Recall that $\omega=\lceil\log_2(2n)\rceil$.
For $i\ne h$, let $c_{ih}$ be the smallest power of two at least $\|p_i-p_h\|_2$.
Apply the greedy algorithm with stretch $2\omega+1$ to the implicit complete graph with edge lengths $c_{ih}$, and denote the resulting graph by $H$.
Every edge permits transportation in both directions.
For $n=1$, use the one-vertex graph.
We next present the guarantee of this sparse graph.

\begin{lemma}
\label{wt:lem:graph}
The graph $H$ has $n$ vertices and $O(n)$ edges, is deterministically constructible in polynomial time using $\widetilde O(nd)$ bits, and satisfies
\[
 \|p_i-p_h\|_2
 \le d_H(p_i,p_h)
 \le (4\omega+2)\|p_i-p_h\|_2
 \qquad\text{for all }i,h \in [n].
\]
Its edge lengths are positive integers of $O(\log(2d\Delta))$ bits.
\end{lemma}
\begin{proof}
By \cite[Theorem~1]{Althofer1993}, the greedy graph has fewer than $n\lceil n^{1/\omega}\rceil\le2n$ edges.
Every edge length dominates the Euclidean distance between its endpoints, so the triangle inequality gives the lower distance bound.
The spanner guarantee and $c_{ih}\le2\|p_i-p_h\|_2$ give the upper bound.

The greedy construction processes candidate pairs in nondecreasing order of distance.
We generate this order without storing all pairs.
Assign each unordered pair $\{p_i,p_h\}$ to the \emph{length class} $c$ with $4^{c-1}<\|p_i-p_h\|_2^2\le4^c$, i.e., the pair's distance rounded up to the power of two $2^c$.
Since the candidates are grid points, $\|p_i-p_h\|_2^2$ is an integer, so its class is determined exactly by comparing it with powers of four, and since distinct grid points are at distance between $1$ and $\sqrt d\,\Delta$, there are $O(\log(2d\Delta))$ classes.
We process the classes in increasing order.
For each class, we scan all pairs and process those in the class in lexicographic order.
This rescans the candidates once per class, but never stores a list of pairs.
The greedy graph has $O(n)$ edges, so each shortest-path test runs on $O(n)$ records of $O(\log(2nd\Delta))$ bits.
Together with the stored candidates, this uses $\widetilde O(nd)$ bits and polynomial time in the worst case.
\end{proof}

\paragraph{Minimum-Cost Flow.}
We use the minimum-cost-flow algorithm of \cite{GoldbergTarjan1990} as a black box.
It runs in polynomial time and uses storage linear in the graph size, up to bit-length factors.
Replace each edge of $H$ by two opposite arcs of the same cost, and add a source $s$ and a sink $t$.
For $\widehat\sigma_i>0$, add a zero-cost arc $s\to p_i$ of capacity $\widehat\sigma_i$, and for $\widehat\sigma_i<0$, add a zero-cost arc $p_i\to t$ of capacity $-\widehat\sigma_i$.
Every arc of $H$ has capacity $\sum_i\widehat\sigma_i^+$, and we find a minimum-cost flow of value $\sum_i\widehat\sigma_i^-$.
Such a flow exists, since the total positive supply is at least the total deficit by $|\widehat\sigma|=M\ge0$.
The unused positive supplies form the output:
\[
 \sigma_i:=
 \begin{cases}
 \widehat\sigma_i-\operatorname{flow}(s,p_i),
     &\widehat\sigma_i>0,\\
 0,  &\widehat\sigma_i\le0.
 \end{cases}
\]
If no estimate is negative, then $\sigma=\widehat\sigma$.
The costs are already integers; scale the capacities by $\gamma^{-1}$ before invoking the black box, so rescaling its integral output gives weights in $\gamma\Z$.

\begin{lemma}
\label{poly:lem:masses}
The procedure returns $\sigma\ge0$ on $P$ with $|\sigma|=M$, and a certificate $U$ satisfying
\[
 \mathcal K(\mu-\sigma)
 \le U
 \le16\cost(\mu,P)+\eta S.
\]
It uses $\widetilde O(kd\,\eta^{-O(1)})$ bits and polynomial time.
\end{lemma}
\begin{proof}
Put $\omega_{\max}:=\lceil\log_2(2n_{\max})\rceil$, where $n_{\max}=\widetilde O(k/\eta)$ is the bound on $|P|$ from \Cref{poly:lem:locations}.
Choose the accuracy $\beta$ before the stream as the largest power of two satisfying $(4\omega_{\max}+11)\beta\le\eta/4$, i.e., $(4\omega+11)\beta\le\eta/4$ for all possible candidate locations sets.
Fix the finest level $L$, the rounding mesh $\gamma$, and the inner-product accuracy using this $\beta$.
Apply \Cref{wt:lem:direct} with this accuracy and output $U:=8\widehat D_P+\eta S/2$.

\textbf{Cost of the Correction.}
Let $\mathcal K_H$ denote transportation cost under $d_H$, and fix an optimal plan for $\widehat\sigma-\sigma^*$ under $d_H$.
Its supply at $p_i$ is $(\widehat\sigma_i-\sigma_i^*)^+\le\widehat\sigma_i^+$, since $\sigma^*\ge0$.
At each $p_i$ with $\widehat\sigma_i<0$, it delivers $\sigma_i^*-\widehat\sigma_i\ge-\widehat\sigma_i$.
Keep only enough of these deliveries to fill the deficit $-\widehat\sigma_i$, discard all other shipments, and route the kept shipments along shortest paths in $H$.
This is a feasible flow in the correction network, and by \Cref{wt:lem:graph}, its cost is at most
\[
 \mathcal K_H(\widehat\sigma-\sigma^*)
 \le(4\omega+2)\mathcal K(\widehat\sigma-\sigma^*)
 \le(4\omega+2)\beta S.
\]
The minimum-cost flow is no more expensive.
Moving weight from each positive entry along its flow paths turns $\widehat\sigma$ into $\sigma$, and since $d_H$ dominates Euclidean distance, $\mathcal K(\sigma-\widehat\sigma)$ is at most the flow cost.
Hence
\[
 \mathcal K(\sigma-\sigma^*)
 \le\mathcal K(\sigma-\widehat\sigma)
     +\mathcal K(\widehat\sigma-\sigma^*)
 \le(4\omega+3)\beta S.
\]

\textbf{Error and Certificate.}
Every deficit is filled and the unused supplies are nonnegative, so $\sigma\ge0$ and $|\sigma|=|\widehat\sigma|=M$.
By the triangle inequality and \Cref{wt:clm:smooth,wt:lem:direct},
\[
 \mathcal K(\mu-\sigma)
 \le8\widehat D_P+8mr_0+(4\omega+11)\beta S
 \le U.
\]
The last inequality uses $8mr_0\le2\eta S/25$ and the choice of $\beta$.
Conversely, \Cref{wt:clm:smooth,wt:lem:direct} and $\beta\le\eta/60$ give
\[
 U\le8D_P+8\beta S+\eta S/2
 \le16\cost(\mu,P)+\eta S.
\]

\textbf{Complexities.}
For a feasible network with $N$ vertices, $E$ arcs, and $B$-bit integer data, the cost-scaling algorithm of \cite{GoldbergTarjan1990} returns an integral minimum-cost flow in polynomial bit time, and its sparse implementation stores $O(N+E)$ arc and vertex records of $O(B+\log(N+E))$ bits each by~\cite[Sections~3.2--3.3]{KiralyKovacs2012}.
By \Cref{wt:lem:graph}, our network has $O(n)$ vertices and arcs with polylogarithmic bit length, i.e., $\widetilde O(k/\eta)$ bits in total.
Together with \Cref{wt:lem:direct}, this gives the claimed space and polynomial time.
The graph construction and flow correction are deterministic for every candidate pool, so they introduce no additional failure event under adaptive updates.
\end{proof}

Finally, we combine the two steps to prove the reconstruction lemma.

\begin{proof}[Proof of \Cref{poly:lem:reconstruction}]
Apply \Cref{poly:lem:locations,poly:lem:masses} with accuracy $\eta/64$, and with the finest scale $\ell_{\min}\le\eta/(65536m)$.
Since $S\ge1$,
\[
 \mathcal K(\mu-\sigma)\le U\le16\Bigl(7\opt(\mu)+\frac{\eta S}{64}+4m\ell_{\min}\Bigr)+\frac{\eta S}{64}\le112\opt(\mu)+\eta S.
\]
The success events of both steps hold uniformly over all inputs and candidate pools, so the guarantee also holds under adaptive updates.
At fixed reconstruction accuracy, the space is $\widetilde O(kd)$ bits.
\end{proof}

\section*{AI Disclosure} 
The research direction, including identifying the problems studied and formulating the central research questions, 
was conceived entirely by the human authors without AI assistance. 
The technical results were subsequently developed in conjunction with OpenAI’s GPT Astra, 
which assisted through many iterations in exploring mathematical approaches, developing arguments, and refining proofs. 
All mathematical claims and proofs were independently examined and verified by the human authors. 
The manuscript was written entirely by the human authors, 
who take full responsibility for the correctness and content of the work.

\bibliographystyle{alphaurl}

\begin{thebibliography}{GLW{\etalchar{+}}26b}

\bibitem[ABJ{\etalchar{+}}22]{AjtaiBJSSWZ22}
Mikl{\'{o}}s Ajtai, Vladimir Braverman, T.~S. Jayram, Sandeep Silwal, Alec Sun, David~P. Woodruff, and Samson Zhou.
\newblock The white-box adversarial data stream model.
\newblock In {\em {PODS} '22: International Conference on Management of Data}, pages 15--27, 2022.

\bibitem[AC24]{AhmadianC24}
Sara Ahmadian and Edith Cohen.
\newblock Unmasking vulnerabilities: Cardinality sketches under adaptive inputs.
\newblock In {\em Forty-first International Conference on Machine Learning, {ICML}}, 2024.

\bibitem[ACGS23]{AssadiCGS23}
Sepehr Assadi, Amit Chakrabarti, Prantar Ghosh, and Manuel Stoeckl.
\newblock Coloring in graph streams via deterministic and adversarially robust algorithms.
\newblock In {\em Proceedings of the 42nd {ACM} {SIGMOD-SIGACT-SIGAI} Symposium on Principles of Database Systems, {PODS}}, pages 141--153, 2023.

\bibitem[ACKS15]{socg/AwasthiCKS15}
Pranjal Awasthi, Moses Charikar, Ravishankar Krishnaswamy, and Ali~Kemal Sinop.
\newblock The hardness of approximation of euclidean k-means.
\newblock In {\em 31st International Symposium on Computational Geometry, SoCG}, volume~34, pages 754--767, 2015.

\bibitem[ADD{\etalchar{+}}93]{Althofer1993}
Ingo Alth{\"o}fer, Gautam Das, David Dobkin, Deborah Joseph, and Jos{\'e} Soares.
\newblock On sparse spanners of weighted graphs.
\newblock {\em Discrete \& Computational Geometry}, 9(1):81--100, 1993.

\bibitem[AGK{\etalchar{+}}04]{siamcomp/AryaGKMMP04}
Vijay Arya, Naveen Garg, Rohit Khandekar, Adam Meyerson, Kamesh Munagala, and Vinayaka Pandit.
\newblock Local search heuristics for $k$-median and facility location problems.
\newblock {\em SIAM J. Comput.}, 33(3):544--562, 2004.

\bibitem[AMYZ19]{AvdiukhinMYZ19}
Dmitrii Avdiukhin, Slobodan Mitrovic, Grigory Yaroslavtsev, and Samson Zhou.
\newblock Adversarially robust submodular maximization under knapsack constraints.
\newblock In {\em Proceedings of the 25th {ACM} {SIGKDD} International Conference on Knowledge Discovery {\&} Data Mining, {KDD}}, pages 148--156, 2019.

\bibitem[Bar96]{focs/Bartal96}
Yair Bartal.
\newblock Probabilistic approximations of metric spaces and its algorithmic applications.
\newblock In {\em {FOCS}}, pages 184--193. {IEEE} Computer Society, 1996.

\bibitem[BCP{\etalchar{+}}24]{focs/0001CPSS24}
Nikhil Bansal, Vincent Cohen{-}Addad, Milind Prabhu, David Saulpic, and Chris Schwiegelshohn.
\newblock Sensitivity sampling for k-means: Worst case and stability optimal coreset bounds.
\newblock In {\em {FOCS}}, pages 1707--1723. {IEEE}, 2024.

\bibitem[BDP{\etalchar{+}}25]{BravermanDPSW25}
Vladimir Braverman, Prathamesh Dharangutte, Shreyas Pai, Vihan Shah, and Chen Wang.
\newblock Fully dynamic adversarially robust correlation clustering in polylogarithmic update time.
\newblock In {\em International Conference on Artificial Intelligence and Statistics, {AISTATS}}, pages 1477--1485, 2025.

\bibitem[BEO22]{Ben-EliezerEO22}
Omri Ben{-}Eliezer, Talya Eden, and Krzysztof Onak.
\newblock Adversarially robust streaming via dense-sparse trade-offs.
\newblock In {\em 5th Symposium on Simplicity in Algorithms, SOSA@SODA}, pages 214--227, 2022.

\bibitem[BFL{\etalchar{+}}17]{icml/BravermanFLSY17}
Vladimir Braverman, Gereon Frahling, Harry Lang, Christian Sohler, and Lin~F. Yang.
\newblock Clustering high dimensional dynamic data streams.
\newblock In {\em {ICML}}, volume~70 of {\em Proceedings of Machine Learning Research}, pages 576--585. {PMLR}, 2017.

\bibitem[BHM{\etalchar{+}}21]{BravermanHMSSZ21}
Vladimir Braverman, Avinatan Hassidim, Yossi Matias, Mariano Schain, Sandeep Silwal, and Samson Zhou.
\newblock Adversarial robustness of streaming algorithms through importance sampling.
\newblock In {\em Advances in Neural Information Processing Systems 34: Annual Conference on Neural Information Processing Systems, NeurIPS}, pages 3544--3557, 2021.

\bibitem[BIRW16]{BackursIRW16}
Arturs Backurs, Piotr Indyk, Ilya~P. Razenshteyn, and David~P. Woodruff.
\newblock Nearly-optimal bounds for sparse recovery in generic norms, with applications to \emph{k}-median sketching.
\newblock In {\em Proceedings of the Twenty-Seventh Annual {ACM-SIAM} Symposium on Discrete Algorithms, {SODA}}, pages 318--337, 2016.

\bibitem[BJWY22]{jacm/Ben-EliezerJWY22}
Omri Ben{-}Eliezer, Rajesh Jayaram, David~P. Woodruff, and Eylon Yogev.
\newblock A framework for adversarially robust streaming algorithms.
\newblock {\em J. {ACM}}, 69(2):17:1--17:33, 2022.

\bibitem[BKM{\etalchar{+}}22]{BeimelKMNSS22}
Amos Beimel, Haim Kaplan, Yishay Mansour, Kobbi Nissim, Thatchaphol Saranurak, and Uri Stemmer.
\newblock Dynamic algorithms against an adaptive adversary: generic constructions and lower bounds.
\newblock In {\em {STOC} '22: 54th Annual {ACM} {SIGACT} Symposium on Theory of Computing}, pages 1671--1684, 2022.

\bibitem[BOS26]{Ben-EliezerOS26}
Omri Ben{-}Eliezer, Krzysztof Onak, and Sandeep Silwal.
\newblock Robust streaming against low-memory adversaries.
\newblock In {\em 17th Innovations in Theoretical Computer Science Conference, {ITCS}}, pages 16:1--16:23, 2026.

\bibitem[BY20]{pods/Ben-EliezerY20}
Omri Ben{-}Eliezer and Eylon Yogev.
\newblock The adversarial robustness of sampling.
\newblock In Dan Suciu, Yufei Tao, and Zhewei Wei, editors, {\em Proceedings of the 39th {ACM} {SIGMOD-SIGACT-SIGAI} Symposium on Principles of Database Systems, {PODS} 2020}, pages 49--62. {ACM}, 2020.

\bibitem[CEM{\etalchar{+}}15]{CohenEMMP15}
Michael~B. Cohen, Sam Elder, Cameron Musco, Christopher Musco, and Madalina Persu.
\newblock Dimensionality reduction for k-means clustering and low rank approximation.
\newblock In {\em Proceedings of the Forty-Seventh Annual {ACM} on Symposium on Theory of Computing, {STOC}}, pages 163--172, 2015.

\bibitem[CGS22]{ChakrabartiGS22}
Amit Chakrabarti, Prantar Ghosh, and Manuel Stoeckl.
\newblock Adversarially robust coloring for graph streams.
\newblock In {\em 13th Innovations in Theoretical Computer Science Conference, {ITCS}}, pages 37:1--37:23, 2022.

\bibitem[Che09]{siamcomp/Chen09}
Ke~Chen.
\newblock On coresets for $k$-median and $k$-means clustering in metric and {Euclidean} spaces and their applications.
\newblock {\em {SIAM} J. Comput.}, 39(3):923--947, 2009.

\bibitem[CKL22]{cohen2022johnson}
Vincent Cohen{-}Addad, C.~S. Karthik, and Euiwoong Lee.
\newblock The {Johnson} coverage hypothesis and geometric $k$-means and $k$-median hardness.
\newblock In {\em Proceedings of the 54th Annual ACM SIGACT Symposium on Theory of Computing (STOC)}, pages 1393--1406. ACM, 2022.

\bibitem[CLMS13]{soda/CzumajLMS13}
Artur Czumaj, Christiane Lammersen, Morteza Monemizadeh, and Christian Sohler.
\newblock $(1 + \epsilon)$-approximation for facility location in data streams.
\newblock In {\em {SODA}}, pages 1710--1728. {SIAM}, 2013.

\bibitem[CLN{\etalchar{+}}22]{CohenLNSSS22}
Edith Cohen, Xin Lyu, Jelani Nelson, Tam{\'{a}}s Sarl{\'{o}}s, Moshe Shechner, and Uri Stemmer.
\newblock On the robustness of countsketch to adaptive inputs.
\newblock In {\em International Conference on Machine Learning, {ICML}}, pages 4112--4140, 2022.

\bibitem[CLSS22]{stoc/Cohen-AddadLSS22}
Vincent Cohen{-}Addad, Kasper~Green Larsen, David Saulpic, and Chris Schwiegelshohn.
\newblock Towards optimal lower bounds for k-median and k-means coresets.
\newblock In {\em {STOC}}, pages 1038--1051. {ACM}, 2022.

\bibitem[CM06]{ciss/Cormode006}
Graham Cormode and S.~Muthukrishnan.
\newblock Combinatorial algorithms for compressed sensing.
\newblock In {\em {CISS}}, pages 198--201. {IEEE}, 2006.

\bibitem[CN20]{CherapanamjeriN20}
Yeshwanth Cherapanamjeri and Jelani Nelson.
\newblock On adaptive distance estimation.
\newblock In {\em Advances in Neural Information Processing Systems 33: Annual Conference on Neural Information Processing Systems, NeurIPS}, 2020.

\bibitem[CSS21]{stoc/Cohen-AddadSS21}
Vincent Cohen{-}Addad, David Saulpic, and Chris Schwiegelshohn.
\newblock A new coreset framework for clustering.
\newblock In {\em {STOC}}, pages 169--182. {ACM}, 2021.

\bibitem[CSW{\etalchar{+}}23]{CherapanamjeriSWZZZ23}
Yeshwanth Cherapanamjeri, Sandeep Silwal, David~P. Woodruff, Fred Zhang, Qiuyi Zhang, and Samson Zhou.
\newblock Robust algorithms on adaptive inputs from bounded adversaries.
\newblock In {\em The Eleventh International Conference on Learning Representations, {ICLR}}, 2023.

\bibitem[CW25]{CharikaW25}
Moses Charikar and Erik Waingarten.
\newblock The {Johnson-Lindenstrauss} lemma for clustering and subspace approximation: From coresets to dimension reduction.
\newblock In {\em Proceedings of the 2025 Annual {ACM-SIAM} Symposium on Discrete Algorithms, {SODA}}, pages 3172--3209, 2025.

\bibitem[CWZ23]{Cohen-AddadWZ23}
Vincent Cohen{-}Addad, David~P. Woodruff, and Samson Zhou.
\newblock Streaming {Euclidean} $k$-median and $k$-means with $o(\log n)$ space.
\newblock In {\em 64th {IEEE} Annual Symposium on Foundations of Computer Science, {FOCS}}, pages 883--908, 2023.

\bibitem[FIS08]{FrahlingIS08}
Gereon Frahling, Piotr Indyk, and Christian Sohler.
\newblock Sampling in dynamic data streams and applications.
\newblock {\em Int. J. Comput. Geom. Appl.}, 18(1/2):3--28, 2008.

\bibitem[FJW24]{FengJW24}
Ying Feng, Aayush Jain, and David~P. Woodruff.
\newblock Fast white-box adversarial streaming without a random oracle.
\newblock In {\em Forty-first International Conference on Machine Learning, {ICML}}, 2024.

\bibitem[FL11]{stoc/FeldmanL11}
Dan Feldman and Michael Langberg.
\newblock A unified framework for approximating and clustering data.
\newblock In {\em {STOC}}, pages 569--578. {ACM}, 2011.

\bibitem[FS05]{FrahlingS05}
Gereon Frahling and Christian Sohler.
\newblock Coresets in dynamic geometric data streams.
\newblock In {\em Proceedings of the 37th Annual {ACM} Symposium on Theory of Computing}, pages 209--217, 2005.

\bibitem[FSS20]{siamcomp/FeldmanSS20}
Dan Feldman, Melanie Schmidt, and Christian Sohler.
\newblock Turning big data into tiny data: Constant-size coresets for k-means, pca, and projective clustering.
\newblock {\em {SIAM} J. Comput.}, 49(3):601--657, 2020.

\bibitem[GKSY26]{GalKSY26}
Anna G{\'{a}}l, Gillat Kol, Raghuvansh~R. Saxena, and Huacheng Yu.
\newblock Optimal white-box adversarial streaming lower bounds for approximating {LIS} length.
\newblock In {\em 17th Innovations in Theoretical Computer Science Conference, {ITCS}}, pages 64:1--64:17, 2026.

\bibitem[GLW{\etalchar{+}}24]{GribelyukLWYZ24}
Elena Gribelyuk, Honghao Lin, David~P. Woodruff, Huacheng Yu, and Samson Zhou.
\newblock A strong separation for adversarially robust $l_0$ estimation for linear sketches.
\newblock In {\em 65th {IEEE} Annual Symposium on Foundations of Computer Science, {FOCS}}, pages 2318--2343, 2024.

\bibitem[GLW{\etalchar{+}}25]{GribelyukLWYZ25}
Elena Gribelyuk, Honghao Lin, David~P. Woodruff, Huacheng Yu, and Samson Zhou.
\newblock Lifting linear sketches: Optimal bounds and adversarial robustness.
\newblock In {\em Proceedings of the 57th Annual {ACM} Symposium on Theory of Computing, {STOC}}, pages 395--406, 2025.

\bibitem[GLW{\etalchar{+}}26a]{focs/GLWYZ26}
Elena Gribelyuk, Honghao Lin, David~P. Woodruff, Huacheng Yu, and Samson Zhou.
\newblock Adversarial robustness for small frequency moments and a weak equivalence theorem for turnstile streams.
\newblock In {\em {FOCS}}. {IEEE}, 2026.

\bibitem[GLW{\etalchar{+}}26b]{stoc/GLWYZ26}
Elena Gribelyuk, Honghao Lin, David~P. Woodruff, Huacheng Yu, and Samson Zhou.
\newblock Adversarial robustness on insertion-deletion streams.
\newblock In {\em Proceedings of the 57th Annual {ACM} Symposium on Theory of Computing, {STOC}}, pages 2278--2289, 2026.

\bibitem[GS02]{isr/GibbsS02}
Alison~L. Gibbs and Francis~Edward Su.
\newblock On choosing and bounding probability metrics.
\newblock {\em International Statistical Review}, 70(3):419--435, 2002.

\bibitem[GT90]{GoldbergTarjan1990}
Andrew~V. Goldberg and Robert~E. Tarjan.
\newblock Finding minimum-cost circulations by successive approximation.
\newblock {\em Mathematics of Operations Research}, 15(3):430--466, 1990.

\bibitem[Har11]{HP11}
Sariel Har{-}Peled.
\newblock {\em Geometric Approximation Algorithms}.
\newblock American Mathematical Society, USA, 2011.

\bibitem[HJKY25]{stoc/HuangJKY25}
Lingxiao Huang, Shaofeng~H.{-}C. Jiang, Robert Krauthgamer, and Di~Yue.
\newblock Near-optimal dimension reduction for facility location.
\newblock In {\em {STOC}}, pages 665--676. {ACM}, 2025.

\bibitem[HK07]{dcg/Har-PeledK07}
Sariel Har{-}Peled and Akash Kushal.
\newblock Smaller coresets for $k$-median and $k$-means clustering.
\newblock {\em Discret. Comput. Geom.}, 37(1):3--19, 2007.

\bibitem[HLW24]{stoc/Huang0024}
Lingxiao Huang, Jian Li, and Xuan Wu.
\newblock On optimal coreset construction for {Euclidean} $(k, z)$-clustering.
\newblock In {\em {STOC}}, pages 1594--1604. {ACM}, 2024.

\bibitem[HM04]{stoc/Har-PeledM04}
Sariel Har{-}Peled and Soham Mazumdar.
\newblock On coresets for k-means and k-median clustering.
\newblock In {\em {STOC}}, pages 291--300. {ACM}, 2004.

\bibitem[HW13]{stoc/HardtW13}
Moritz Hardt and David~P. Woodruff.
\newblock How robust are linear sketches to adaptive inputs?
\newblock In {\em {STOC}}, pages 121--130. {ACM}, 2013.

\bibitem[Ind04]{stoc/Indyk04}
Piotr Indyk.
\newblock Algorithms for dynamic geometric problems over data streams.
\newblock In {\em {STOC}}, pages 373--380. {ACM}, 2004.

\bibitem[ISZ21]{IzzoSZ21}
Zachary Izzo, Sandeep Silwal, and Samson Zhou.
\newblock Dimensionality reduction for wasserstein barycenter.
\newblock In {\em Advances in Neural Information Processing Systems 34: Annual Conference on Neural Information Processing Systems, NeurIPS}, pages 15582--15594, 2021.

\bibitem[JKS24]{compgeom/JiangKS24}
Shaofeng~H.{-}C. Jiang, Robert Krauthgamer, and Shay Sapir.
\newblock Moderate dimension reduction for k-center clustering.
\newblock In {\em 40th International Symposium on Computational Geometry, SoCG}, pages 64:1--64:16, 2024.

\bibitem[JKS{\etalchar{+}}26]{innovations/JiangKSSY26}
Shaofeng~H.{-}C. Jiang, Robert Krauthgamer, Shay Sapir, Sandeep Silwal, and Di~Yue.
\newblock Dimension reduction for clustering: The curious case of discrete centers.
\newblock In {\em 17th Innovations in Theoretical Computer Science Conference, {ITCS}}, pages 82:1--82:23, 2026.

\bibitem[JL84]{JL84}
William Johnson and Joram Lindenstrauss.
\newblock Extensions of {Lipschitz} maps into a {Hilbert} space.
\newblock {\em Contemporary Mathematics}, 26:189--206, 01 1984.

\bibitem[JW21]{siamcomp/JayaramW21}
Rajesh Jayaram and David~P. Woodruff.
\newblock Perfect l\({}_{\mbox{p}}\) sampling in a data stream.
\newblock {\em {SIAM} J. Comput.}, 50(2):382--439, 2021.

\bibitem[KC22]{kuchibhotla2022moving}
Arun~Kumar Kuchibhotla and Abhishek Chakrabortty.
\newblock Moving beyond sub-gaussianity in high-dimensional statistics: Applications in covariance estimation and linear regression.
\newblock {\em Information and Inference: A Journal of the IMA}, 11(4):1389--1456, 2022.

\bibitem[KK12]{KiralyKovacs2012}
Zolt{\'a}n Kir{\'a}ly and P{\'e}ter Kov{\'a}cs.
\newblock Efficient implementations of minimum-cost flow algorithms.
\newblock {\em Acta Universitatis Sapientiae, Informatica}, 4(1):67--118, 2012.

\bibitem[KNW10]{soda/KaneNW10}
Daniel~M. Kane, Jelani Nelson, and David~P. Woodruff.
\newblock On the exact space complexity of sketching and streaming small norms.
\newblock In {\em {SODA}}, pages 1161--1178. {SIAM}, 2010.

\bibitem[LSW{\etalchar{+}}26]{LinSWXZ26}
Honghao Lin, Zhao Song, David~P. Woodruff, Shenghao Xie, and Samson Zhou.
\newblock $l_p$ sampling in distributed data streams with applications to adversarial robustness.
\newblock In {\em Proceedings of the 2026 Annual {ACM-SIAM} Symposium on Discrete Algorithms, {SODA}}, pages 4342--4409, 2026.

\bibitem[MBN{\etalchar{+}}17]{MitrovicBNTC17}
Slobodan Mitrovic, Ilija Bogunovic, Ashkan Norouzi{-}Fard, Jakub Tarnawski, and Volkan Cevher.
\newblock Streaming robust submodular maximization: {A} partitioned thresholding approach.
\newblock In {\em Advances in Neural Information Processing Systems 30: Annual Conference on Neural Information Processing Systems}, pages 4557--4566, 2017.

\bibitem[MMR19]{siamcomp/MakarychevMR19}
Konstantin Makarychev, Yury Makarychev, and Ilya~P. Razenshteyn.
\newblock Performance of {Johnson-Lindenstrauss} transform for $k$-means and $k$-medians clustering.
\newblock {\em {SIAM} J. Comput.}, 52(on):269--297, 2019.

\bibitem[Nis92]{combinatorica/Nisan92}
Noam Nisan.
\newblock Pseudorandom generators for space-bounded computation.
\newblock {\em Comb.}, 12(4):449--461, 1992.

\bibitem[NSIZ21]{NarayananSIZ21}
Shyam Narayanan, Sandeep Silwal, Piotr Indyk, and Or~Zamir.
\newblock Randomized dimensionality reduction for facility location and single-linkage clustering.
\newblock In {\em Proceedings of the 38th International Conference on Machine Learning, {ICML}}, pages 7948--7957, 2021.

\bibitem[PR23]{PengR23}
Binghui Peng and Aviad Rubinstein.
\newblock Near optimal memory-regret tradeoff for online learning.
\newblock In {\em 64th {IEEE} Annual Symposium on Foundations of Computer Science, {FOCS}}, pages 1171--1194, 2023.

\bibitem[SWZ25]{focs/SwartworthWZ25}
William Swartworth, David~P. Woodruff, and Samson Zhou.
\newblock Perfect {$L_p$} sampling with polylogarithmic update time.
\newblock In {\em 66th {IEEE} Annual Symposium on Foundations of Computer Science, {FOCS}}, pages 1936--1960, 2025.

\bibitem[SYZ18]{SongYZ18}
Zhao Song, Lin~F. Yang, and Peilin Zhong.
\newblock Nearly optimal dynamic $k$-means clustering for high-dimensional data.
\newblock {\em CoRR}, abs/1802.00459, 2018.

\bibitem[Vil03]{books/Villani03}
C{\'e}dric Villani.
\newblock {\em Topics in Optimal Transportation}, volume~58 of {\em Graduate Studies in Mathematics}.
\newblock American Mathematical Society, 2003.

\bibitem[Waj20]{Wajc20}
David Wajc.
\newblock Rounding dynamic matchings against an adaptive adversary.
\newblock In {\em Proccedings of the 52nd Annual {ACM} {SIGACT} Symposium on Theory of Computing, {STOC}}, pages 194--207, 2020.

\bibitem[WZ24]{WoodruffZ24}
David~P. Woodruff and Samson Zhou.
\newblock Adversarially robust dense-sparse tradeoffs via heavy-hitters.
\newblock In {\em Advances in Neural Information Processing Systems 38: Annual Conference on Neural Information Processing Systems, NeurIPS}, 2024.

\bibitem[WZ26]{WoodruffZ26}
David~P. Woodruff and Samson Zhou.
\newblock The adversarial robustness of sketching and streaming algorithms.
\newblock {\em CoRR}, abs/2607.14432, 2026.

\bibitem[WZZ23]{WoodruffZZ23b}
David~P. Woodruff, Fred Zhang, and Samson Zhou.
\newblock On robust streaming for learning with experts: Algorithms and lower bounds.
\newblock In {\em Advances in Neural Information Processing Systems 36: Annual Conference on Neural Information Processing Systems, NeurIPS}, 2023.

\bibitem[ZTHH25]{tit/ZhuTHH25}
Xiaoyi Zhu, Yuxiang Tian, Lingxiao Huang, and Zengfeng Huang.
\newblock Space complexity of {Euclidean} clustering.
\newblock {\em {IEEE} Trans. Inf. Theory}, 71(6):4515--4536, 2025.

\bibitem[ZW22]{zhang2022sharper}
Huiming Zhang and Haoyu Wei.
\newblock Sharper sub-weibull concentrations.
\newblock {\em Mathematics}, 10(13):2252, 2022.

\end{thebibliography}
\newcommand{\etalchar}[1]{$^{#1}$}

\appendix

\section{Proof of \texorpdfstring{\Cref{lemma:inner_prod_non}}{Lemma}}
\label{sec:proof_inner_prod_non}

\lemmainnerprodnon*

\begin{proof}
Fix an oblivious stream and a query $g\in[-1,1]^U$ independently of the algorithm's randomness.
We prove that the required accuracy holds simultaneously at all times for this $g$.
The algorithm itself does not depend on $g$.

We use the $\ell_1$ sampler of~\cite{siamcomp/JayaramW21}, including its estimate of the signed frequency of the sampled coordinate.
For a fixed nonzero vector $x$, the sampler returns $\bot$ with probability at most $1/4$.
Conditioned on not returning $\bot$, it returns a pair $(i,\sigma)$, where $\sigma\in\{-1,1\}$, whose distribution is within $\varepsilon/32$ in total variation of the ideal distribution
\[
 \Pr[i=j]=\frac{|x_j|}{\|x\|_1},
 \qquad \sigma=\sgn(x_i).
\]
To obtain this guarantee, use the sampler's frequency estimate with relative error $1/2$, which determines the sign whenever it is accurate.
Its distributional error and the probability of an inaccurate frequency estimate can both be made inverse-polynomially small.
Zero-padding the universe to size $\operatorname{poly}(m|U|r^+/\varepsilon)$ makes their combined total variation error at most $\varepsilon/32$.
We extend $g$ by zero on the padded coordinates.
Each copy uses $\polylog(m|U|r^+/\varepsilon)$ space.

Let
\[
 s:= \Theta\left( C\varepsilon^{-2}\log\frac{2m}{\delta}\right),
\]
where $C>0$ is a sufficiently large absolute constant.
Maintain $4s$ independent copies of this sampler, and send every update to all copies.
After each update, take the first $s$ copies that return a pair rather than $\bot$, and write their outputs as $(i_r,\sigma_r)$ for $r\in[s]$.
If there are at least $s$ such copies, define
\[
 \widehat q(g):=\frac1s\sum_{r=1}^s\sigma_r g(i_r).
\]
Otherwise, set $\widehat q(g):=0$ for every query $g$.
In particular, $|\widehat q(g)|\le1$ always holds for $g\in[-1,1]^U$.

Independently, maintain a standard non-adaptive $\ell_1$-norm sketch~\cite{soda/KaneNW10} with relative error $\varepsilon/4$ and failure probability $\delta/(2m)$ at each fixed time.
Writing $L=\|x\|_1$, let its output be $\widehat L$, and define
\[
 \widehat I(g):=\widehat L\,\widehat q(g).
\]
We use a linear norm sketch whose decoder returns zero on the zero sketch.
Consequently, when $x=0$, we have $\widehat L=0$ and the answer is exactly zero.

The successful copies are selected independently of $g$, and the same samples and scalar $\widehat L$ are used for every query.
Thus $\widehat I(g)=\sum_{r=1}^s(\widehat L\sigma_r/s)g(i_r)$ is a linear functional of $g$ for every stored state.
If fewer than $s$ copies succeed, it is the zero functional.
This property holds even when an accuracy event fails.

We next verify the accuracy at a fixed time with $L>0$.
For one sampler conditioned on returning a pair, put $Y=\sigma g_i$.
Under ideal signed $\ell_1$ sampling its expectation is
\[
 \sum_{i\in U}\frac{|x_i|}{L}\sgn(x_i)g_i
 =\frac{\langle x,g\rangle}{L}.
\]
Since $|Y|\le1$, the total variation guarantee changes this expectation by at most $\varepsilon/16$.
Thus, writing $q(g)=\langle x,g\rangle/L$,
\[
 |\E Y-q(g)|\le\varepsilon/16,
 \qquad \operatorname{var}(Y)\le1.
\]

Each of the $4s$ copies returns a pair independently with probability at least $3/4$.
A Chernoff bound shows that fewer than $s$ copies return a pair with probability at most $e^{-\Omega(s)}\le\delta/(4m)$.
Conditional on any pattern of which copies return pairs, their returned pairs are independent and have the same conditional distribution described above.
Hence, conditional on obtaining at least $s$ pairs, Hoeffding's inequality gives
\[
 \Pr\!\left[
 \left|\widehat q(g)-\E Y\right|>\varepsilon/8
 \,\middle|\,\text{at least $s$ pairs}
 \right]
 \le2e^{-s\varepsilon^2/128}
 \le\frac{\delta}{4m}.
\]
Together with the bias bound, this implies
\[
 \Pr\!\left[|\widehat q(g)-q(g)|>\varepsilon/4\right]
 \le\frac{\delta}{2m}.
\]

Also, $|\widehat L-L|\le(\varepsilon/4)L$ except with probability $\delta/(2m)$.
When both guarantees hold, we obtain
\[
 \begin{aligned}
 |\widehat I(g)-\langle x,g\rangle|
 &\le |\widehat L-L|\,|\widehat q(g)|
       +L|\widehat q(g)-q(g)|\\
 &\le\frac{\varepsilon}{4}L+\frac{\varepsilon}{4}L
 \le\varepsilon\|x\|_1.
 \end{aligned}
\]
Thus the failure probability at each fixed time is at most $\delta/m$.
Since the stream is oblivious, each prefix is independent of the sketches' randomness, and a union bound over the $m$ times proves the claimed simultaneous guarantee for the fixed $g$.
This union bound does not require independence between different times.

Finally, there are $4s=O(\varepsilon^{-2}\log(2m/\delta))$ sampler copies, the same asymptotic number as before.
The norm sketch, amplified by independent repetition, fits within the same total space bound.
Consequently, the space is
\[
 O\!\left(\varepsilon^{-2}\log\frac{m}{\delta}\right)
 \cdot\polylog\!\left(\frac{m|U|r^+}{\varepsilon}\right),
\]
using the usual convention that logarithms in complexity bounds are bounded below by one.

The maintained samples and norm estimate are shared by all queries; evaluating $\widehat I(g)$ only requires the coordinates of $g$ at sampled indices.
Their single average is evaluated directly, so no table of query answers is stored.
\end{proof}

\end{document}